\documentclass[preprint, times]{elsarticle}

\usepackage{amsthm}
\usepackage{mathrsfs}
\usepackage{amsmath}
\usepackage{jcomp}
\usepackage{framed,multirow}
\usepackage{amssymb}
\usepackage{latexsym}
\usepackage{subfigure}
\usepackage{caption}
\usepackage{url}
\usepackage{xcolor}
\usepackage{appendix}
\definecolor{newcolor}{rgb}{.8,.349,.1}

\journal{Journal of Computational Physics}

\begin{document}

\verso{Xinwei Cai \textit{et al}}

\begin{frontmatter}

\title{Arbitrary slip length for fluid-solid interface of arbitrary geometry in smoothed particle dynamics}%

\author[1]{Xinwei \snm{Cai}\fnref{fn1}}
\ead{xcai@zju.edu.cn}
\author[2]{Zhen \snm{Li}}
\ead{zli7@clemson.edu}
\author[1]{Xin \snm{Bian}\corref{cor1}}
\cortext[cor1]{Corresponding author}
\ead{bianx@zju.edu.cn}

\address[1]{State Key Laboratory of Fluid Power and Mechatronic Systems, Zhejiang University, Hangzhou 310027, P. R. China}
\address[2]{Department of Mechanical Engineering, Clemson University, Clemson, SC 29634, USA}

\received{XX XX 2023}
\finalform{XX XX 2023}
\accepted{XX XX 2023}
\availableonline{XX XX 2023}
\communicated{editor prof.}

\begin{abstract}
We model a slip boundary condition at fluid-solid interface of an arbitrary geometry in
smoothed particle hydrodynamics and smoothed dissipative particle dynamics simulations.  Under an assumption of linear profile of the tangential velocity at quasi-steady state near the interface, an arbitrary slip length $b$ can be specified and correspondingly, an artificial velocity for every boundary particle can be calculated. Therefore,  $b$ as an input parameter affects the calculation of dissipative and random forces near the interface. For $b \to 0$, the no-slip is recovered while for $b \to \infty$, the free-slip is achieved. Technically, we devise two different approaches to calculate the artificial velocity of any boundary particle. The first has a succinct principle and is competent for simple geometries, while the second is subtle and affordable for complex geometries. Slip lengths in simulations for both steady and transient flows coincide with the expected ones. As demonstration, we apply the two approaches extensively to simulate curvy channel flows, dynamics of an ellipsoid in pipe flow and flows within complex microvessels, where desired slip lengths at fluid-solid interfaces are prescribed.
The proposed methodology may apply equally well to other particle methods such as dissipative particle dynamics and moving particle semi-implicit methods.
\end{abstract}


\end{frontmatter}


\section{Introduction}
For more than two centuries, the no-slip hypothesis at fluid-solid interface has been often accepted, especially for continuum flows at macroscale~\cite{landau1959}. In many cases, however, the no-slip is no longer valid and debates on how to specify a correct boundary condition filled the whole 20th-century~\cite{neto2005boundary}. As a matter of fact, a slip is ubiquitous at interfaces in nature. For example, the special surface of lotus leaves makes them water-repellent, which not only allows droplets to slide effortlessly but also reduces significantly adhesion of contaminated particles~\cite{barthlott1997purity}; Aphids exploit hydrophobic wax coated on their excreted honeydew for protection and transport~\cite{pike2002aphids}; Shark skins have drag-reducing properties~\cite{reif1982hydrodynamics}, which are leveraged for artificial materials~\cite{oeffner2012hydrodynamic}. At meso-/micro-scale, a slip is even more common. For example, flows in hydrophobic capillaries~\cite{rothstein2010slip} and ultrafast transport of water in carbon nanotubes~\cite{majumder2005enhanced, holt2006fast, papadopoulou_nanopumps_2022} exhibit apparent slip at interfaces. Above examples no longer respect the no-slip condition at the fluid-solid interface. Evidently, a slip is normal and the no-slip is only a special case often adopted for convenience.

\begin{figure}[h!]
    \centering \includegraphics[width=70mm]{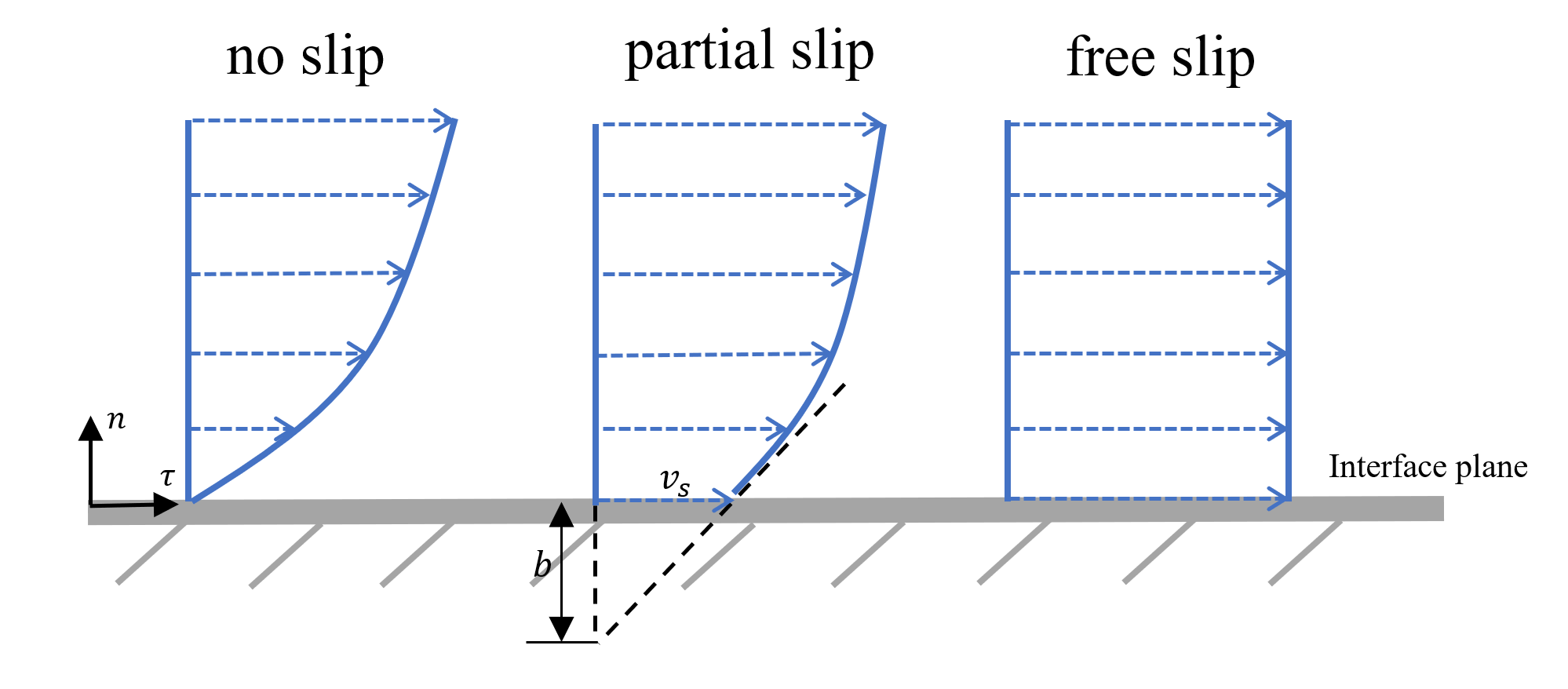} \caption{Schematic of velocity field near fluid-solid interface: slip lengths are $b=0$ for no-slip, $b>0$ for partial-slip, and $b=\infty$ for free-slip.}\label{boundary_classification}
\end{figure}

A linear slip boundary condition was first proposed by Navier~\cite{navier1823memoire} and augmented by Maxwell with a slip length~\cite{maxwell1879vii}. This was rigorously proved by Tolstoi~\cite{tolstoi1954molecular} and further by Blake~\cite{blake1990slip}. In Navier's assumption, a slip velocity is defined at the fluid-solid interface as
\begin{equation}
    {\bf v}_s=b\frac{\partial \mathbf{v}} {\partial \mathbf{n}},
\end{equation}
where $b$ is the slip length, $\mathbf{v}$ is the velocity of the fluid, and $\mathbf{n}$ is the unit normal at the interface. Fig.~\ref{boundary_classification} sketches three different slip lengths at the interface, where $b=0$ and $b\to \infty$ are two special cases corresponding to the no-slip and free-slip, respectively.
The slip boundary has seen further developments in this century.
Ou et al.~\cite{ou2004laminar} demonstrated in experiment that superhydrophobic surfaces reduce flow resistance by up to $40\%$ of pressure drop in laminar flow; Kamrin et al.~\cite{kamrin2010effective} derived a general expression for slip boundaries with periodic fluctuations of height on the surface in shear flow; Zampogna et al.~\cite{zampogna2019generalized} presented a generalized slip expression over rough surfaces in turbulent flows. Beyond linear theories, nonlinear analyses on the slip boundary have also been developed in the last decade. Sander et al.~\cite{sanders2006bubble} proposed an exhaust slip boundary in place of a continuous layer of actively released bubbles at the outer edge of the boundary layer; Inspired by the water-repellent property of salvinia leaves, Xiang et al.~\cite{xiang2020superrepellency} imitated a slip surface containing a continuous air mattress; Bottaro~\cite{bottaro2019flow} proposed a concept of deformable surface, which consists of linearly elastic material placed periodically on a rigid substrate.

The rapid development of numerical simulations allows one to observe phenomena due to slip boundaries readily. At micro-/nano-scale, there has been a large number of simulations by molecular dynamics manifesting a slip boundary~\cite{koplik1998continuum,priezjev2009shear,papadopoulou_nanopumps_2022}. For macro-/meso-scopic flow problems, lattice Boltzmann and smoothed particle dynamics (SPD) are two popular methods. The former is mesh-based and an implementation of slip boundary is straightforward, which is evidenced by several relevant works~\cite{verhaeghe2009lattice,sundin2019interaction,yang2020analysis}. However, for mesh-less methods such as SPD,  it remains obscure how to model an arbitrary slip length, especially at an interface of arbitrary geometry.

SPD represents either smoothed particle hydrodynamics (SPH) or smoothed dissipative particle dynamics (SDPD) for solving macroscopic or mesoscopic flow problems, respectively. 
Its Lagrangian feature is advantageous to deal with complex interfaces of dynamic nature, which involve coupling, interface tracking and moving boundary.
SPH was originally intended to simulate phenomena in astrophysics and has since then been largely extended to flow problems of various kinds ~\cite{monaghan_smoothed_2005, price2012smoothed, zhang2017, ye2019smoothed, zhang2022smoothed}. 
SDPD was proposed by Espa{\~n}ol and Revenga~\cite{espanol2003smoothed} by introducing stochastic forces into SPH under the GENERIC framework of thermodynamics~\cite{Grmela1997} so that it is an effective solver for the Landau-Lifshitz-Navier-Stokes equations~\cite{landau1959, bian2015fluctuating, bian2016pre, Bian2018, ellero2018everything}. 
It has been extensively applied to study physics of various mesoscopic flows~\cite{hu_multi-phase_2006, litvinov2008smoothed, Vazquez-Quesada2009a, bian_multiscale_2012, Lei2015, Muller2015a, bian2015fluctuating, bian2016pre, Ye2017}.
So far, there has been many striving efforts for a sensible implementation of no-slip or arbitrary flow boundary condition in SPD~\cite{takeda1994numerical, morris_modeling_1997, Monaghan2009, Litvinov2010a, bian_multiscale_2012, adami_generalized_2012, Bian2015jcp, marrone2013accurate, Valizadeh2015, Moreno2021}.  For a slip boundary, Hu et al.~\cite{hu_multi-phase_2006} mentioned briefly an implementation for flow over a flat plate; Pan et al.~\cite{pan_smoothed_2014} proposed a Robin boundary condition for planar and circular interfaces. 

In this work, we propose a slip boundary condition with arbitrary slip length at an interface of arbitrary geometry in both SPH and SDPD methods. It is characterized by only one input parameter, that is, the slip length $b$, which also effectively equals to the output one in simulations. We devise two novel algorithms to achieve the desired slip length: the first one is an extension of Morris et al.'s interpolation method between a pair of interacting fluid particle and boundary particle~\cite{morris_modeling_1997}, 
while the second one is based on Adami et al.'s summation method by employing the average effects of neighboring fluid particles~\cite{adami_generalized_2012}.
The structure of the following parts is as follows. 
In Section \ref{section_method},  we present the equations of fluid mechanics in
both continuum and discrete forms. 
In Section \ref{section_slip}, we describe the two key algorithms.
In Section \ref{section_numerics}, we simulate multiple flow problems by SPD, which demonstrate the reliability of the proposed two algorithms.
Finally in Section \ref{section_conclusion}, we summarize this work.

\section{The method}
\label{section_method}
\subsection{Lagrangian hydrodynamic equations and its boundary conditions}
We consider an isothermal Newtonian fluid governed by the continuity and Navier-Stokes equations
in Lagrangian form as follows
\begin{eqnarray}
   \frac{{\mathrm{d}}\rho}{{\mathrm{d}} t} & =&  -\rho \nabla \cdot \mathbf{v},\label{continuity_eq} \\
    \rho \frac{{\mathrm{d}}\mathbf{v}}{{\mathrm{d}}t} &=& -\nabla p + \eta \nabla^2\mathbf{v}+\frac{\eta}{3}\nabla\nabla\cdot\mathbf{v} + \rho\mathbf{f}, \label{ns_eq}
\end{eqnarray}
where $\rho$, $\mathbf{v}$, $p$, $\eta$, and $\mathbf{f}$ are material density, velocity, pressure, dynamic viscosity and body force per unit mass, respectively.
An equation of state (EOS) relating the pressure to the density is necessary to provide a closure for
a weakly compressible description. Both an perfect-gas-like EOS and liquid-like stiff
EOS are widely used in the literature~\citep{morris_modeling_1997, monaghan_smoothed_2005}, 
and they can be expressed in a general form:
\begin{equation}
p=\frac{c_{s}^{2} \rho_{0}}{\gamma}\left[\left(\frac{\rho}{\rho_{0}}\right)^{\gamma}-1\right]+\chi \label{eoc_eq},
\end{equation}
where $\rho_0$ is the equilibrium density. An artificial sound speed $c_s$ is chosen based on a scale analysis~\citep{morris_modeling_1997,monaghan_smoothed_2005} such that the pressure field reacts strongly to small deviations in the density, and therefore a quasi-incompressibility is fulfilled. 
In this case, the last second term in Eq.~(\ref{ns_eq}) may be negligible.
Here, $\chi$ as a positive constant is introduced to enforce the non-negativity of pressure on discrete particles.

We characterize the slip boundary by a slip length $b$, namely the Navier slip length, at which the velocity profile linearly extrapolated to be zero inside the solid,  as shown in Fig.~\ref{boundary_classification}.  
The shear stress at the fluid-solid interface can be described as
\begin{eqnarray}
\sigma^{\tau n} = \kappa v_{s}\label{sur_friction},
\end{eqnarray}
where $\kappa$ is the surface friction coefficient and $v
_{s}=|{\bf v}_s|$ is the magnitude of slip velocity or relative velocity between the fluid and solid.  Assuming a Newtonian fluid we further have
\begin{eqnarray}
\sigma^{\tau n} = \eta \frac{\partial v^{\tau} }{\partial n}\label{shear_vis}.
\end{eqnarray}
Due to the assumption of the linear velocity profile of $v^\tau$ inside the solid near the interface,
combining Eq.~(\ref{sur_friction}) and Eq.~(\ref{shear_vis}), we get
\begin{eqnarray}
v_{s}=\frac{\eta}{\kappa} \frac{\partial v^{\tau}}{\partial n} = b \frac{\partial v^{\tau}}{\partial n}\label{v_slip},
\end{eqnarray}
where slip length $b$ is equivalent to the ratio of the dynamic viscosity and surface friction coefficient.
This is the key relation, upon which two numerical algorithms are built.
Gradient of velocity in the normal direction at the interface vanishes $\partial v^n / \partial n=0$, corresponding to an impermeable boundary condition,
which should also be appreciated in the numerical algorithms.
Although we elucidate the boundary condition in two dimensions, its extension to three dimensions is trivial.

\subsection{Smoothed particle dynamics}
For comprehensive descriptions of SPH/SDPD, we refer to recent reviews articles~\citep{monaghan_smoothed_2005, price2012smoothed, ellero2018everything}. 
For convenience, we define some simple notations as reference
\begin{eqnarray}
        \mathbf{r}_{i j} &=&\mathbf{r}_{i}-\mathbf{r}_{j}, \\
        \mathbf{v}_{i j} &=&\mathbf{v}_{i}-\mathbf{v}_{j}, \\
        \mathbf{e}_{i j} &=&\mathbf{r}_{i j} / r_{i j}, \quad r_{i j}=\left|\mathbf{r}_{i j}\right|,
\end{eqnarray}
where $\mathbf{r}_i$, $\mathbf{v}_i$ are position and velocity of particle i; 
$\mathbf{r}_{ij}$, $\mathbf{v}_{ij}$ are relative position and velocity of particles $i$ and $j$; $r_{ij}$ is the distance of the two and $\mathbf{e}_{ij}$ is the unit vector pointing from j to i. 
Each particle's position is updated according to
\begin{eqnarray}
        \frac{\mathrm{d}\mathbf{r}_i}{dt} = \mathbf{v}_i. \label{position_eq}
\end{eqnarray}
The density field is computed as~\citep{espanol2003smoothed}
\begin{equation}
    \begin{aligned}
        \sigma_{i} &=\frac{\rho_{i}}{m_{i}}=\sum_{j} W\left(r_{i j}\right)=\sum_{j} W_{i j}\label{sigma_eq},
    \end{aligned}
\end{equation}
where $\sigma$ is number density defined as the ratio of $\rho$ and particle mass $m$ (constant).
Note that the density summation in Eq.~(\ref{sigma_eq}) together with the position update in Eq.~(\ref{position_eq}) already account for the continuity equation in Eq.~(\ref{continuity_eq}), which does not need to be discretized separately~\cite{espanol2003smoothed}.
The weight function $W(r)$, also known as kernel, has at least two properties:
\begin{equation}
    \begin{aligned}
    \lim _{h \rightarrow 0} W\left(\mathbf{r}-\mathbf{r}^{\prime}, h\right)=\delta\left(\mathbf{r}-\mathbf{r}^{\prime}\right), \quad \int W\left(\mathbf{r}-\mathbf{r}^{\prime}, h\right) d \mathbf{r}^{\prime}=1,
\end{aligned}
\end{equation}
where $h$ is quoted as smoothing length.
This indicates that any kernel adopted should converge to the Dirac delta function $\delta$ as $h\to0$ and its integral must be normalized.  To balance the computational efficiency and accuracy, a finite support domain described by a cutoff radius $r_c$ is usually adopted. 
When two particles' distance is larger than $r_c$, $W(r_{ij} \geq r_c,h)=0$ and there is no direct contribution to each other's dynamics. 
In this work we adopt the quintic spline function with $r_c = 3h$, which has been proven to be accurate~\citep{morris_modeling_1997}:
\begin{equation}
    \begin{aligned}
W(s, h)=C_D\frac{1}{h^D}\left\{
    \begin{array}{ll}
(3-s)^{5}-6(2-s)^{5}+15(1-s)^{5}, & 0 \leq s<1 ; \\
(3-s)^{5}-6(2-s)^{5}, & 1 \leq s<2 ; \\
(3-s)^{5}, & 2 \leq s<3 ; \\
0, & s \geq 3 .
    \end{array}\right.
\end{aligned}
\end{equation}
Here $s = {r_{ij}}{/h}$ and $D$ is the number of dimension. The normalization coefficients are $C_2 = 7/478\pi$, and $C_3 = 3/359\pi$ in two and three dimensions, respectively.

The momentum equation of every particle in SPD can be expressed succinctly as follows
\begin{eqnarray}
    m_i\frac{{\mathrm{d}}\mathbf{v}_i}{{\mathrm{d}}t} = \sum_j(\mathbf{F}_{ij}^C + \mathbf{F}_{ij}^D + \mathbf{F}_{ij}^A + \mathbf{F}_{ij}^R) + \mathbf{F}^B.
\label{SPD_eq}
\end{eqnarray}
Here $\mathbf{F}_{ij}^C$ and $\mathbf{F}_{ij}^D$ are conservative and dissipative forces between a pair of neighbouring particles, the sum of which correspond to a discretization of forces due to pressure and viscous stress in the Navier-Stokes equations in Eq.~(\ref{ns_eq}).
In this work, we shall consider two sets of formulations.
The first set is as follows~\citep{hu_multi-phase_2006}
\begin{eqnarray}
        \mathbf{F}_{ij}^{C1} &=& -(\frac{1}{\sigma_i^2} + \frac{1}{\sigma_j^2})\frac{\rho_jp_i+\rho_ip_j}{\rho_i+\rho_j}\frac{\partial{W}}{\partial{r_{ij}}}\mathbf{e}_{ij}, \label{sph_conservative_eq}\\
        \mathbf{F}_{ij}^{D1} &=& (\frac{1}{\sigma_i^2}+\frac{1}{\sigma_j^2})\frac{2\eta_i\eta_j}{\eta_i+\eta_j}\frac{\mathbf{v}_{ij}}{r_{ij}}\frac{\partial{W}}{\partial{r_{ij}}}\label{sph_dissipative_eq}.
\end{eqnarray}
Here the particle-averaged pressure and viscosity are employed,
which are suitable for handling multiphase problems.
In addition, we shall consider another set of formulations,
which respect the $1st$ and $2nd$ laws of thermodynamics explicitly~\cite{espanol2003smoothed}
\begin{eqnarray}
            \mathbf{F}_{ij}^{C2} &=& -\left(\frac{p_i}{\sigma_i^2} + \frac{p_j}{\sigma_j^2}\right)\frac{\partial{W}}{\partial{r_{ij}}}\mathbf{e}_{ij}\label{sdpd_cons}, \\
            \mathbf{F}_{ij}^{D2} &=& \frac{\eta}{\sigma_i \sigma_j r_{ij}}\frac{\partial{W}}{\partial{r_{ij}}}\left(\frac{2D-1}{D}\mathbf{v}_{ij}+\frac{D+2}{D}\mathbf{e}_{ij}\cdot\mathbf{v}_{ij}\mathbf{e}_{ij}\right)\label{sdpd_dissipative_eq}.
\end{eqnarray}

To minimize the so-called tensile instability, we adopt the transport-velocity formulation from Adami et al.~\citep{adami_transport-velocity_2013}, which is one of the particle shifting techniques originated from XSPH \cite{monaghan_smoothed_2005}. 
Its spirit is to adjust dynamically the irregular distributions of particles so that numerical errors
are diminished.
During this process, an additional force $\mathbf{F}_{ij}^A$ is generated as follows
\begin{eqnarray}
        \mathbf{F}_{ij}^{A1} &=& \frac{1}{2}(\frac{1}{\sigma_i^2}+\frac{1}{\sigma_j^2})(\mathbf{A}_i+\mathbf{A}_j)\cdot\frac{\partial{W}}{\partial{r_{ij}}}\mathbf{e}_{ij},\label{addition_eq}
\end{eqnarray}
where $\mathbf{A} = \rho\mathbf{v}(\Tilde{\mathbf{v}} - \mathbf{v})$ is a tensor from the dyadic product of the two vectors.
Moreover, $\mathbf{v}$ is the velocity for the momentum and force calculations,
while $\mathbf{\Tilde{v}}$ is the modified transport velocity utilized to update the position of each particle as follows,
\begin{eqnarray}
\tilde{\mathbf{v}}_i &=& \mathbf{v}_i(t) + \delta t(\frac{d\mathbf{v}_i}{dt}-\frac{1}{\rho_i}\nabla \chi),\label{transport_eq} \\
\frac{{\mathrm{d}}\mathbf{r}}{{\mathrm{d}}t} &=& \mathbf{\tilde{v}}_i.
\end{eqnarray}
Here, $\delta t$ is the time step, the constant pressure $\chi$ appears only in Eq.~(\ref{transport_eq}) instead of momentum equation. 
Since the transport velocity $\mathbf{\tilde{v}}$ is only used in the position evolution, linear momentum is consequently strictly conserved. 
Following the same discrete form in Eq.~(\ref{sph_conservative_eq}),
Eq.~(\ref{transport_eq}) unfolds as follows
\begin{equation}
    \tilde{\mathbf{v}}_i = \mathbf{v}_i(t) + \delta t(\frac{d\mathbf{v}_i}{dt}-\frac{\chi}{m_i}\sum_j(\frac{1}{\sigma_i^2}+\frac{1}{\sigma_j^2})\frac{\partial{W}}{\partial{r_{ij}}}\mathbf{e}_{ij}).
\end{equation}

At mesoscale, the molecular entities and their incessant movements manifest themselves as random stresses
in the fluid dynamics equations. 
To have a local thermodynamic equilibrium, the pair of random stress and dissipative stress
are inherently related and must respect the fluctuation-dissipation theorem. 
In a discrete setting, for a given expression of the dissipative force $\mathbf{F}^D_{ij}$, 
we may resort to the GENERIC framework to obtain the corresponding random force $\mathbf{F}^R_{ij}$.
For example, given $\mathbf{F}^{D2}_{ij}$ in Eq.~(\ref{sdpd_dissipative_eq}), the rational expression of the random force is as follows
\begin{equation}
    \mathbf{F}_{ij}^{R2}=\left(-\frac{20\eta}{3}\frac{k_BT}{\sigma_i\sigma_jr_{ij}}\frac{\partial{W}}{\partial{r_{ij}}}\right)^{1/2}d\overline{\mathscr{W}}_{ij}\cdot \mathbf{e}_{ij},\label{frandom_eqn}
\end{equation}
where $d\mathscr{W}$ is a matrix of independent increments of the Wiener process, and $d\overline{\mathscr{W}}$ is the symmetric part of it
\begin{equation}
    d\overline{\mathscr{W}}_{ij} = (d\mathscr{W}_{ij}+d\mathscr{W}_{ij}^T)/2.
\end{equation}
Furthermore, the following symmetry between particles $i$ and $j$ is preserved
\begin{equation}
    d\mathscr{W}_{ij} = d\mathscr{W}_{ji}.
\end{equation}
The independent increments of the Wiener process satisfy the following mnemotechnical $\mathrm{It\hat{o}}$ rules
\begin{equation}
    d\mathscr{W}_{ij}^{\alpha \beta}d\mathscr{W}_{kl}^{\kappa \lambda} = [\delta_{ik}\delta_{jl}+\delta_{il}\delta_{jk}]\delta^{\alpha \kappa}\delta^{\beta \lambda}dt.
\end{equation}

There are various other formulations for the pairwise forces in the literature 
and different combinations also exist~\citep{monaghan_smoothed_2005, price2012smoothed, ellero2018everything}.
As we shall notice, however, the proposed boundary condition is not restricted to any particular force formulation. Therefore, we consider only two sets of them given above.
More specifically, $\mathbf{F}^{C1}_{ij}$, $\mathbf{F}^{D1}_{ij}$, $\mathbf{F}^{A1}_{ij}$
are employed for SPH simulations;
$\mathbf{F}^{C2}_{ij}$, $\mathbf{F}^{D2}_{ij}$, $\mathbf{F}^{R2}_{ij}$ 
are utilized for SDPD simulations.
It is apparent that the transport-velocity formulation with $\mathbf{F}^{A}_{ij}$ is only applicative in SPH, but not in SDPD, as the latter has already random forces to redistribute particles. 
$\mathbf{F}^B$ is any body force such as force due to gravity,
which is present whenever necessary for both SPH and SDPD simulations.
Time integration is performed with velocity Verlet method.

\section{Modeling arbitrary slip length/velocity at fluid-solid interface}
\label{section_slip}
With advancement of micro-and nano-fluidics, 
apparent wall slip is frequently observed at small scales. 
We propose two effective alternatives to implement an arbitrary slip length/velocity at the fluid-solid interface, especially of an arbitrary geometry. We employ boundary particles to describe any solid and they are placed statically within a layer of cutoff radius inside the solid surface.

\subsection{Conservative force at interface}
The pressure force of boundary particle has to be determined from the fluid in such a way that
the pressure gradient near the interface is respected.
Consequently, a balance between the forces of a fluid particle $\mathbf{a}_A$ and a boundary particle $\mathbf{a}_B$ near the interface can be defined~\citep{adami_generalized_2012}:
\begin{eqnarray}
\mathbf{a}_A = -\frac{\nabla p_A}{\rho_A} + \mathbf{f} = \mathbf{a}_B.
\end{eqnarray}
By definition the gradient equals to the directional derivative times the distance between the two particles, therefore the pressure on the boundary particle is obtained as
\begin{eqnarray}
p_B = p_A + \rho_A(\mathbf{f} - \mathbf{a}_B)\cdot \mathbf{r}_{BA}.
\end{eqnarray}
After introducing the SPH summation, the discrete pressure of the boundary particle becomes~\citep{adami_generalized_2012}:
\begin{eqnarray}
p_{B}=\frac{\sum_{A} p_{A} W_{BA}+\left(\mathbf{f}-\mathbf{a}_B\right) \cdot \sum_{A} \rho_{A} \mathbf{r}_{B A} W_{B A}}{\sum_{A} W_{B A}}.
\end{eqnarray}
Thereafter, boundary particle $B$ adopts $p_B$ as its pressure 
when calculating the pairwise conservative force with any neighboring fluid particle.
This boundary condition of pressure is important, as it assures a smooth density profile near the fluid-solid interface.

\subsection{Dissipative force at interface}
A slip boundary stems from a weaker viscous force than that of the no-slip boundary at the interface. With this observation, we proceed with a modification on the calculation of the pairwise dissipative force $\mathbf{F}_{AB}^D$ between a fluid particle $A$ and a boundary particle $B$.
To implement Eq.~(\ref{v_slip}) in SPD, we shall describe two approaches as follows.
In the first one, we regard a fluid particle as the center one and calculate its distance to the interface. Thereafter, we consider the distance of each nearby boundary particle
and assign a proper artificial velocity to it.
Therefore, the desired slip length/velocity is achieved at the location of the interface, which is determined by the fluid particle.
For reference, we call this approach as fluid-particle-centric method. 
In the second one, we consider a boundary particle as the center one and
calculate the position/velocity of a virtual particle from the nearby fluid particles.
The virtual particle represents the average effects of the fluid particles on the boundary 
particle and the desired slip length/velocity is achieved at the location of the interface,
which is determined by the boundary particle.
For reference, we name this approach as boundary-particle-centric method.

\subsubsection{Fluid-particle-centric method}
\begin{figure}[ht]
    \centering 
        \includegraphics[scale=0.3]{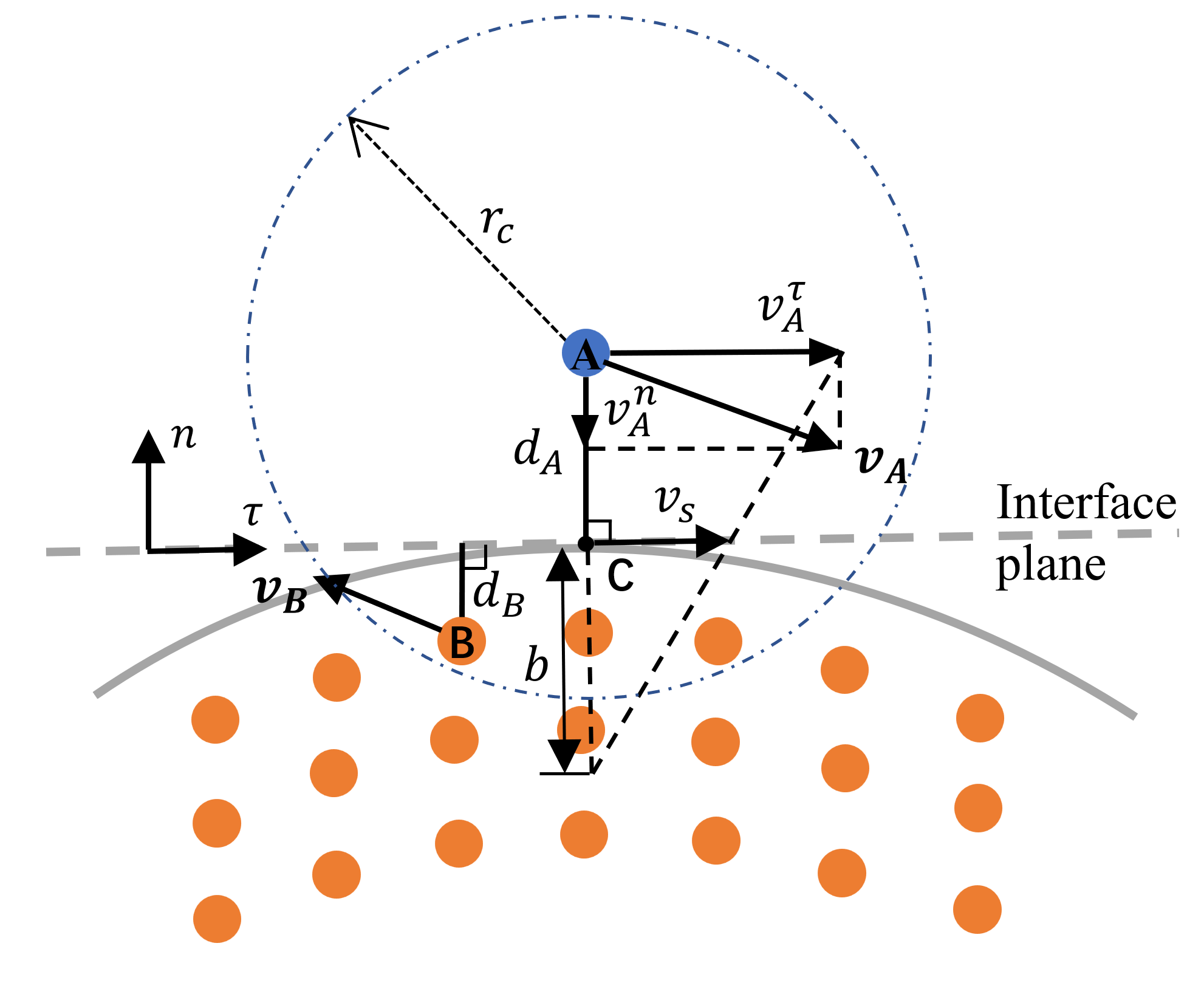} 
        \caption{Schematic of fluid-particle-centric method. Given a fluid particle $A$,
        an interface plane tangential to the solid surface is defined to be perpendicular to line $AC$ with length $d_A$, where $C$ is the intersection point on the surface. A Cartesian coordinate is chosen in such a way that $\tau$ direction is on the interface plane, $n$ direction is perpendicular to it, and meanwhile both $C$ and $\mathbf{v}_A$ are within $\tau n$ plane.
        Accordingly, $b$ is the Navier slip length along $n$ direction and ${\bf v}_s$ is the slip velocity along $\tau$ direction.
        Therefore, we reduce a three dimensional problem to a two dimensional one.
        $B$ is an example of neighboring boundary particles with $d_B$ away from the interface plane.  During the pairwise dissipative force calculation between $A$ and $B$, an artificial velocity $\mathbf{v}_B$ is assigned to $B$ so that the linear interpolation between $v_A$ and $v_B$ has a tangential velocity $v_C^{\tau} = v_s$ and a normal velocity $v_C^{n} = 0 $ at $C$.  Given the same fluid particle $A$, any other nearby boundary particles follow the same procedure of $B$.  Since point $C$ is determined solely by fluid particle $A$, this approach is referred to as fluid-particle-centric method. The solid is at rest in this sketch.}\label{method1_slip_boundary}
\end{figure}

As we assume a linear profile of the tangential velocity $v^\tau$ inside the solid very adjacent to the interface, the slip velocity can be approximated by particle's properties as shown in Fig.~\ref{method1_slip_boundary} as
\begin{eqnarray}
v_s=b\frac{\partial v^\tau}{\partial y} \approx b\frac{v_A^\tau}{d_A + b} \label{v_slip2}.
\end{eqnarray}
Here $\mathbf{v}_A = (v_A^\tau, v_A^n)$ is the velocity of a fluid particle $A$ with a distance $d_A$ to the interface.  For the calculation of $\mathbf{F}_{AB}^D$ between the fluid particle $A$ and any boundary particle $B$,  we need to explicitly respect the slip velocity $v_s$ according to the the slip length $b$ at the interface. To this end, an artificial velocity $\mathbf{v}_B = (v_B^\tau, v_B^n)$ is assigned with tangential and normal components in $\tau$ and $n$ directions to $B$, respectively, as follows:
\begin{eqnarray}
v_B^\tau &=& -\frac{d_B}{d_A}(v_A^\tau - v_s) + v_s = \frac{b - d_B}{b + d_A}v_A^\tau, \notag \\
v_B^n &=& -\frac{d_B}{d_A}v_A^n \label{v_B1},
\end{eqnarray}
where $d_B$ is the distance of $B$ from the interface. With Eq.~(\ref{v_B1}), 
the linear interpolation of velocities between $A$ and $B$ induces a slip velocity $v_s$ 
in the tangential direction and zero (impermeable condition) in the normal direction.
Furthermore, the local Navier slip length $b$ is also well maintained at the interface point $C$.
Since the intersection point $C$ is determined solely by the fluid particle $A$
and does not differ for interacting boundary particles,
we refer this approach as fluid-particle centric (FPC) method.

It is simple to show the relative velocity $\mathbf{v}_{AB} = (v_{AB}^\tau, v_{AB}^n)$ between $A$ and $B$ particles, which is directly involved in computing the pairwise dissipative force $\mathbf{F}_{AB}^D$ in Eq.~(\ref{sph_dissipative_eq}) or Eq.~(\ref{sdpd_dissipative_eq})
\begin{eqnarray}
v_{AB}^\tau &=& \frac{d_A + d_B}{d_A + b}v_A^\tau,\notag \\
v_{AB}^n &=& \frac{d_A + d_B}{d_A}v_A^n \label{v_AB}.
\end{eqnarray}
Considering two limit cases in Eq.~(\ref{v_AB}): at first when $b \to 0$, we obtain
\begin{eqnarray}
v_{AB}^\tau &=& \frac{d_A + d_B}{d_A}v_A^\tau,\notag \\
v_{AB}^n &=& \frac{d_A + d_B}{d_A}v_A^n,  \label{no-slip}
\end{eqnarray}
which is exactly the classical no-slip boundary condition utilized in SPH method~\citep{morris_modeling_1997,takeda1994numerical}; Secondly, when $b \to \infty$
\begin{eqnarray}
v_{AB}^\tau &=& 0,\notag \\
v^n_{AB} &=& \frac{d_A + d_B}{d_A}v_A^n,
\end{eqnarray}
which indicates a free-slip in the tangential direction.
The key difference between the implementations of a slip and the no-slip boundary is that the former must treat tangential component differently from the normal one as in Eq.~(\ref{v_AB}) while the latter has the same scaling constant for both components as in Eq.~(\ref{no-slip}). 
In any case, the impermeable boundary condition in the normal direction remains unchanged.

If a slip takes place at a non-planar interface, the major task is to calculate $d_A$, $d_B$ and the normal vector $\mathbf{n}$ of the interface.
This can still be accomplished for other simple geometries such as a sphere
with a little extra effort~\citep{morris_modeling_1997, bian_multiscale_2012}.
For complex geometries, we adopt a boundary value fraction (BVF) method \cite{holmes2011smooth, li2018dissipative} to acquire $d_A$, as described in \ref{appendixB}.
The unit normal vector $\mathbf{n}$ of the boundary can be obtained from
\begin{eqnarray}
\mathbf{n} = \frac{\nabla \varphi(\mathbf{r} - \mathbf{r}')}{|\nabla \varphi(\mathbf{r} - \mathbf{r}')|}, \label{normal_vector1}
\end{eqnarray}
and the discrete gradient of $\varphi$ is based on particle $A$
\begin{eqnarray}
\nabla \varphi_A= \frac{1}{\rho_A}\sum_j\frac{\partial W}{\partial r_{Aj}}\mathbf{e}_{Aj}, \label{normal_vector_phi1}
\end{eqnarray}
where $j$ is index of boundary particles. Thus, we obtain the normal distance from the  fluid particle $A$ and boundary particle $B$ to the interface using Eq.~(\ref{distance_boundary}) and (\ref{normal_vector1}) as
\begin{eqnarray}
d_A &=& d_{BVF}, \\
d_B &=& \mathbf{r}_{AB}\cdot\mathbf{n} - d_A.
\end{eqnarray}

Furthermore, if a slip boundary takes place at the surface of a mobile solid object, 
we need to take into account the rigid motion at the intersection point $C$ shown in Fig.~\ref{method1_slip_boundary},  as how it is done for a no-slip boundary~\citep{bian_multiscale_2012}. The slip velocity in Eq.~(\ref{v_slip2}) is modified to be
\begin{eqnarray}
v_s \approx b\frac{v_A^\tau - v_C^\tau}{d_A+b}.
\end{eqnarray}
Te artificial velocity for the boundary particle $B$ in this case is
\begin{eqnarray}
v_B^\tau & = & -\frac{d_B}{d_A}(v_A^\tau - v_s - v_C^\tau) + v_s + v_C^\tau \notag \\
& = & \frac{b - d_B}{b + d_A}v_A^\tau + \frac{d_A + d_B}{b + d_A}v_C^\tau, \notag\\
v_B^n &=& -\frac{d_B}{d_A}(v_A^n - v_C^n) + v_C^n.
\end{eqnarray}

The algorithm described in this approach represents a simple modification 
to an existing SPH code to control the slip boundary at the fluid-solid interface with Navier slip length $b$ as input parameter.

\subsubsection{Boundary-particle-centric method}
\begin{figure}[h!]
\centering \includegraphics[scale=0.3]{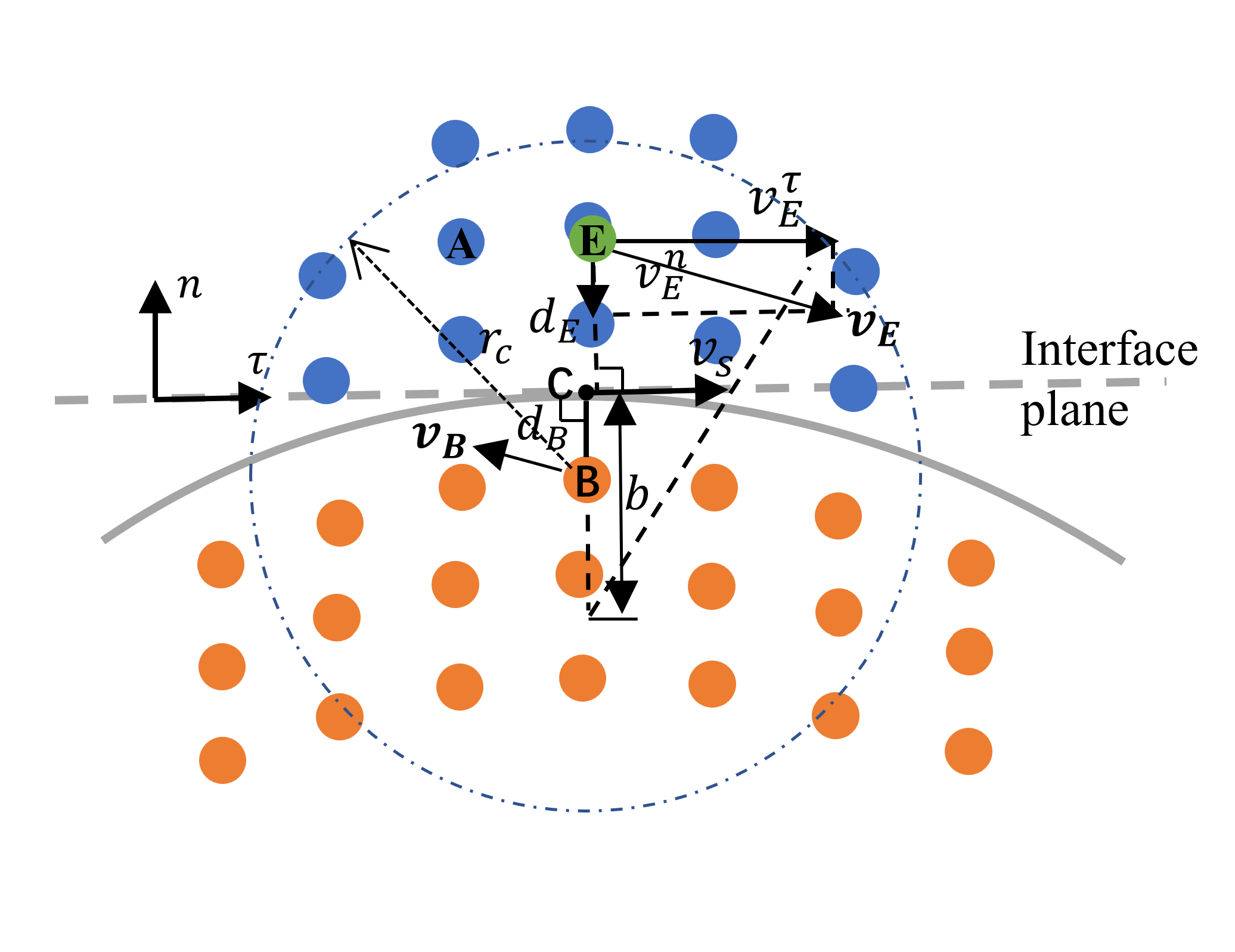} 
\caption{Schematic of boundary-particle-centric method. Given a boundary particle $B$,
        an interface plane tangential to the solid surface is defined to be perpendicular to line $BC$ with length $d_B$, where $C$ is the intersection point on the surface.
        A virtual particle $E$ represents the average effects of the neighboring fluid particles on $B$ and its position ${\bf x}_E$ and velocity ${\bf v}_E$ are calculated by SPH interpolations. The distance of $E$ to the interface plane is $d_E$. A Cartesian coordinate is chosen in such a way that $\tau$ direction is on the interface plane, $n$ direction is perpendicular to it, and meanwhile both $C$ and $\mathbf{v}_E$ are within $\tau n$ plane.
        Accordingly, $b$ is the Navier slip length along $n$ direction and ${\bf v}_s$ is the slip velocity along $\tau$ direction.
        Therefore, we reduce a three dimensional problem to a two dimensional one.
        An artificial velocity $\mathbf{v}_B$ is assigned to $B$ so that the linear interpolation between $v_E$ and $v_B$ has a tangential velocity $v_C^{\tau} = v_s$ and a normal velocity $v_C^{n} = 0 $ at $C$.  
        During the pairwise dissipative force calculation between the boundary particle $B$ and any fluid particle $A$, the artificial velocity $\mathbf{v}_B$ remains the same until ${\bf x}_E$ or ${\bf v}_E$ is updated. Since point $C$ is determined solely by boundary particle $B$, this approach is referred to as boundary-particle-centric method. The solid is at rest in this sketch.}
\label{method2_slip_boundary}
\end{figure}

The previous approach involves frequent calculations of distance from any fluid particle
to the interface and becomes tedious especially for complex geometries.
To avoid this deficiency, we may take a boundary particle as the center one
and consider a virtual particle $E$,
which has the average effects of the neighboring fluid particles on the boundary particle,
as shown in Fig.~\ref{method2_slip_boundary}.
Given a boundary particle $B$, the position and velocity of the virtual particle $E$ 
are calculated by SPH interpolation as 
\begin{eqnarray}
\mathbf{x}_{E} &=& \frac{\sum_{A}\mathbf{x}_{A}W_{BA}}{\sum_AW_{BA}}, \\
\mathbf{v}_{E} &=& \frac{\sum_{A}\mathbf{v}_{A}W_{BA}}{\sum_AW_{BA}}.
\end{eqnarray}
Thereafter, we assign an artificial velocity ${\bf v}_B=(v^{\tau}_B, v^n_B)$ to $B$ as
\begin{eqnarray}
v_{B}^{\tau} &=&  -\frac{d_B}{d_E}(v_E^{\tau}-v_s)+v_s=\frac{b-d_B}{b+d_E}v_E^{\tau}, \notag \\
v_B^n&=&-\frac{d_B}{d_E}v_E^{n}, \label{v_B2}
\end{eqnarray}
where $d_E$ and $d_B$ the distances of $E$ and $B$ from the interface, respectively.
With Eq.~(\ref{v_B2}), 
the linear interpolation of velocities between $E$ and $B$ induces a slip velocity $v_s$ 
in the tangential direction and zero (impermeable condition) in the normal direction.
Furthermore, the local Navier slip length $b$ is also well maintained at the interface point $C$.
Since the intersection point $C$ is determined solely by the boundary particle $B$
and does not differ for interacting fluid particles,
we refer this approach as boundary-particle centric (BPC) method.

If a slip takes place at a non-planar interface, we adopt the BVF method as described in \ref{appendixB} to calculate the distance $d_B$ and normal vector ${\bf n}$ of each boundary particle to the interface {\it only one time} at the beginning,
\begin{eqnarray}
d_B&=&d_{BVF},\\
\mathbf{n} &=& \frac{\nabla \varphi(\mathbf{r} - \mathbf{r}')}{|\nabla \varphi(\mathbf{r} - \mathbf{r}')|}, \label{normal_vector2}
\end{eqnarray}
and the discrete gradient of $\varphi$ is based on particle $B$
\begin{eqnarray}
\nabla \varphi_B= \frac{1}{\rho_B}\sum_j\frac{\partial W}{\partial r_{Bj}}\mathbf{e}_{Bj}, \label{normal_vector_phi2}
\end{eqnarray}
where $j$ is index of boundary particles.
During simulations, based on position of ${\bf x}_E$, we obtain
\begin{eqnarray}
d_E&=&\mathbf{r}_{EB}\cdot\mathbf{n}-d_B.
\end{eqnarray}
In practice, the directions of the normal vector $\mathbf{n}$ and $\mathbf{r}_{EB}$ are almost aligned, the normal distance of $E$ may also be approximated as $d_E 
\approx |\mathbf{r}_{EB}|-d_B$. 

Furthermore, for a mobile solid the rigid body motion of the intersection point $C$ is taken into account as
\begin{eqnarray}
v_B^\tau & = & \frac{b - d_B}{b + d_E}v_E^\tau + \frac{d_E + d_B}{b + d_E}v_C^\tau, \notag\\
v_B^n &=& -\frac{d_B}{d_E}(v_E^n - v_C^n) + v_C^n.
\end{eqnarray}
For calculation of ${\bf F}^D_{AB}$ between any neighboring fluid particle $A$ and
boundary particle $B$, the relative velocity ${\bf v}_{AB}$ is needed.
The procedure is the same as the previous approach, we omit it here.
For these calculations, the artificial velocity of $B$ remains the same until ${\bf x}_E$ or ${\bf v}_E$ is updated.

\subsubsection{Comments on the two algorithms}
Some comments are in order. 
The first approach is an extension of Morris et al.'s method~\cite{morris_modeling_1997}.
It relies on the distances of a pair of interacting fluid particle and boundary particle 
to the interface to determine an artificial velocity of the latter.
In this way, an desired value of the interpolated velocity between the two particles
is achieved at the interface, corresponding to the slip velocity.
This approach is accurate and can be directly applied to interfaces with simple geometries such as flat wall and cylindrical/spheric objects.
However, it becomes tedious for a complex interface, 
as the distance from any fluid particle to the interface has to be calculated frequently.
The second approach is based on Adami et al.'s method~\cite{adami_generalized_2012}.
It avoids frequent calculations of distances of fluid particles to the interface.
Instead, the distance of each boundary particle to the interface is
calculated only one time before simulation starts.
During simulations a virtual particle is created for each boundary particle
and its position and velocity are interpolated from the nearby fluid particles using SPH kernel.
This virtual particle represents the average effects of the fluid particles on the 
corresponding boundary particle so that an artificial velocity of the latter
is calculated based on the former.
The virtual particle's position and velocity can be done in the density summation loop as in Eq.~(\ref{sigma_eq}).
Superficially, the second approach seems to sacrifice accuracy.
However, we shall demonstrate with numerical examples that the second approach
has negligible errors compared to the first approach.

The artificial velocity of a boundary particle described in both approaches is employed only in the calculation of dissipative force,  but not intended for the kinematics of the solid. 
If the solid is mobile, all constituent boundary particles move along together
and follow rigid body dynamics as described in \ref{appendixA}. 

\subsection{Random force at interface}
In SDPD, the random force has to be modified according to the dissipative force so that the fluctuation-dissipation theorem is satisfied also at the interface. 
We first recall the special case of $b=0$ at the interface, 
namely the no-slip boundary condition, for a solid object at rest. 
Introducing the relative velocity between a fluid particle and a boundary particle of Eq.~(\ref{no-slip}) into dissipative force of Eq.~(\ref{sdpd_dissipative_eq}), we obtain
\begin{equation}
    \mathbf{F}_{AB}^{D2}=\left(\frac{d_A+d_B}{d_A}\right)\frac{\eta}{\sigma_A \sigma_B r_{AB}}\frac{\partial{W}}{\partial{r_{AB}}}\left(\frac{2D-1}{D}\mathbf{v}_{AB}+\frac{D+2}{D}\mathbf{e}_{AB}\cdot\mathbf{v}_{AB}\mathbf{e}_{AB}\right)\label{diss_noslip}.
\end{equation}
Comparing to the dissipative force between two fluid particles, 
the dissipative force between a fluid particle and a boundary particle
induces an effective viscosity as~\cite{bian2015fluctuating}
\begin{equation}
    \eta_{AB}=\frac{d_A+d_B}{d_A}\eta \label{effecitve_viscosity}.
\end{equation}
It is simple to see that this holds also for a solid object in a rigid motion.

Following the GENERIC structure~\citep{Grmela1997}, we obtain directly random force between $A$ and $B$ as
\begin{equation}
    \mathbf{F}_{AB}^{R2}=\left(\frac{d_A+d_B}{d_A}\right)^{1/2} \left(-\frac{20\eta}{3}\frac{k_BT}{\sigma_A\sigma_Br_{AB}}\frac{\partial{W}}{\partial{r_{AB}}}\right)^{1/2}d\overline{\mathscr{W}}_{AB}\cdot \mathbf{e}_{AB}\label{noslip_fran_eqn}
\end{equation}
so that the fluctuation-dissipation theorem is appreciated at the interface.
Eq.~(\ref{noslip_fran_eqn}) is the proper random force at interface by simply replacing $\eta$ by $\eta_{AB}$ in Eq.~(\ref{frandom_eqn}).

For a general partial slip at the fluid-solid interface, the correction for random force
can be done according to an effective viscosity with anisotropy. 
Correspondingly, the anisotropic scaling factors for the random force of $A$ and $B$ with $b > 0$ are as follows
\begin{eqnarray}
    \tau&:& \left(\frac{d_A+d_B}{d_A+b}\right)^{1/2}, \notag \\
    n&:& \left(\frac{d_A+d_B}{d_A}\right)^{1/2},  \label{fran_slip_eqn}
\end{eqnarray}
where $\tau$ and $n$ are tangential and normal direction on the interface plane, respectively. It is simple to see that random force determined by Eq.~(\ref{fran_slip_eqn}) at interface recovers Eq.~(\ref{noslip_fran_eqn}), if $b \to 0$.

\section{Numerical results}
\label{section_numerics}
To demonstrate the competency of the two proposed approaches for slip boundary condition,
we select multiple examples to cover a wide range of scenarios from simple to complex flows:
flows around flat and circular interface, as well as complex interface of arbitrary geometry;
flows in both transient and steady states;
macroscopic flows and mesoscopic flows with thermal fluctuations.
Whenever possible we derive analytical solutions to compare with results of SPD simulations.
If analytical solutions are formidable, we construct solutions with finite difference method
or finite volume method as references.
Since the difference in results of the two approaches is mostly negligible,
we present only the results from the boundary-particle-centric method to avoid redundancy. 
If no other reference is available, we present results of both approaches
and compare one against the other.

\subsection{Couette flow}
\begin{figure}[h!]
\centering \includegraphics[scale=0.6]{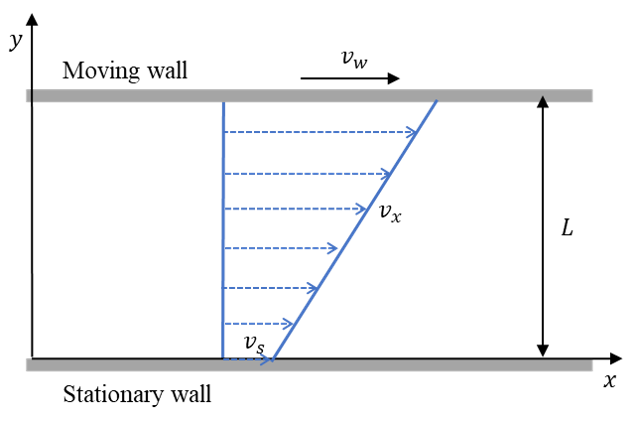}
\caption{Schematic of Couette flow with upper wall moving. Either upper or lower wall is with partial slip boundary condition.}\label{couette_sche}
\end{figure}

We first consider the two-dimensional Couette flow with a linear distribution of steady velocity as shown in Fig.~\ref{couette_sche}. The distance between the upper and lower walls is $L= 10^{-3}\mathrm{m}$. The lower wall at $y=0$ always remains still and the upper wall drives at constant velocity $v_w=1.25\times10^{-5}\mathrm{ms^{-1}}$ in $x$ direction. We set the fluid kinematic viscosity $\nu = 10^{-6}\mathrm{m^2s^{-1}}$ and density $\rho = 1000\mathrm{kgm^{-3}}$.
Therefore, the corresponding Reynolds number is $Re=v_wL/\nu=0.0125$. 
When flow reaches steady state, the analytical solution of velocity with slip boundary conditions on the two walls is
\begin{equation}
    v_x(y)=\frac{v_w(y+b^{lo})}{L+b^{up}+b^{lo}},\label{couette_state_vel}
\end{equation}
where $b^{up}$ and $b^{lo}$ are slip lengths specified on the upper and lower walls, respectively.

\begin{figure*}[h!]
\centering
\subfigure[Simulation results compared with analytical solutions at steady state: different  slip lengths $b$ prescribed on the lower static wall and no-slip on the upper moving wall. 
Absolute values of y-intercept of the velocity profile indicate resulted slip lengths of the simulations. ]{
\includegraphics[scale=0.48]{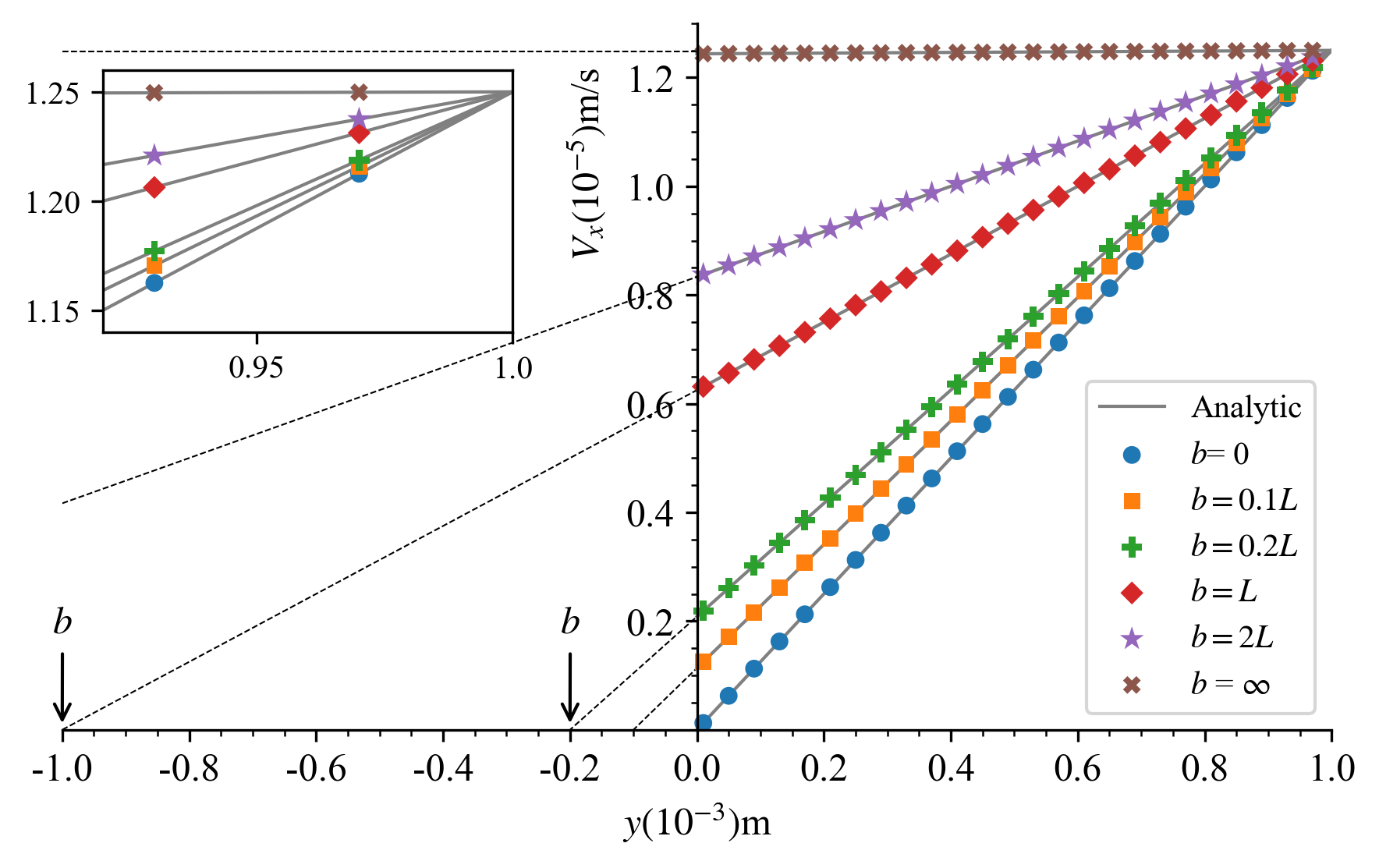}
\label{couette_fig1}}
\quad
\subfigure[Convergence study on $\Delta x$ for $b = 0.5L$ and $b = L$ in Fig.~\ref{couette_fig1}.]{%
\includegraphics[scale=0.48]{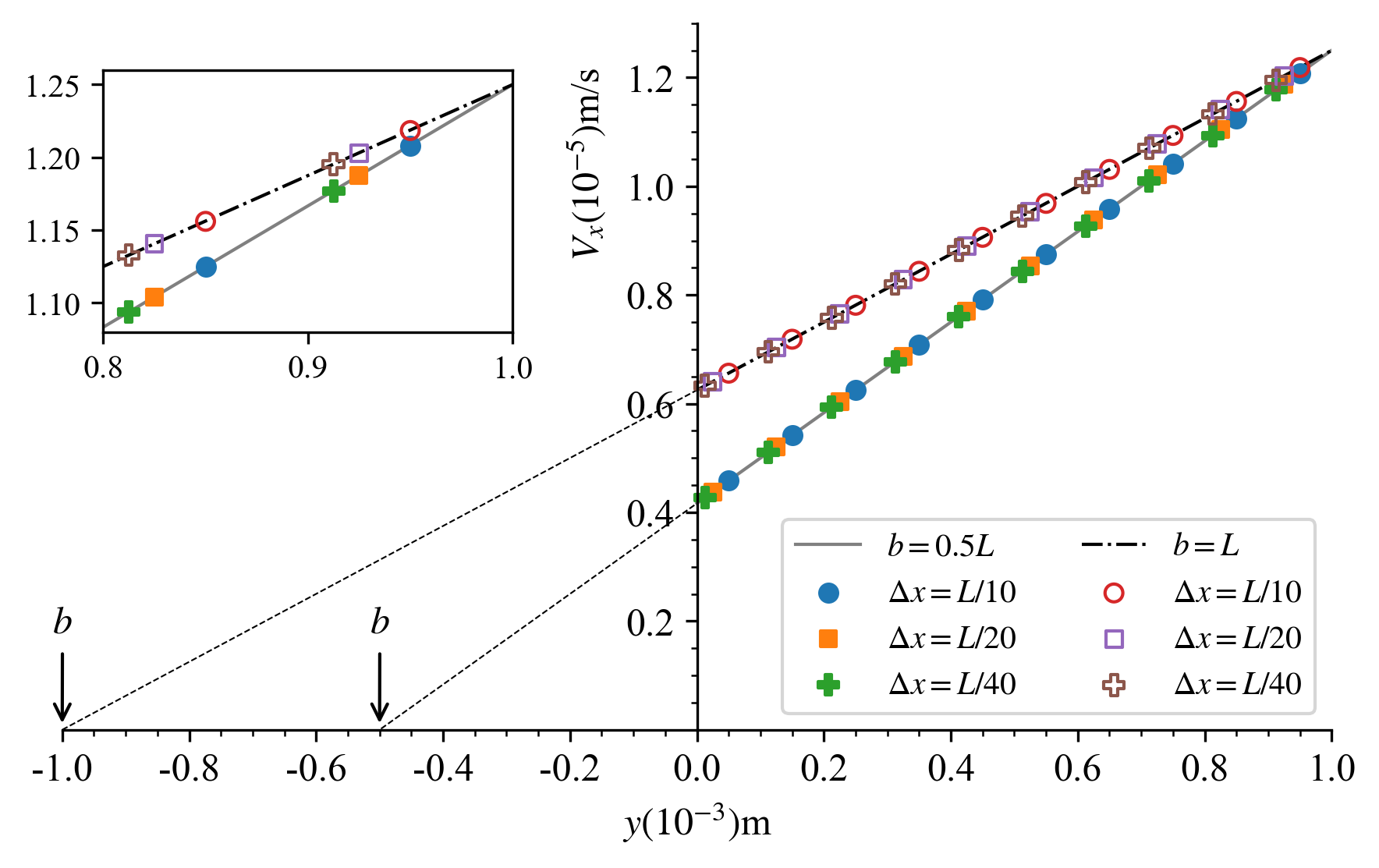}
\label{couette_fig2}}
\subfigure[Simulation results compared with analytical solutions at steady state: different slip lengths $b$ prescribed on the upper moving wall and no-slip on the lower static wall.]{%
\includegraphics[scale=0.48]{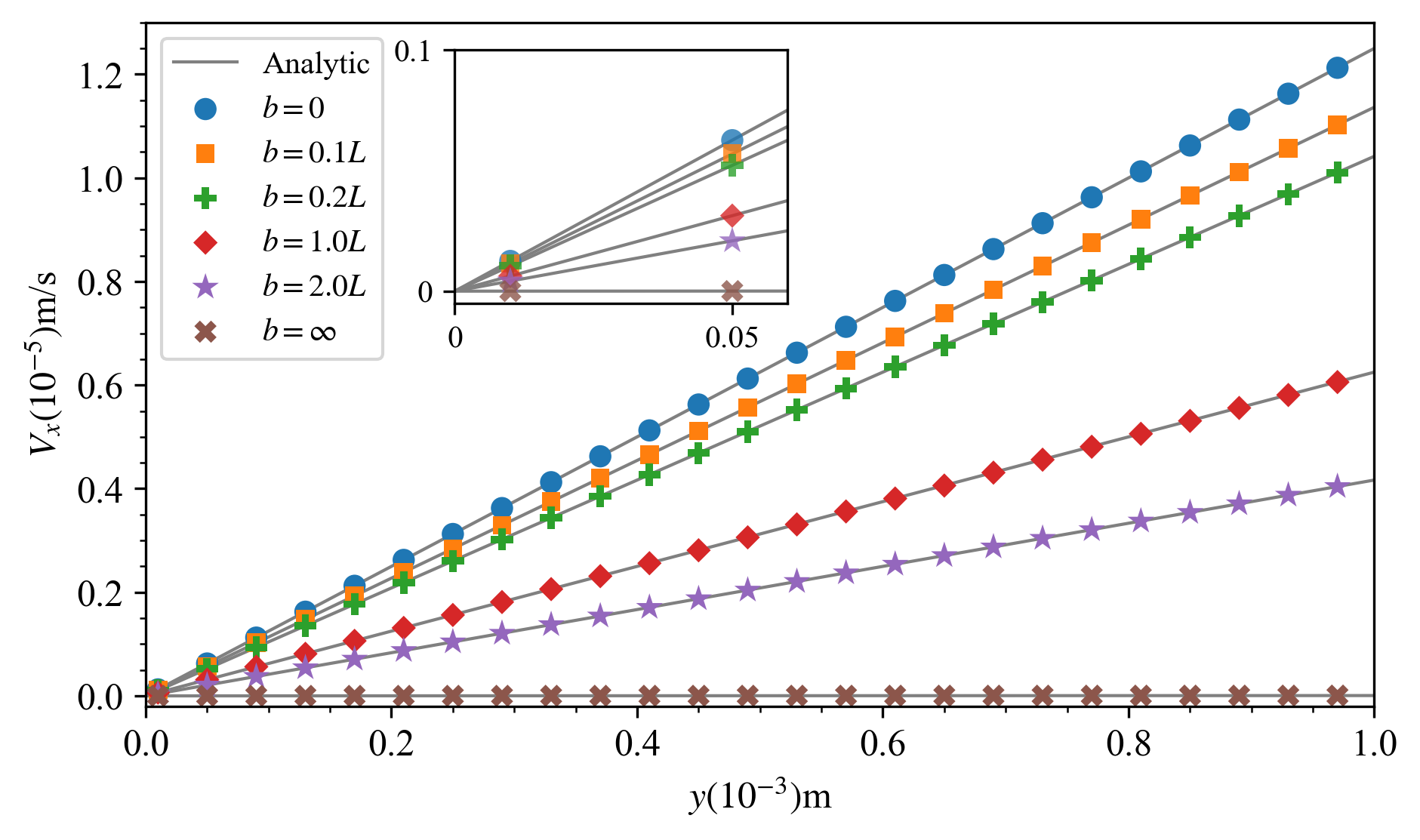}
\label{couette_fig3}}
\quad
\subfigure[Simulation results compared with solutions of finite difference method at transient states:  lower static wall has slip length $b = 0.1L$ and
upper moving wall has no slip boundary condition.]{%
\includegraphics[scale=0.48]{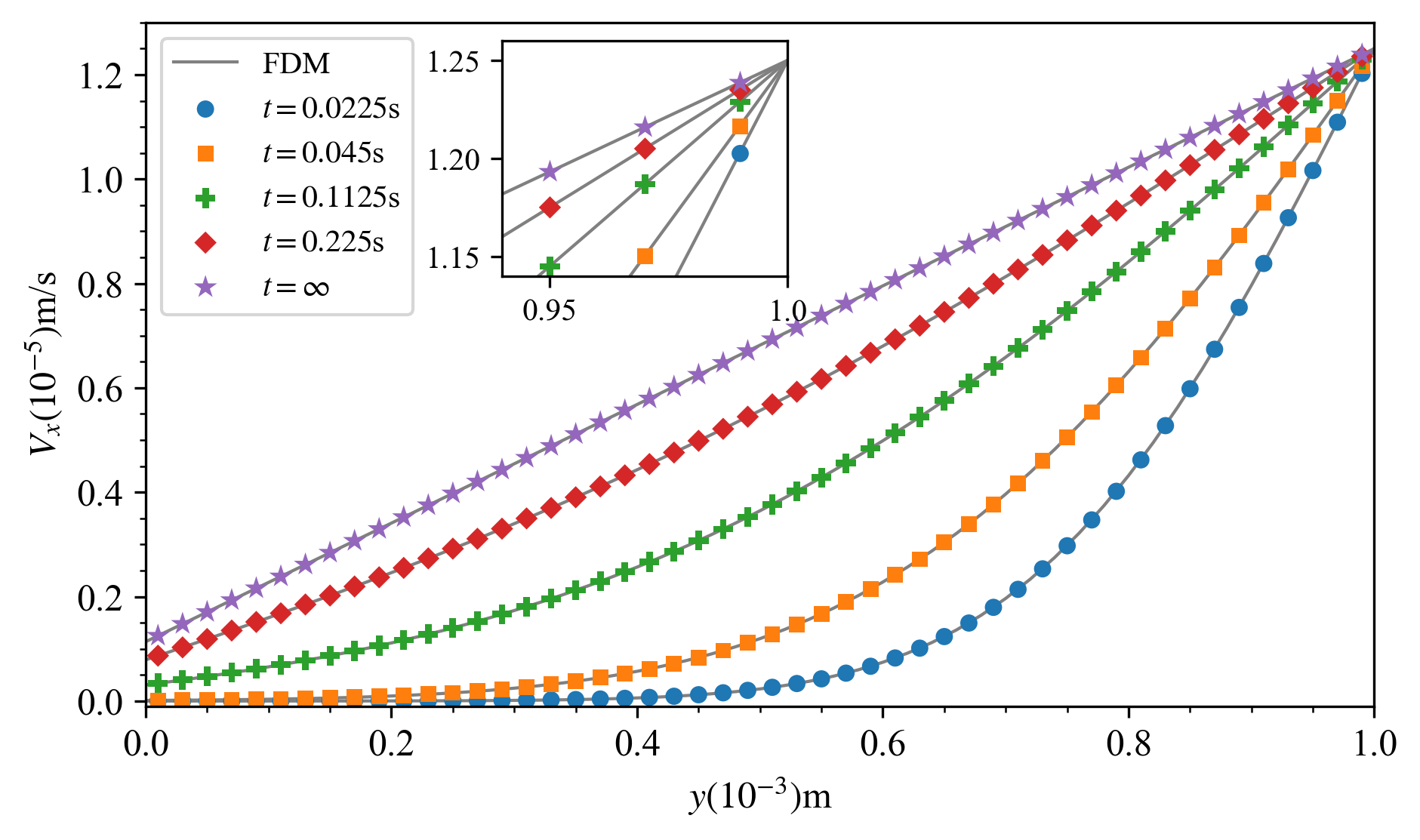}
\label{couette_fig4}}
\caption{Velocity profiles of SPH simulations with $r_c = 3.0\Delta x$ for Couette flow. Resolution $\Delta x = L/50$ in Fig.~\ref{couette_fig1}, \ref{couette_fig3} and \ref{couette_fig4}.}
\label{couette_fig}
\end{figure*}

Fig.~\ref{couette_fig1} shows a comparison between the velocity profiles by SPH simulations and analytical solution in Eq.~(\ref{couette_state_vel}).
The lower static wall has a specified slip length while the upper moving wall has no-slip boundary condition.
The results of simulations represented by symbols fall on top of the corresponding individual analytical lines.
The lines on the far left represent extrapolations of the velocity into the boundary and their intercepts with the horizontal coordinate indicate various slip lengths.
It is apparent that simulation results are reliable for
 no slip, any partial slip, and free slip boundary conditions. 
In addition, we perform convergence study on the particle resolution, as shown in Fig.~\ref{couette_fig2}. It can be seen that simulations
with as low as $10$ particles across the channel is sufficient for the one directional flow.

Furthermore, we switch to allow slip boundary on the upper moving wall
while keep lower static wall with no-slip boundary. Fig.~\ref{couette_fig3} presents the agreement between simulation results and analytical solutions,
where the upper wall is specified with different slip lengths. 
The velocity gradient at the upper wall is inversely proportional to the slip length.
In the limit of infinite slip length, the fluid is not driven by the upper wall
and remains stationary.

Finally, we validate SPH simulations by finite difference method (FDM) for time-dependent velocity profiles. In this case, the lower static wall has a slip length of $b=0.1L$,
while the upper moving wall drives gradually the fluid from rest to flow. Selective velocity profiles at different moments are shown in Fig.~\ref{couette_fig4},
where results of SPH simulations have an excellent agreement with solutions of FDM.
This indicates that the proposed approach is competent for one directional time-dependent flows.

\subsection{Poiseuille flow}
\begin{figure}[h!]
\centering \includegraphics[scale=0.6]{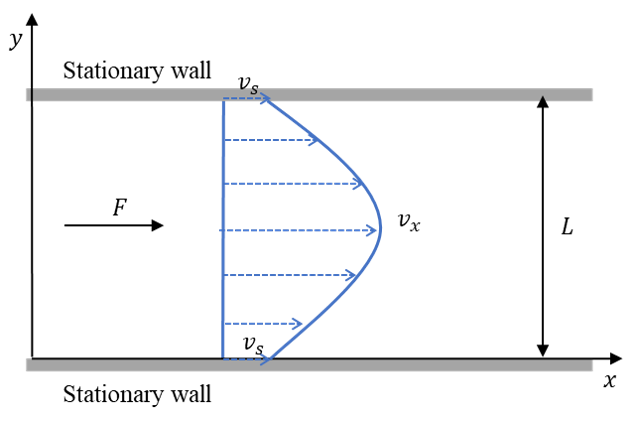}
\caption{Schematic of Poiseuille flow driven by a body force: either upper or lower wall has slip boundary condition.}\label{poiseuille_sche}
\end{figure}

We further consider a two-dimensional Poiseuille flow with a nonlinear velocity distribution as shown in Fig.~\ref{poiseuille_sche}. 
The distance between the upper and lower walls is $L= 10^{-3}\mathrm{m}$. The lower wall at $y=0$
and both walls remain still.
The fluid kinematic viscosity $\nu = 10^{-6}\mathrm{m^2s^{-1}}$ and density $\rho = 1000\mathrm{kgm^{-3}}$.
The flow is driven by a constant body force $F=10^{-4}\mathrm{N{kg}^{-1}}$ in $x$ direction.
Therefore, for no-slip boundary conditions the maximum velocity of the flow  $v_{max}=1.25\times10^{-5}\mathrm{ms^{-1}}$, which define a Reynolds number as $Re=v_{max}L/\nu=0.0125$.
When the flow arrives at steady state,
the analytical solution of velocity with slip boundary conditions is as follows
\begin{equation}
    v_x(y)=-\frac{F}{2\nu}y^2+\frac{FH(2b^{up}+L)}{2\nu(L+b^{lo}+b^{up})}y+\frac{Fb^{lo}H(2b^{up}+L)}{2\nu(L+b^{lo}+b^{up})}.\label{poiseuille_state_vel}
\end{equation}

\begin{figure*}[h!]
\centering
\subfigure[Simulation results compared with analytical solutions at steady state: identical slip lengths $b$ are specified on both walls. Absolute values of $y$-intercept of the velocity profile indicate resulted slip lengths of the simulations.]{
\includegraphics[scale=0.48]{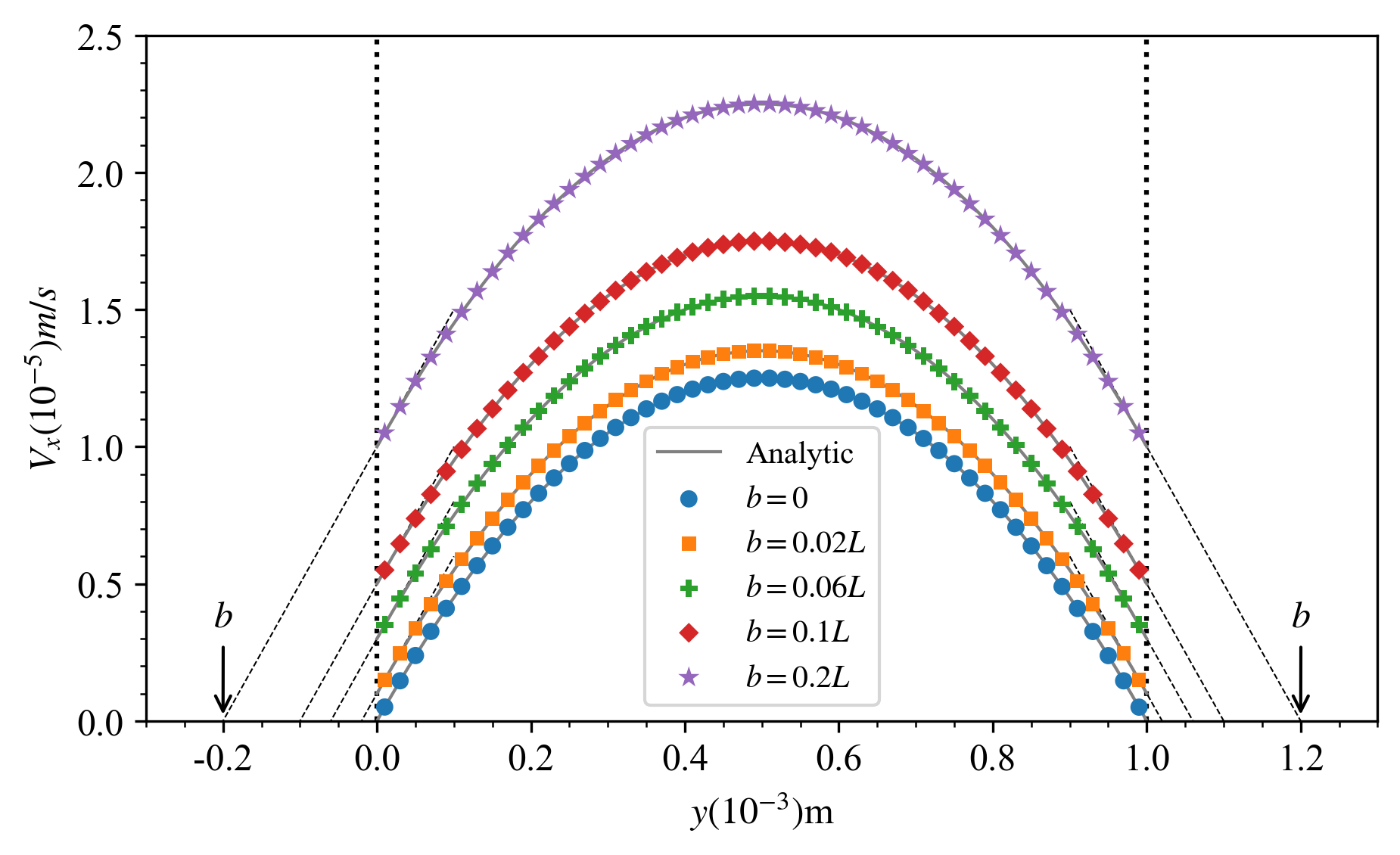}
\label{poiseuille_fig1}}
\quad
\subfigure[Convergence study on $\Delta x$ for $b = 0.1L$ and $b = 0.2L$ in Fig.~\ref{poiseuille_fig1}.]{%
\includegraphics[scale=0.48]{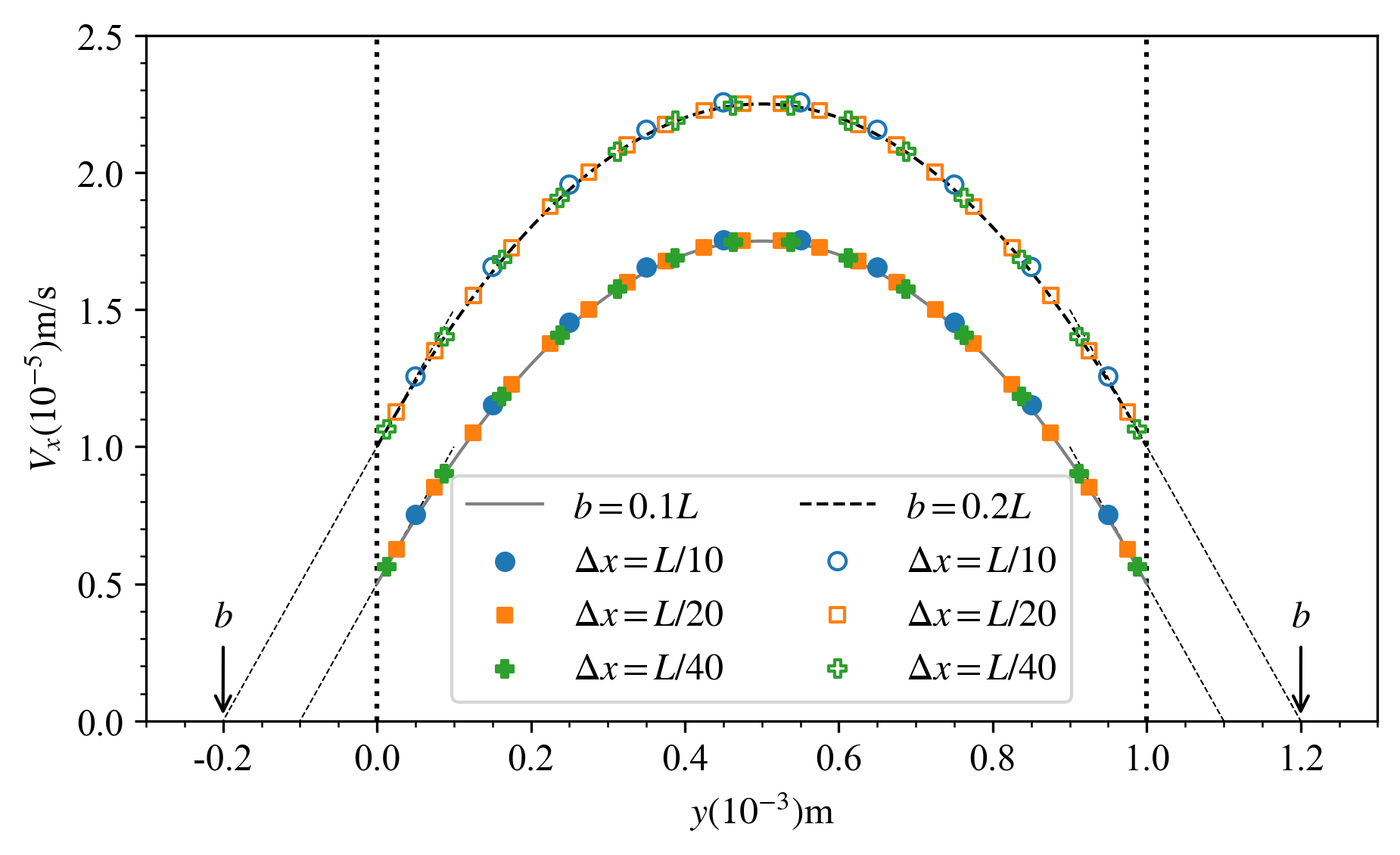}
\label{poiseuille_fig2}}
\subfigure[Simulation results compared results compared  with analytical solutions at steady state: different slip lengths $b$ specified on the upper wall and no-slip on the lower wall. Absolution values of $y$-intercept of the velocity profile indicate resulted slip lengths of the simulations.]{
\includegraphics[scale=0.48]{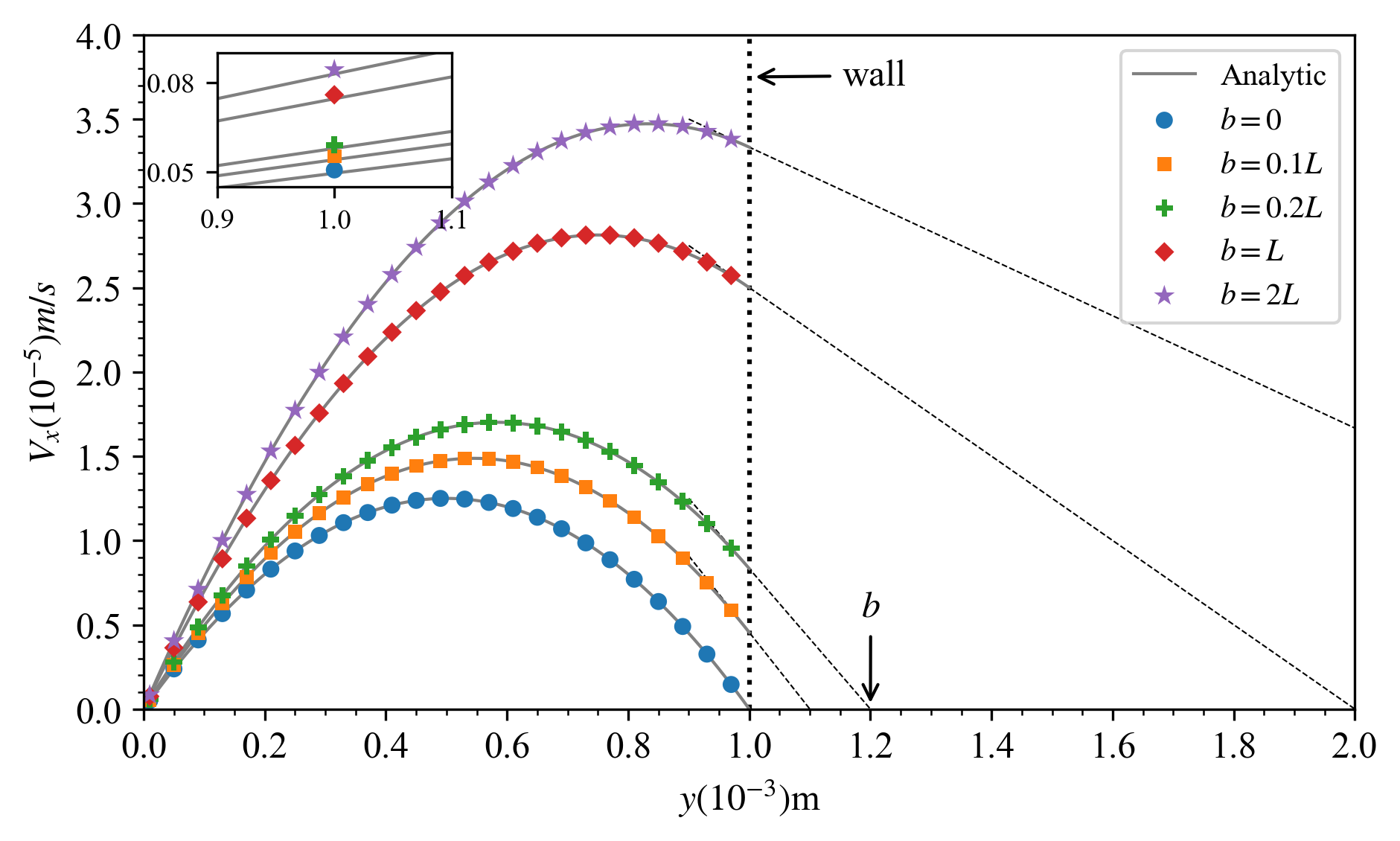}
\label{poiseuille_fig3}}
\quad
\subfigure[Simulation results compared with solutions of finite difference method at transient states. Both walls have slip boundary condition with identical slip with $b = 0.1L$.]{%
\includegraphics[scale=0.48]{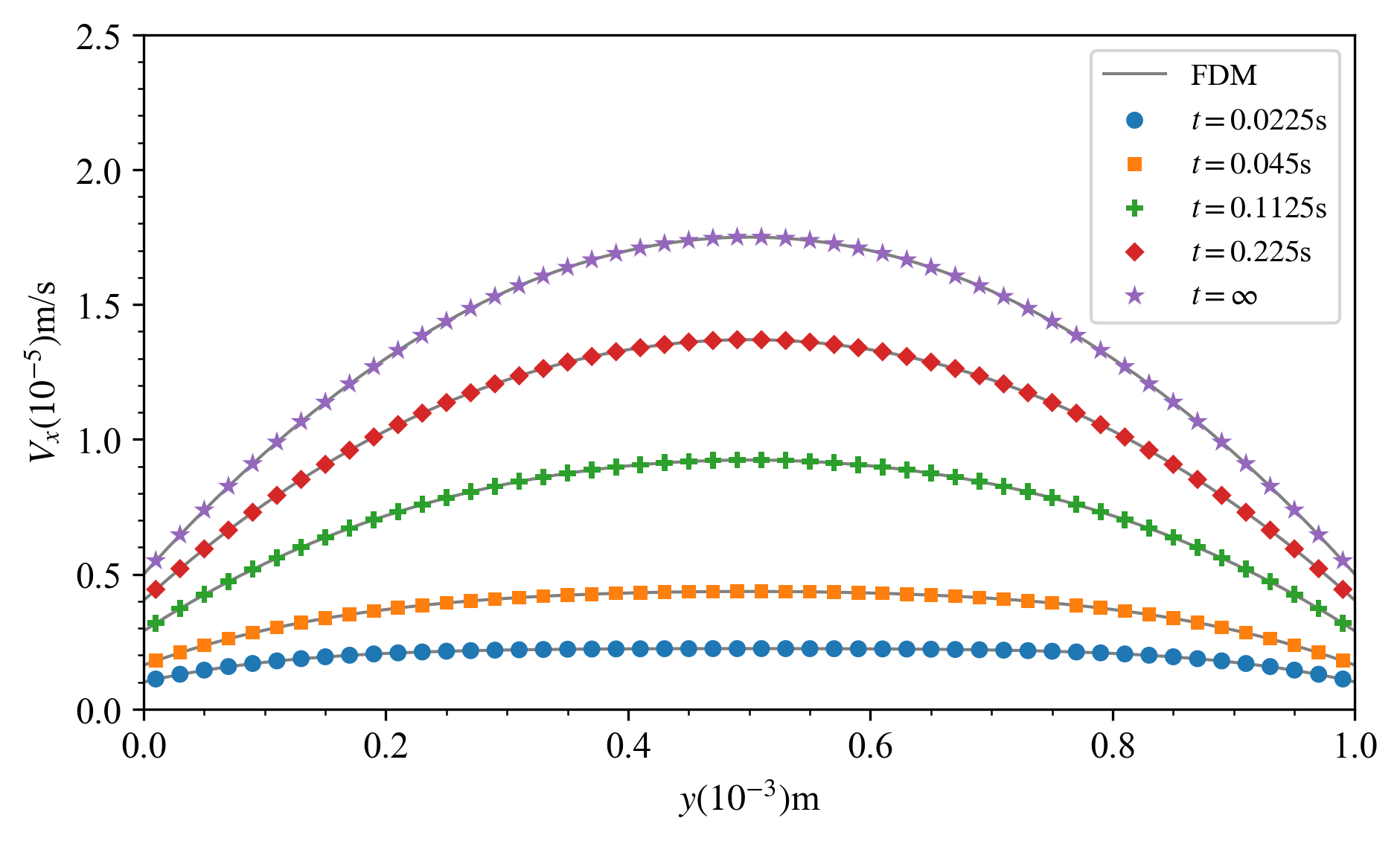}
\label{poiseuille_fig4}}
\caption{Velocity profiles of SPH simulations with $r_c=3\Delta x$ for Poiseuille flow. Resolution $\Delta x = L/50$ in Fig.~\ref{poiseuille_fig1}, \ref{poiseuille_fig3} and \ref{poiseuille_fig4}.}
\label{poiseuille_fig}
\end{figure*}

Fig.~\ref{poiseuille_fig1} shows a comparison of the velocity profiles by SPH and the analytical solution in Eq.~(\ref{poiseuille_state_vel}).
Both lower and upper walls have specified identical slip lengths. 
Initially, $50$ particles are uniformly distributed across the channel.
The results of SPH simulations are in good agreement with analytical solutions.
At fluid-solid interface the velocity distribution is non-linear, 
therefore, the velocity gradient of the fluid at the interface is exploited to extrapolate the velocity linearly into the interior of the solid. The magnitudes of $y-$intercepts represent the slip lengths
from the simulations, which are consistent with the specified ones. 
In addition, we perform a resolution study for two selected slip lengths as shown in Fig.~\ref{poiseuille_fig2},
where the velocity profiles of simulations with as low as $10$ particles match the analytical solutions.
In another the case, we consider no slip boundary on the lower wall and different slip lengths on the upper wall, results of which are shown in Fig.~\ref{poiseuille_fig3}.
Again the SPH results of velocity are consistent with analytical solutions
and the $y$-intercepts of the velocity profile into the wall reproduce the desired slip lengths. 
Moreover, we also show the time-dependent velocity profiles in Fig.~\ref{poiseuille_fig4},
where simulation results agree very well with solutions of FDM at selective moments.

\begin{figure*}[h!]
\centering
\subfigure[Particle configurations for Poiseuille flow at steady state: left one started from a square lattice
versus right one started from a disordered state.
The color represents velocity magnitude with blue as $0$ and red as maximum.
Both walls have slip length $b=0.1L$.]{
\includegraphics[scale=0.45]{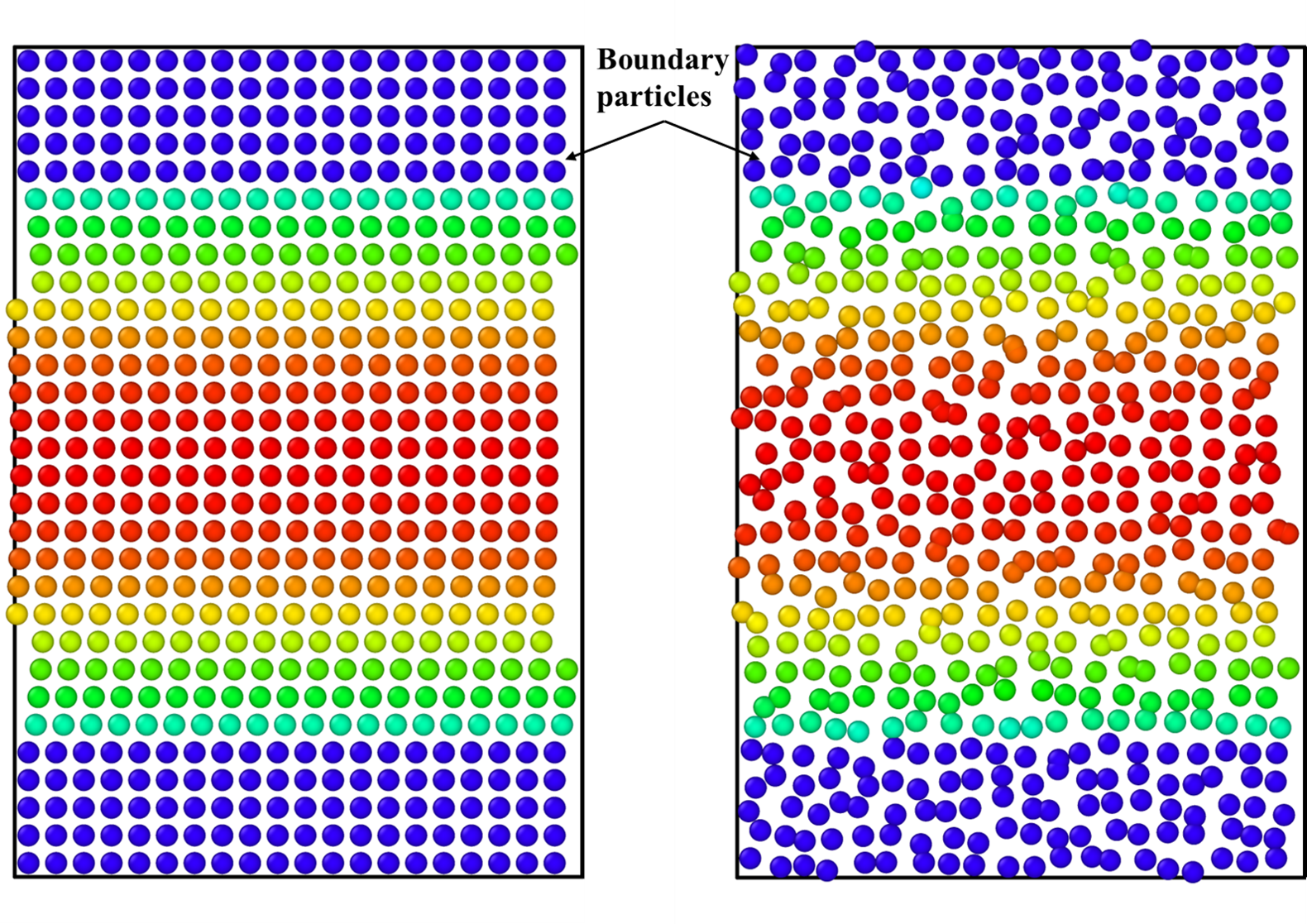}
\label{ordered_disordered_particles}}
\subfigure[$\Delta x = L/40$]{%
\includegraphics[scale=0.48]{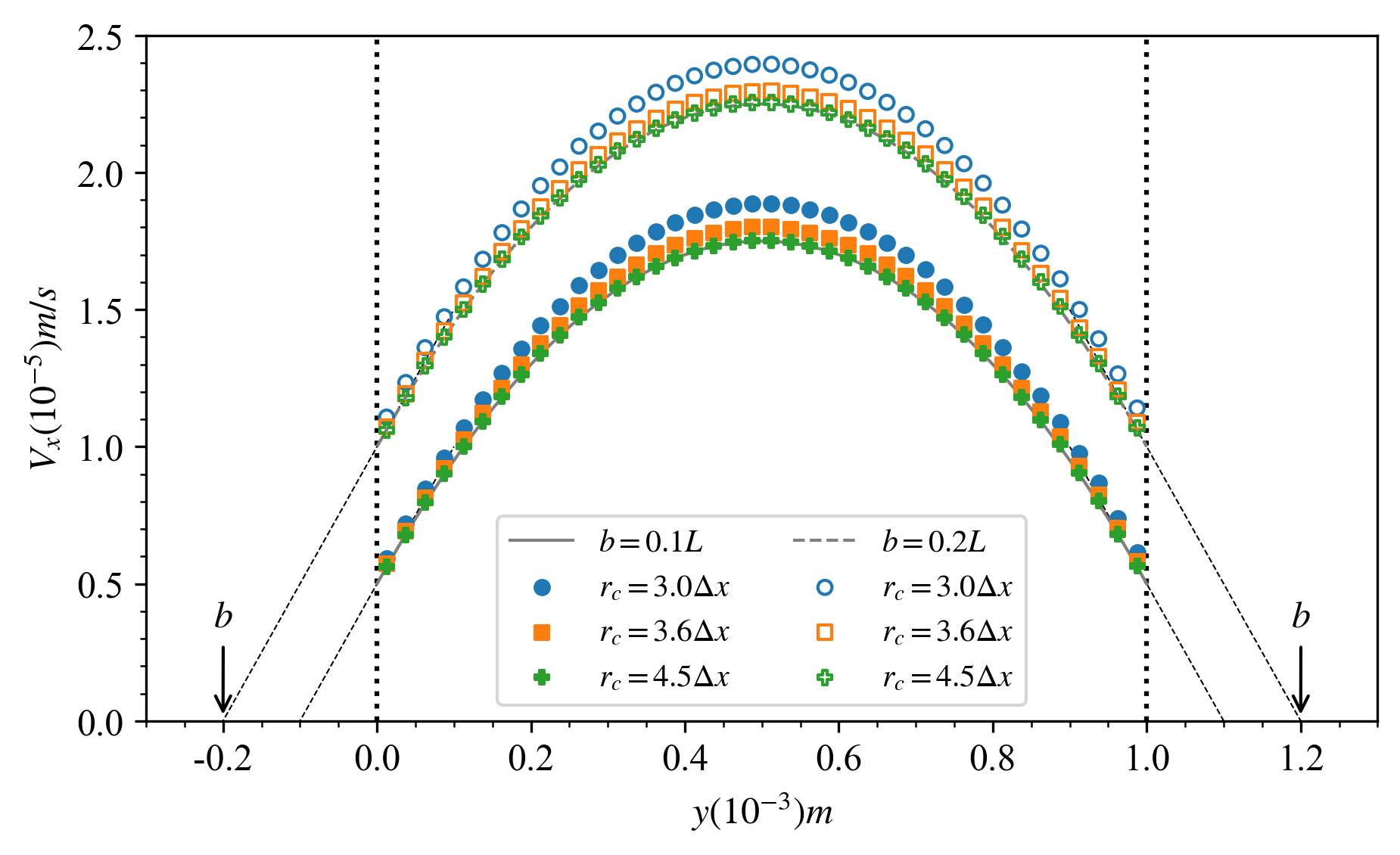}
\label{poiseuille_disordered_convergence1}}
\quad
\subfigure[$r_c=4.5\Delta x$]{%
\includegraphics[scale=0.48]{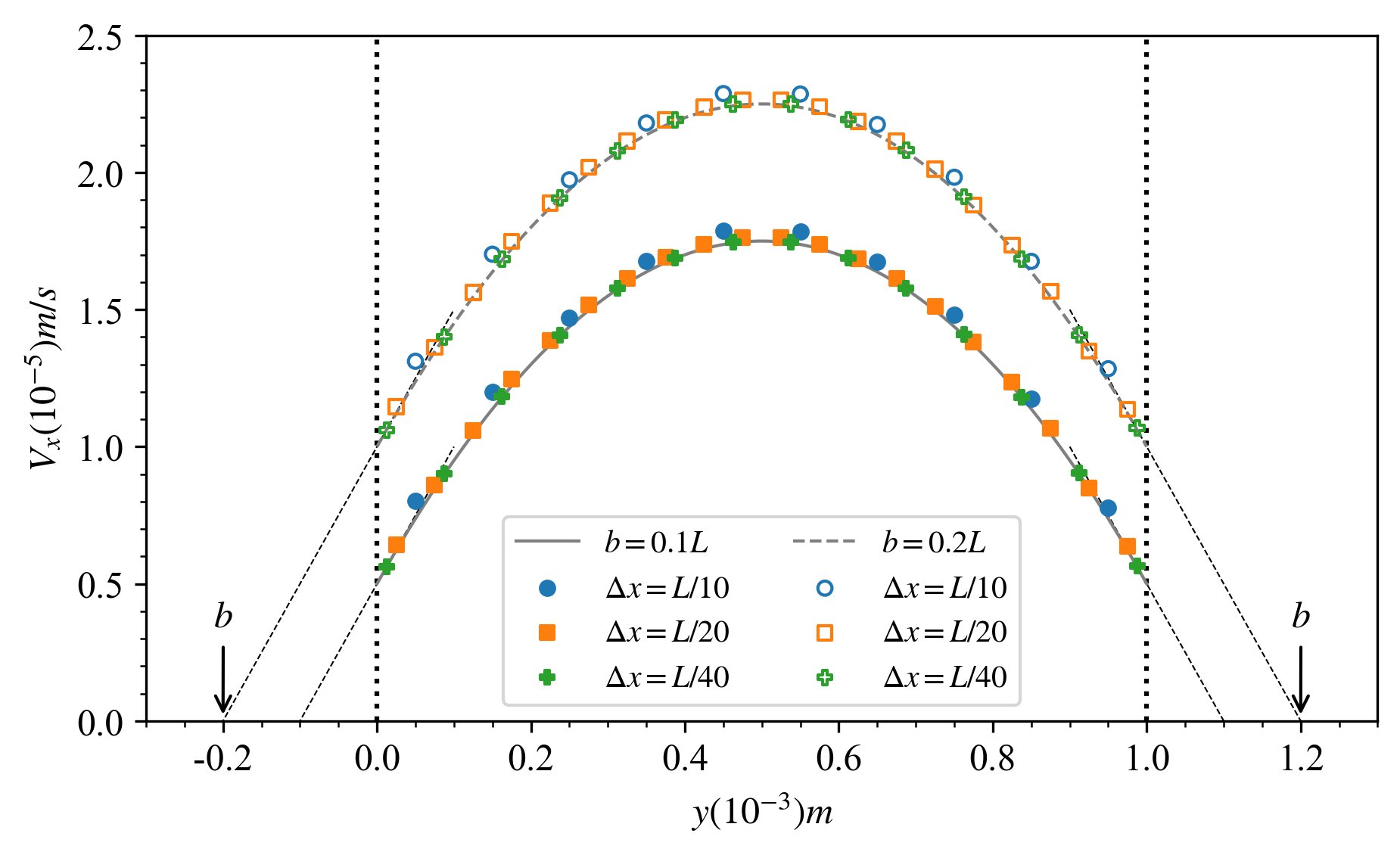}
\label{poiseuille_disordered_convergence2}}
\caption{SPH simulation results of Poiseuille flow using disordered particles with identical slip length on both walls: $b = 0.1L$ and $0.2L$ }
\label{poiseuille_disordered_convergence}
\end{figure*}

To examine the proposed approaches on disorder particle configurations,
we pre-run a simulation with SDPD method with $r_c=3\Delta x$ so that the particles are redistributed
by thermal fluctuations.
Thereafter, we perform a SPH simulation with the disordered particle configuration as initial condition.
When flow reaches steady state, the new simulation has significantly different
particle configuration as the previous simulation with particles on square lattice as initial condition,
as shown in Fig.~\ref{ordered_disordered_particles}. 
The left snapshot shows configuration of particles started from square lattice 
while the right snapshot presents configuration of particles initially disordered. 
Since a highly disordered configuration of particles leads to magnified numerical errors in SPH summation,
a larger cut-off radius of $r_c=4.5\Delta x$ is required to achieve accurate results as shown in Fig.~\ref{poiseuille_disordered_convergence1}.
For this large cut-off, results are insensitive
to $\Delta x$ as shown in Fig.~\ref{poiseuille_disordered_convergence2},
where again as few as $10$ particles across the channel is sufficient.

\begin{figure}[h!]
\centering \mbox{ 
\subfigure[$Re=1$]{\includegraphics[scale=0.48]{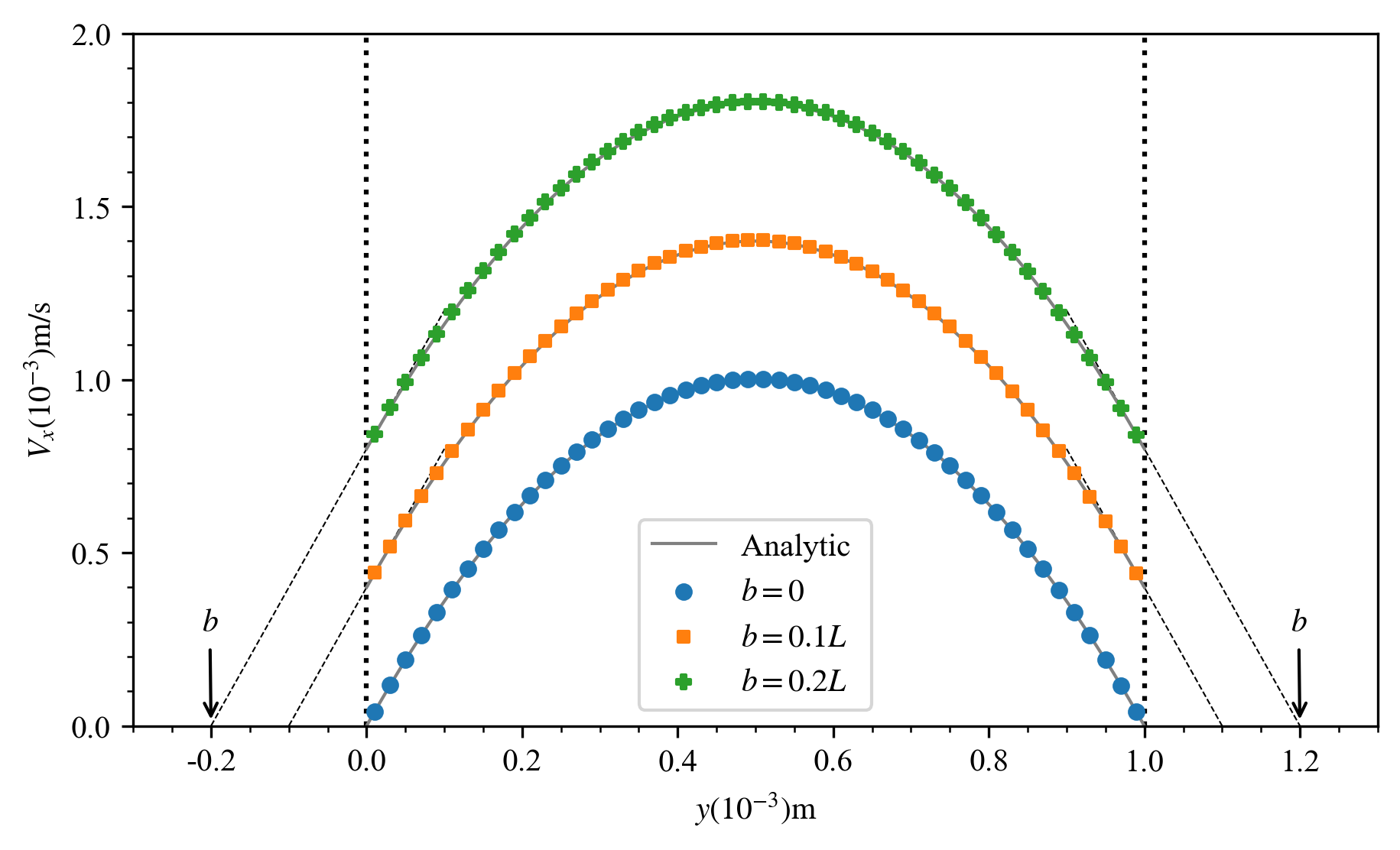}}\quad 
\subfigure[$Re=10$]{\includegraphics[scale=0.48]{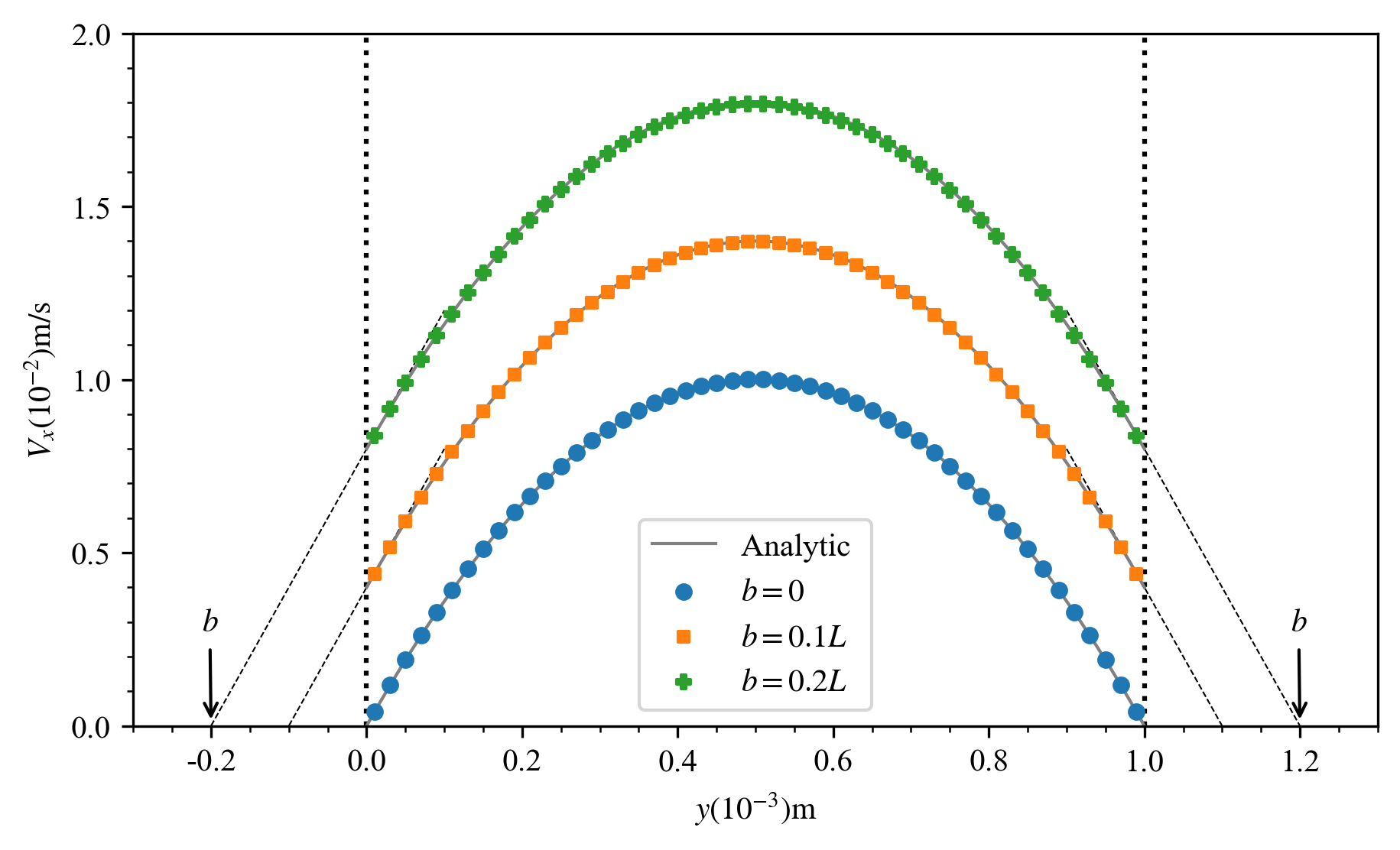}}} 
\caption{SPH simulation results for Poiseuille flow with $Re=1$ and $Re=10$ with particles initially on square lattice: $r_c=3\Delta x$ and $\Delta x=L/40$.}\label{poiseuille_Re_1_10e}
\end{figure}

The proposed approaches work equally well
at moderate Reynolds numbers, as shown in Fig.~\ref{poiseuille_Re_1_10e},
for three selective slip lengths.

\subsection{Taylor-Couette flow} 
\begin{figure}[h!]
\centering \mbox{ \subfigure[Configuration of different radii.]{\includegraphics[width=50mm]{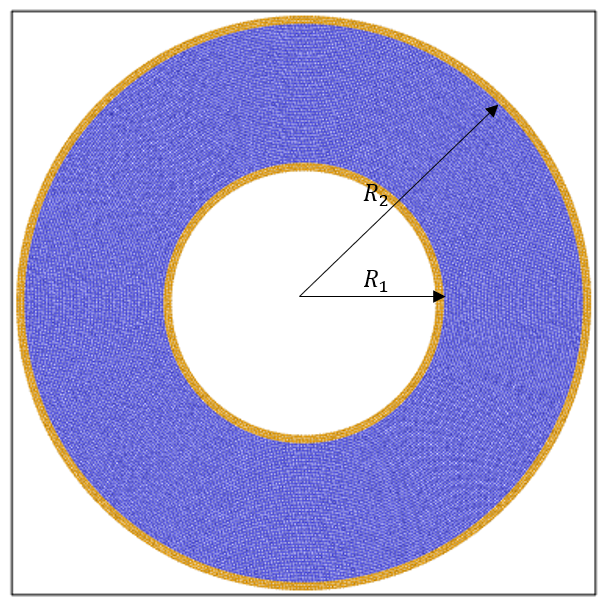}}\quad \subfigure[Velocity field.]{\includegraphics[width=60mm]{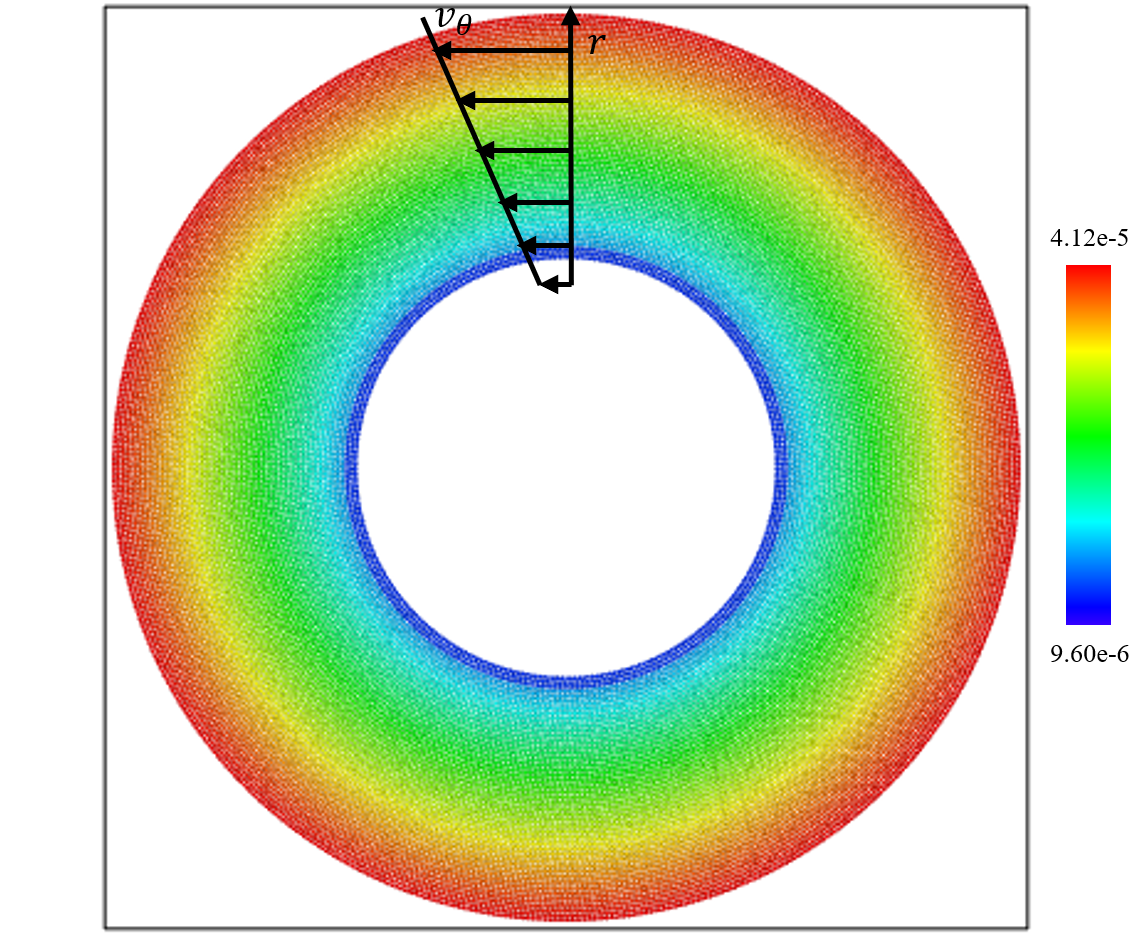}}} \caption{Schematic of Taylor-Couette flow with $R_1=10^{-3}m$, $R_2=2\times 10^{-3}m$ and $L=R_2-R_1=10^{-3}m$. Slip length $b = L/10$ at inner wall and no slip at outer wall.}\label{sche_taylor_fig}
\end{figure}

We have seen that the two approaches work well for planar interfaces.
In the following, we start to consider a slip length taking place at fluid-solid interface of non-planar geometry. 
The first example is the so-called Taylor-Couette flow~\citep{taylor1923viii}.
It consists of two cylinders of different radii $R_1$ and $R_2$, both of which rotate around the same axis with angular velocities $\Omega_1$ and $\Omega_2$, as shown in Fig.~\ref{sche_taylor_fig}(a). 
The fluid is filled between the two cylinders so that there is a convex interface outside the inner wall and a concave interface inside the outer wall.
The general solution for the steady Taylor-Couette flow is known as
\begin{equation}
    v_{\theta}(r)=Ar+\frac{B}{r},\label{taylor_general_solution}
\end{equation}
with $A$ and $B$ as undetermined constants.
We define slip boundary conditions at the two walls as follows
\begin{equation}
\begin{cases}
        v_{\theta}(R_1)-R_1\Omega_1 &= b^{in}\frac{d v_{\theta}}{d\theta},\\
        v_{\theta}(R_2)-R_2\Omega_2 &= -b^{out}\frac{dv_{\theta}}{d\theta},
\end{cases}\label{taylor_boundary_condition}    
\end{equation}
where $b^{in}$ and $b^{out}$ are specified slip lengths
at inner and outer walls, respectively.
Combining Eq.~(\ref{taylor_general_solution}) and Eq.~(\ref{taylor_boundary_condition}), 
we can solve for $A$ and $B$ as
\begin{eqnarray}
   A &=& 
\left | \begin{matrix}
R_1\Omega_1 &   1/R_1+b^{in}/R_1^2 \\
R_2\Omega_2 &   1/R_2-b^{out}/R_2^2 \\
\end{matrix} \right | /
    \left | \begin{matrix}
R_1-b^{in} & 1/R_1+b^{in}/R_2^2 \\
R_2+b^{out} & 1/R_2^2-b^{out}/R_2^2 \\
        \end{matrix} \right |,
\\
B &=& 
    \left | \begin{matrix}
    R_1 - b^{in} &   R_1\Omega_1 \\
    R_2 - b^{out} &   R_2\Omega_2 \\
    \end{matrix} \right | /
        \left | \begin{matrix}
    R_1-b^{in} & 1/R_1+b^{in}/R_2^2 \\
    R_2+b^{out} & 1/R_2^2-b^{out}/R_2^2 \\
            \end{matrix} \right |.
\end{eqnarray}

We take the inner wall with $R_1=10^{-3}\mathrm{m}$ and $\Omega_1=0.01s^{-1}$
while the outer wall with $R_2=2\times10^{-3}\mathrm{m}$ and $\Omega_2=0.02s^{-1}$.
The former has a variable slip length at the interface,
while the latter maintains a no-slip boundary condition at the interface.
Therefore, distance between the inner and outer walls is $L = 10^{-3}\mathrm{m}$.
We arrange uniformly $50$ SPH particles, that is $\Delta x = L/50 = 2\times10^{-5}\mathrm{m}$,
across the channel. The density of the fluid is taken as $\rho=1000\mathrm{kgm^{-3}}$ and kinematic viscosity $\nu = 10^{-6}\mathrm{m^2s^{-1}}$. For no-slip boundary conditions at both walls, the maximum velocity $v_{max}=\Omega R_2=4\times 10^{-5}\mathrm {ms^{-1}}$, which defines a Reynolds number $Re=L v_{max}/\nu=0.04$. 
Fig.~\ref{sche_taylor_fig}(b) shows a typical velocity distribution
for such flow, where for example a slip length $b^{in} = 0.1L$ is specified on the inner wall. Therefore, there is a slip velocity at the outer surface of the inner wall
and the velocity extrapolates to be zero at a distance of $b$ into the inner wall in this example.

\begin{figure*}[h!]
\centering
\subfigure[Transient results at different time with slip length $b = 0.2L$ on the inner wall and no-slip on the outer wall.]{
\includegraphics[scale=0.48]{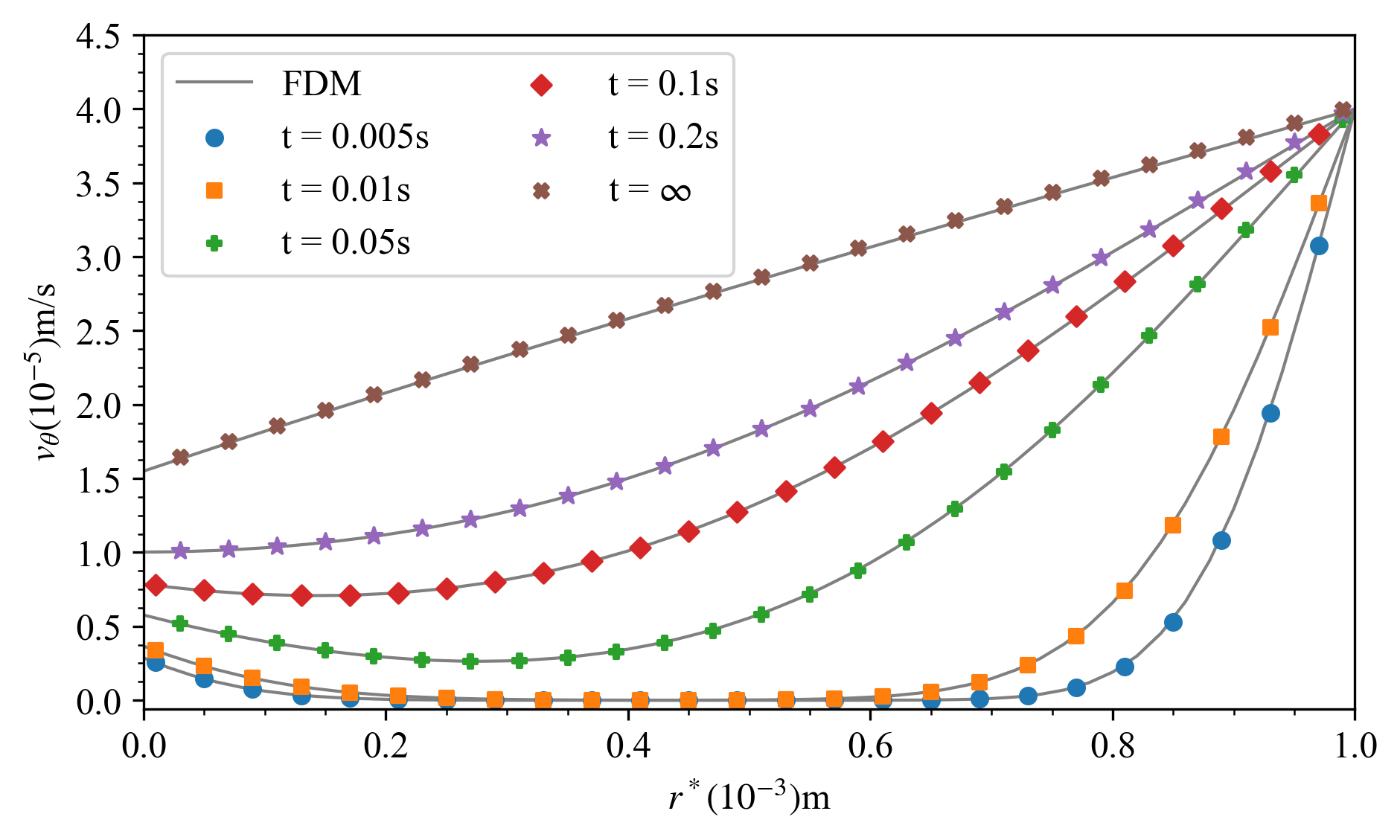}
\label{taylor-couette_transient1}}
\quad
\subfigure[Transient results at $t=0.1s$ with different slip lengths on the inner wall and no-slip on the outer wall.]{%
\includegraphics[scale=0.48]{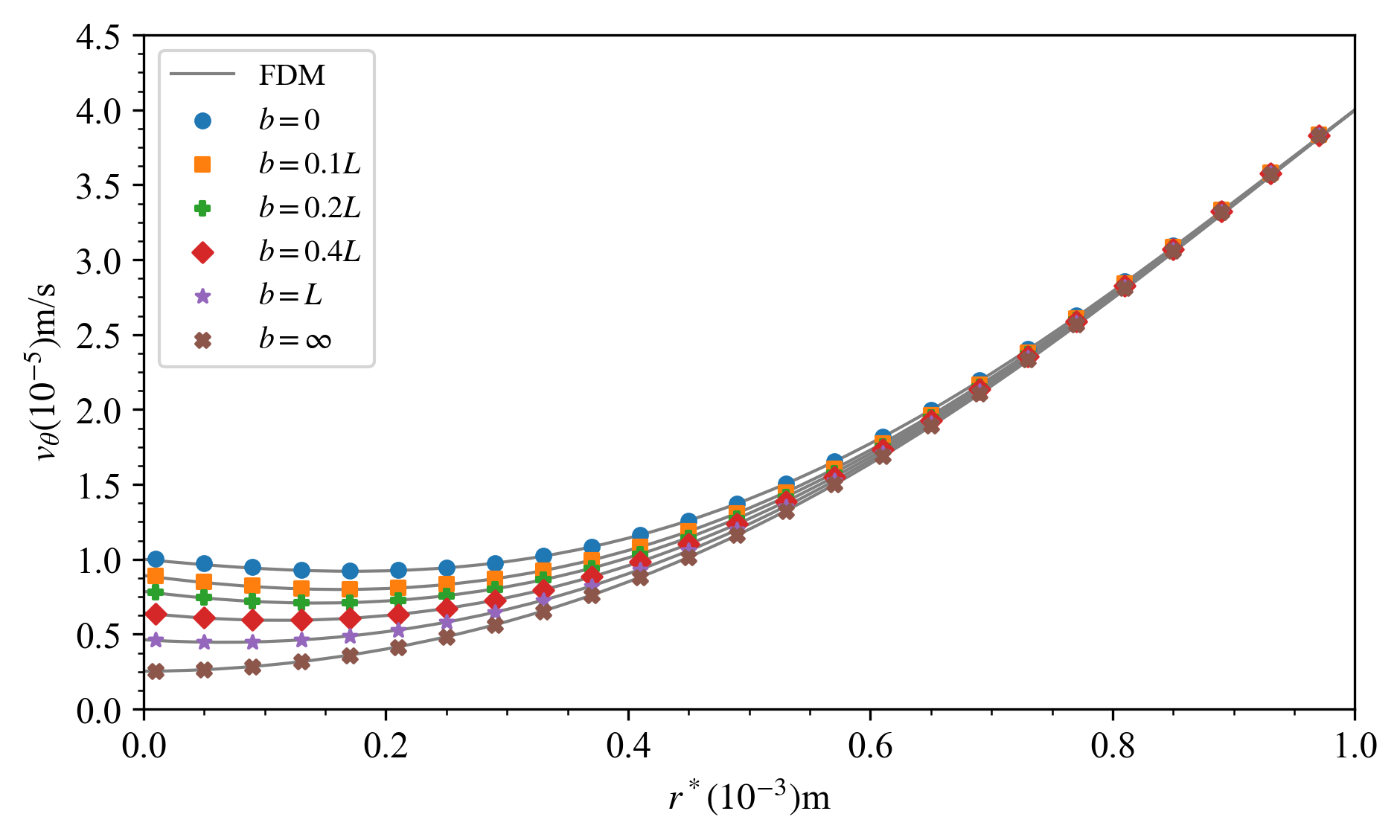}
\label{taylor-couette_transient2}}
\caption{Velocity profiles of SPH simulations for Taylor-Couette flow.}
\label{taylor-couette_transient}
\end{figure*}

For a quantitative evaluation,
we specify a slip length of $b=0.2L$ on the inner wall and no-slip on the outer wall,
and construct a set of solutions by FDM as reference
given in \ref{appendixC}.
We present transient SPH results in comparison with the reference in Fig.~\ref{taylor-couette_transient1},
where the nonlinear velocity profiles between non-planar interfaces
are well captured and the two sets of results follow each other closely.
Moreover, we aim to observe the effects of different slip lengths on the accuracy.
Therefore, we perform SPH simulations with different slip lengths at the inner wall
and correspondingly, construct reference solutions by FDM.
We select velocity profiles at time of $t=0.1s$ to
show in Fig.~\ref{taylor-couette_transient2},
where the two sets of results remain consistent overall.
Results of SPH and FDM agree with each other at other time instants,
we omit them to avoid redundancy.

\begin{figure}[h!]
\centering \mbox{ 
\subfigure[Slip length $b=0.1L$ on the inner wall and no slip on the outer wall]{\includegraphics[scale=0.48]{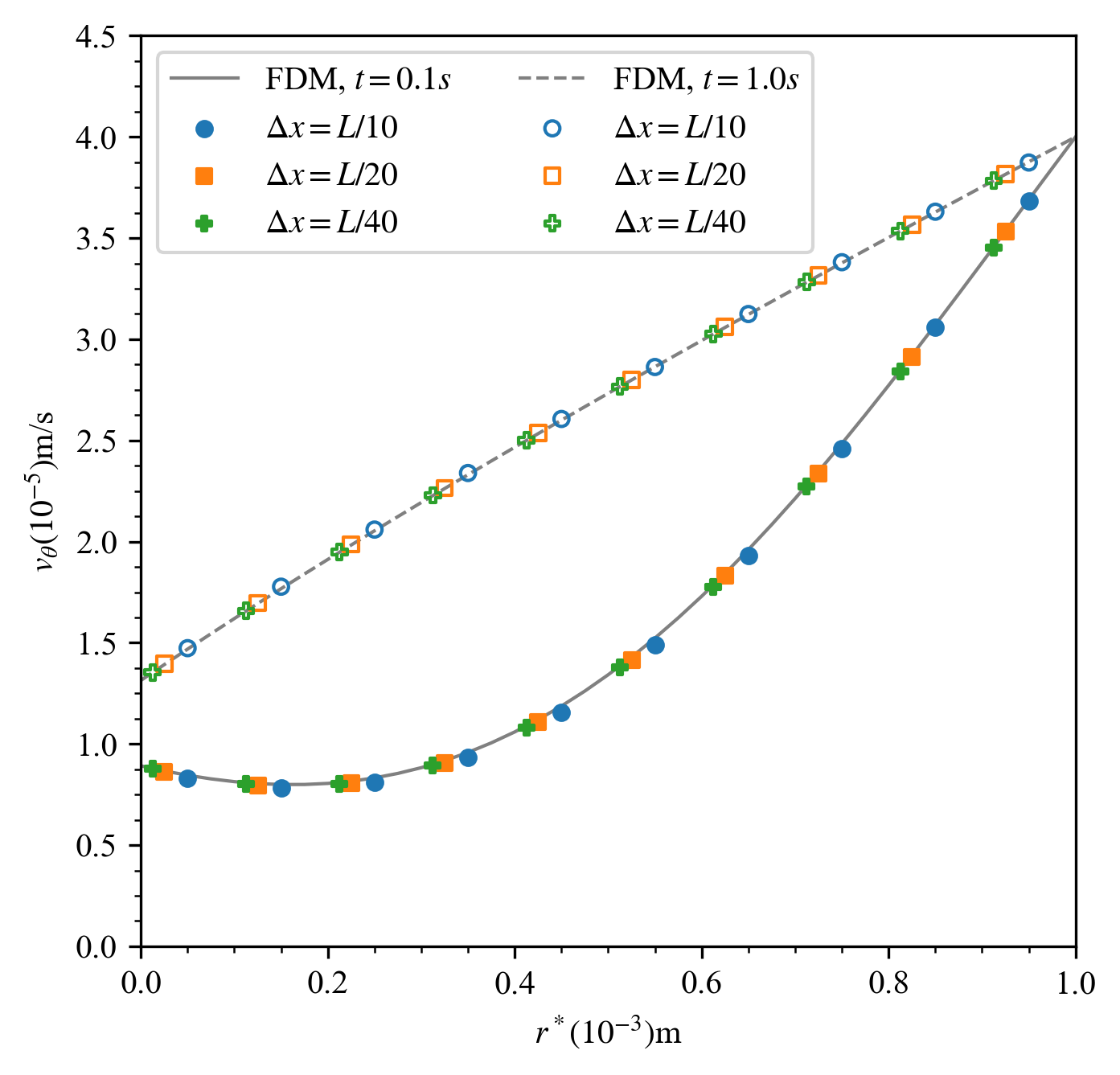}\label{taylor_couette_convergency_study1}}\quad\quad
\subfigure[Identical slip lengths of $b=L$ on both inner wall the outer wall]{\includegraphics[scale=0.48]{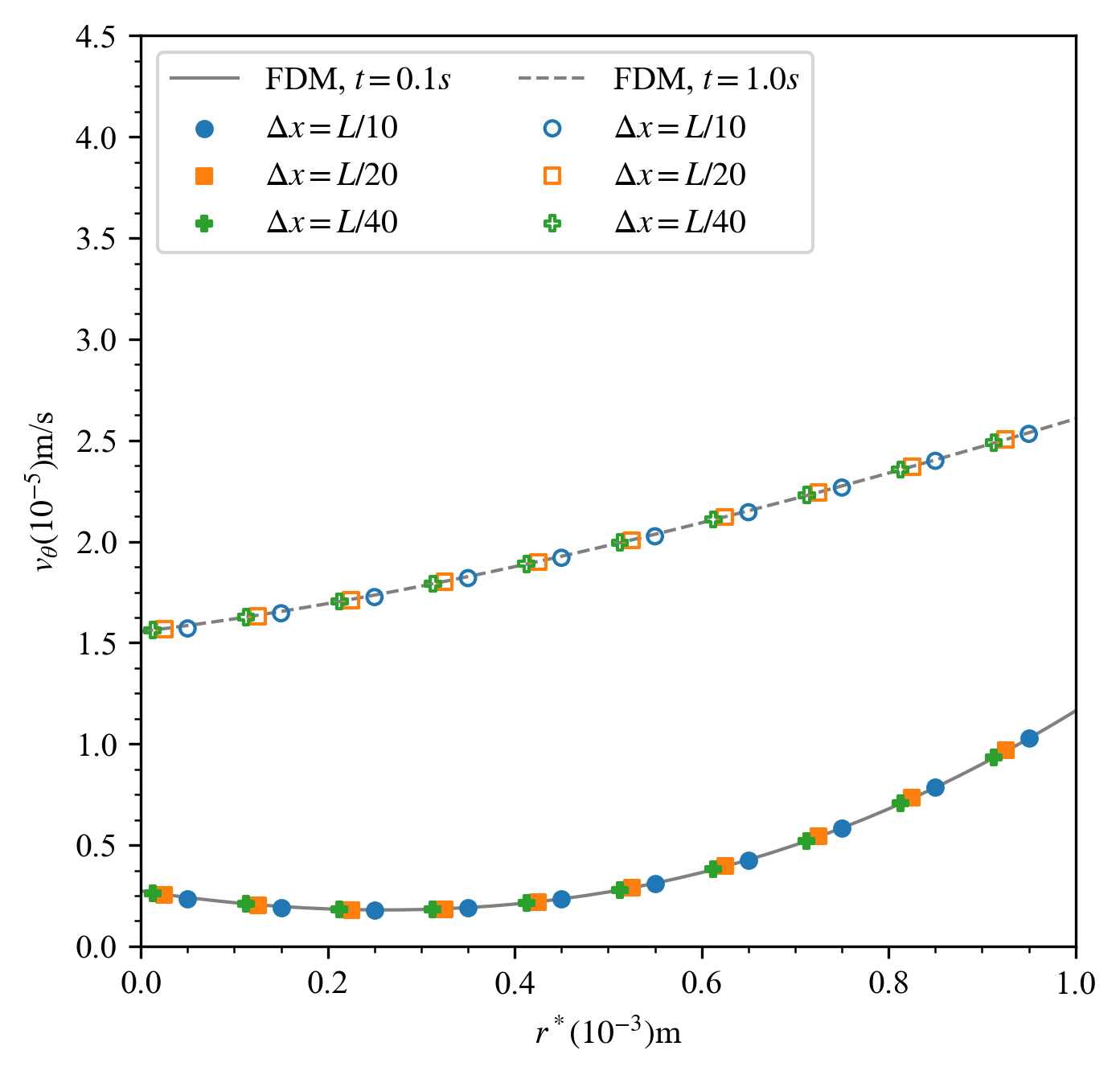}\label{taylor_couette_convergency_study2}}} 
\caption{Convergence study on $\Delta x$ with $r_c=3\Delta x$ for Taylor-Couette flow:  two selective time instants at $t=0.1s$ and $t=1.0s$.}\label{taylor_couette_convergency_study}
\end{figure}

Furthermore, we perform convergence study on the particle resolution.
We present results at two selective time instants $t=0.1\mathrm{s}$ and $1.0\mathrm{s}$ in Fig.~\ref{taylor_couette_convergency_study}.
On Fig.~\ref{taylor_couette_convergency_study1},  
we present show results for a slip length $b=0.1L$ at the inner wall
and no-slip on the outer wall,
while on Fig.~\ref{taylor_couette_convergency_study2} results for identical slip length $b=L$ at both inner and outer walls. 
The results indicate that with resolution as low as $\Delta x=L/10$, that is $10$ particles across the channel, a slip boundary condition is well described.

\begin{figure}[h!]
    \centering \includegraphics[scale=0.6]{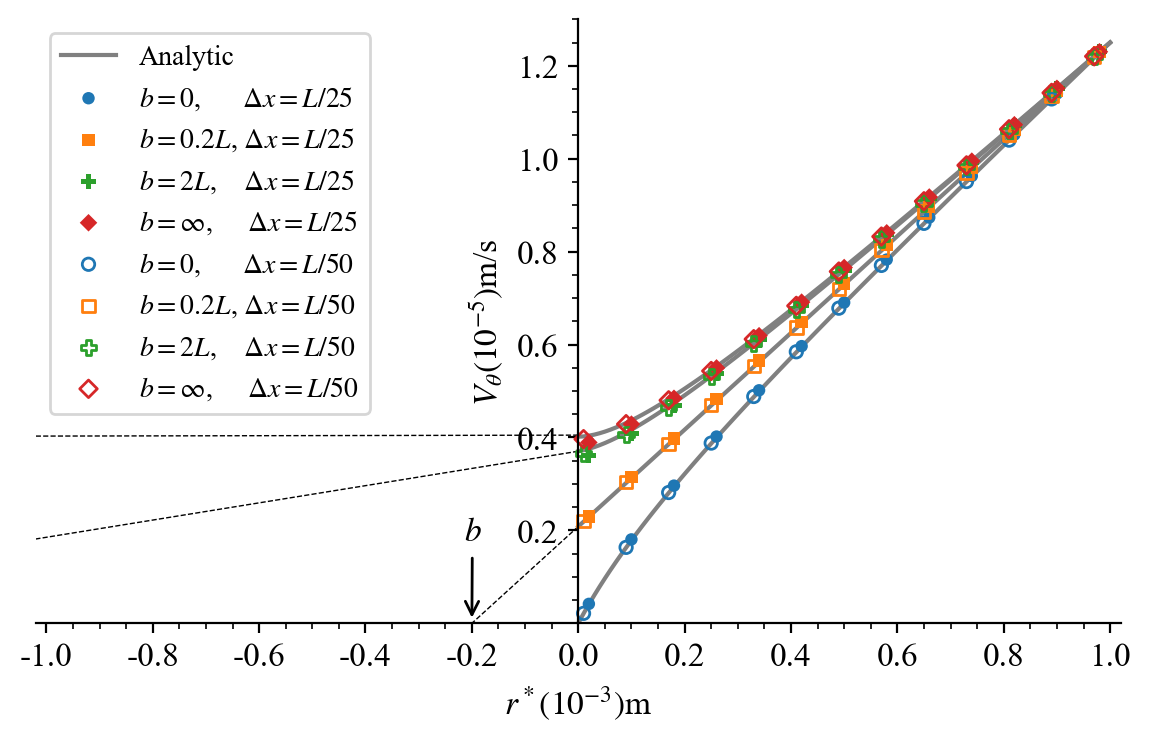} \caption{Velocity profiles of SPH simulations for Taylor-Couette flow at steady states:  different slip lengths $b$ on  the inner wall and no-slip on the outer wall. The inner wall remains still and the outer wall rotates with angular velocity $1.04\times 10^{-2}\mathrm{s^{-1}}$. $R_1=0.2\times 10^{-3} \mathrm{m}$, $R_2=1.2\times 10^{-3} \mathrm{m}$ and channel width $L = 10^{-3}m$; $\Delta x = L/50$ and
    $r_c = 3.0\Delta x$. }\label{taylor_couette_2}
\end{figure}

There is another way to perform the resolution study, that is,
increasing the curvature or reducing the radii of the walls.
Here, we reduce inner radius to be $5$ times smaller and keep the distance of the two walls the same,
that is, $R_1=2\times10^{-4}m$ and $R_2=1.2\times10^{-3}m$.
The outer wall with an angular velocity of $\Omega=1.04\times10^{-2}$ has no slip boundary condition  while the inner wall remains still and may have various slip lengths.
We compare velocity profiles of SPH simulations  at steady states with analytical solutions in Eq.~(\ref{taylor_general_solution}) for various slip lengths, as shown in Fig.~\ref{taylor_couette_2}. 
 We observe that the SPH results with $5$ times more particles, being consistent with increasing inner curvature $5$ times, agree well with analytical solutions.
The dashed lines of velocity extend into the inner wall and their intersections with the horizontal axis represent the slip lengths.

\subsection{Dean flow}
We introduce an azimuthal pressure gradient in Taylor-Couette flow
and this is so called Dean flow~\citep{dean1928fluid}.
In our simulations, we let both the inner and outer walls remain stationary with various identical slip lengths. 
To imitate pressure gradient, an azimuthal body force $F_{\theta}$ is applied everywhere to drive the flow. The general solution for the steady Dean flow is
\begin{equation}
    v_{\theta}=Ar+\frac{B}{r}-\frac{F_{\theta}}{3\nu}r^2,
\end{equation}
which recovers Eq.~(\ref{taylor_general_solution}) without body force.
When the same slip boundary conditions as Eq.~(\ref{taylor_boundary_condition})
with $\Omega_1=\Omega_2=0$ are specified,
we can solver for the coefficients $A$ and $B$ as:
\begin{eqnarray}
    A&=&\frac{F_{\theta}}{3\nu}\frac{(2a-R_1)(R_2-b)R_1^3+(2b+R_2)(R_1+a)R_2^3}{R_1^2(a-R_1)(R_2-b)+R_2^2(b+R_2)(R_1+a)}, \notag \\
    B&=&-\frac{F_{\theta}}{3\nu}\frac{(2a-R_1)R_1^3}{R_1+a}+A_2\frac{R_1^2(a-R_1)}{R_1+a}.
\end{eqnarray}

\begin{figure*}[h!]
\centering
\subfigure[Velocity field  for slip length $b=2\times10^{-4}m$.]{
\includegraphics[scale=0.56]{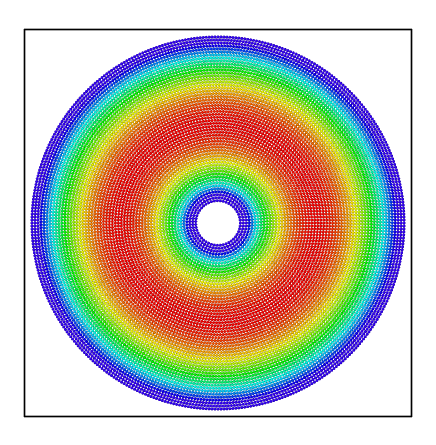}
\label{dean_sche}}
\quad
\subfigure[Velocity profiles for different slip lengths.]{%
\includegraphics[scale=0.48]{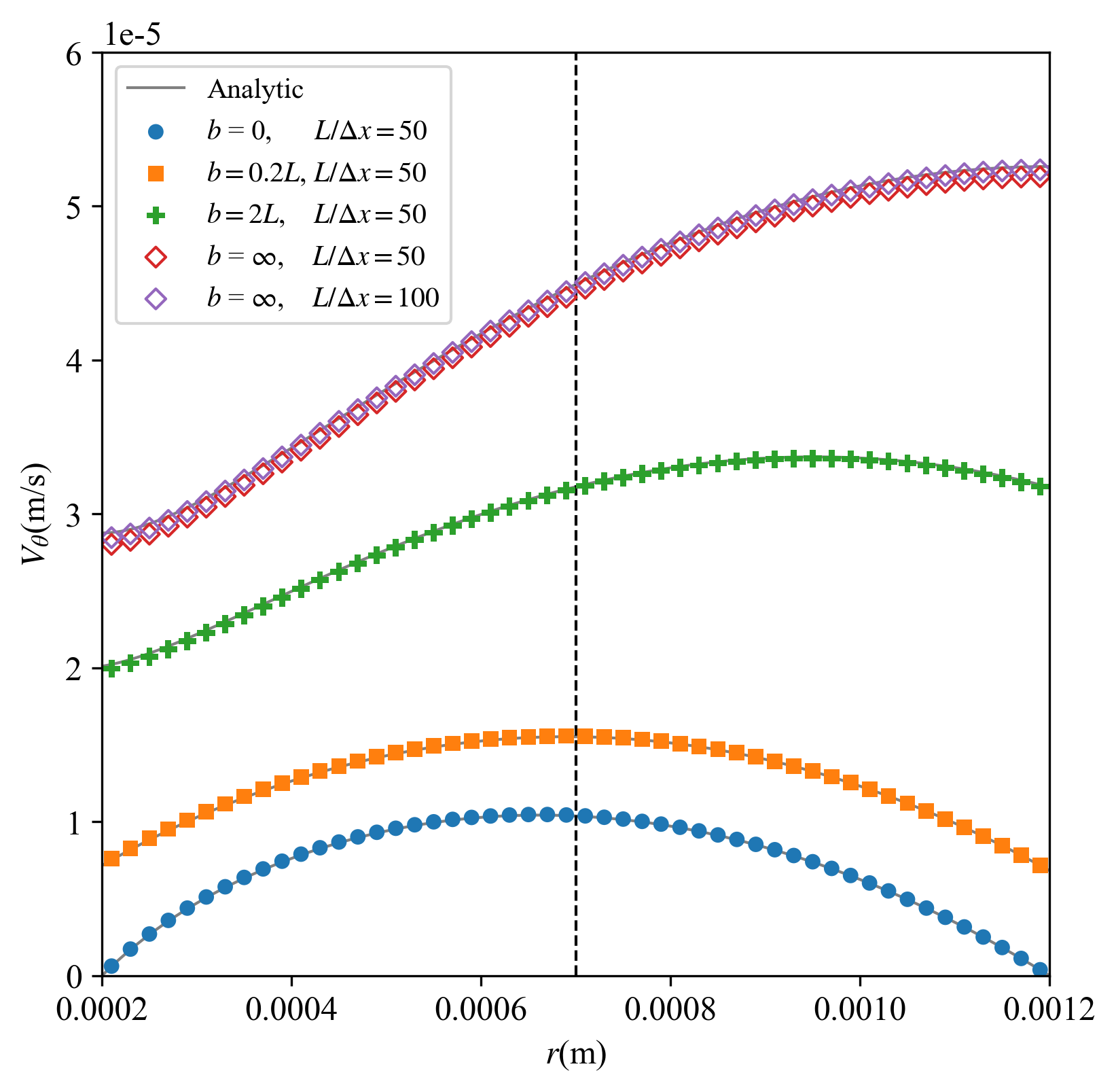}
\label{dean_velocity_profile}}
\caption{Velocity of SPH simulations for Dean flow. $R_1=2\times10^{-4}m$,$R_2=1.2\times10^{-3}m$  and $L=10^{-3}m$. Various identical slip lengths $b$ are specified on both inner and outer walls.}
\label{dean_velocity}
\end{figure*}
 The density of the fluid is taken as $\rho=1000\mathrm{kgm^{-3}}$ and kinematic viscosity $\nu = 10^{-6}\mathrm{m^2s^{-1}}$. The radii of the inner and outer walls are $R_1=2\times10^{-4}\mathrm{m}$ and $R_2=1.2\times10^{-3}\mathrm{m}$, respectively. The body force is $F_{\theta}=10^{-4}\mathrm{N{kg}^{-1}}$ along 
 the tangential direction everywhere. Fig.~\ref{dean_sche} shows the velocity field for  $b=2\times10^{-4}\mathrm{m}$, where the color ranges from minimum in blue inside the boundary to maximum in red. A quantitative comparison for different slip lengths is shown in Fig.~\ref{dean_velocity_profile}, where velocity profiles of SPH simulations agree well with analytical solutions. 
 In particular, for free slip boundary condition when the viscous effects are neglected on the interfaces, results of SPH simulations converge as the resolution increases. 

\subsection{Flow through cylinders}
To examine the slip boundary condition in a more complex flow,
we consider a flow through a periodic lattice of cylinders.
The configuration is taken from Morris et al.~\citep{morris_modeling_1997},
where a cylinder of radius $R=2\times 10^{-2}\mathrm{m}$ is placed in a periodic box of size $L\times L=0.1\mathrm{m}\times 0.1\mathrm{m}$, and the flow is driven by a body force of $F=1.5\times 10^{-7}\mathrm{N{kg}^{-1}}$. The density of the fluid is $\rho=1000\mathrm{kgm^{-3}}$ and kinematic viscosity $\nu=10^{-6}\mathrm{m^2s^{-1}}$. With no slip boundary condition, the velocity of the flow is on the scale of $v=5\times 10^{-5}\mathrm{ms^{-1}}$, which defines a Reynolds number $Re=vL/\nu=1$.

\begin{figure}[h!]
    \centering \mbox{ 
    \subfigure[Path 1]{\includegraphics[scale=0.48]{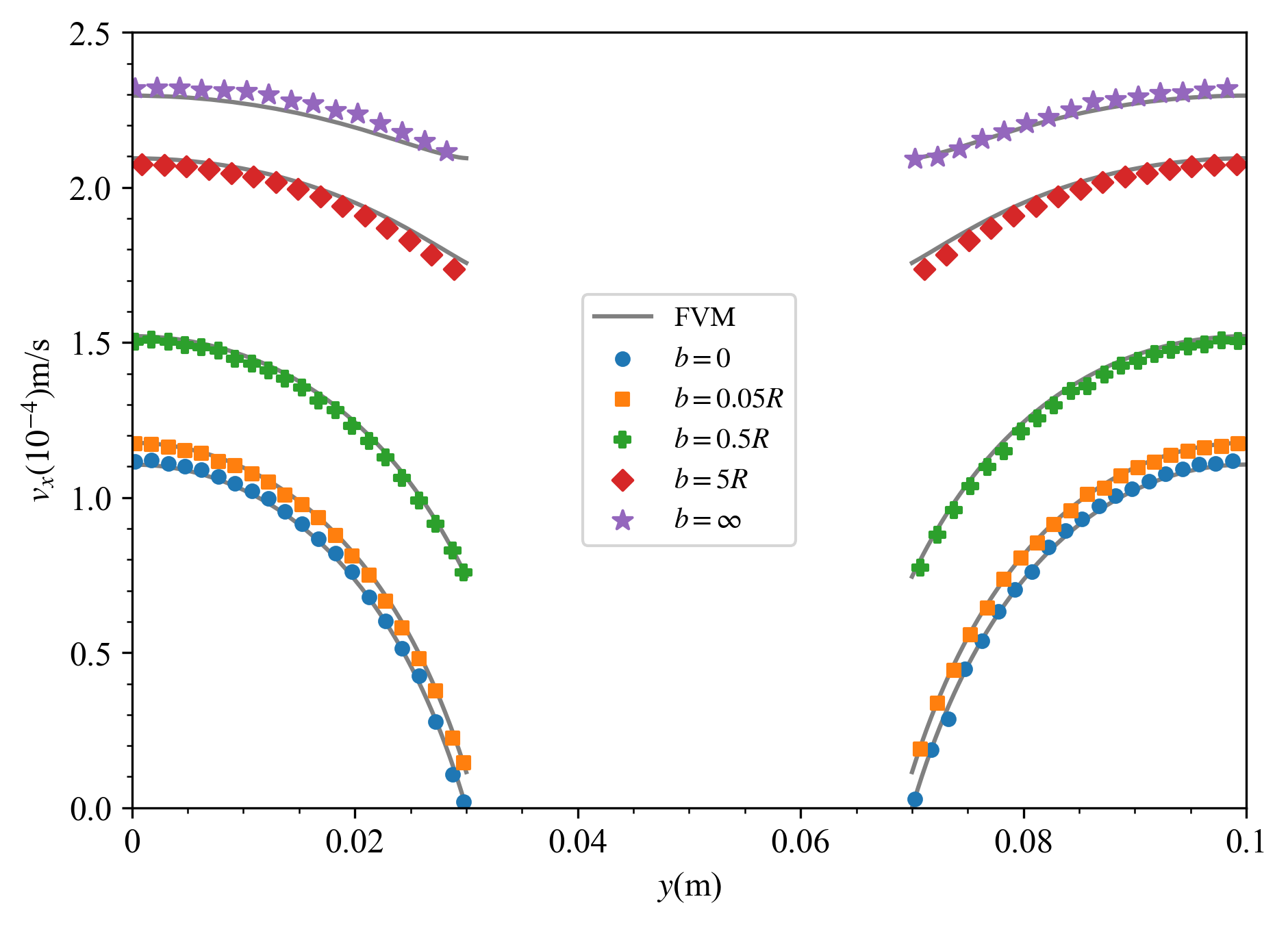}}\quad 
    \subfigure[Path 2]{\includegraphics[scale=0.48]{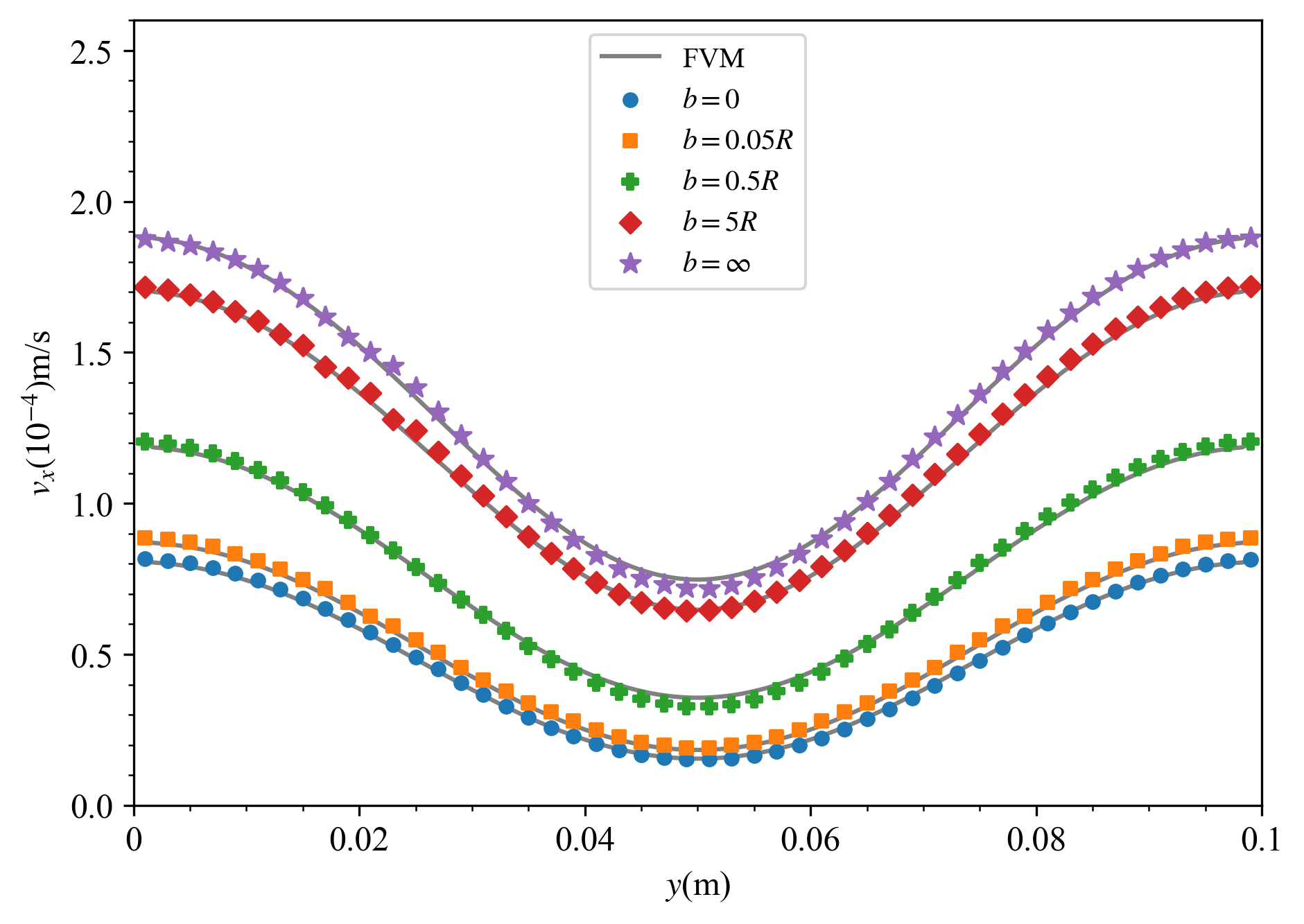}}} 
    \caption{{Velocity profiles with slip boundary condition on the surface of cylinder: SPH with $r_c=3\Delta x$ and $\Delta x=R/40$ versus FVM.}\label{cylinder_path1_2}}
\end{figure}

The flow starts at rest and becomes steady state driven by the body force in $x$ direction. 
To capture free slip ($b \gg 0$) fluid behavior, we adopt a SPH resolution of $\Delta x=R/40=5\times 10^{-4}\mathrm{m}$. 
To validate SPH simulations, we also construct corresponding solutions by finite volume method (FVM) for incompressible flows at steady states.
In particular, we consider velocity profiles across two particular cross section:
path $1$ is a line along $y$ direction and passes through the center of the cylinder;
path $2$ is another line along $y$ direction that is furthest away from the cylinder.
The velocity profiles for path $1$ and $2$ with different slip lengths on the surface of the cylinder are shown in Fig.~\ref{cylinder_path1_2}. Overall, the SPH results are in close agreement with those of FVM. When the slip length $b=0$, the results also are consistent with Morris et al.'s results \citep{morris_modeling_1997}. 

\begin{figure}[h!]
    \centering \mbox{ 
    \subfigure[Path 1]{\includegraphics[scale=0.48]{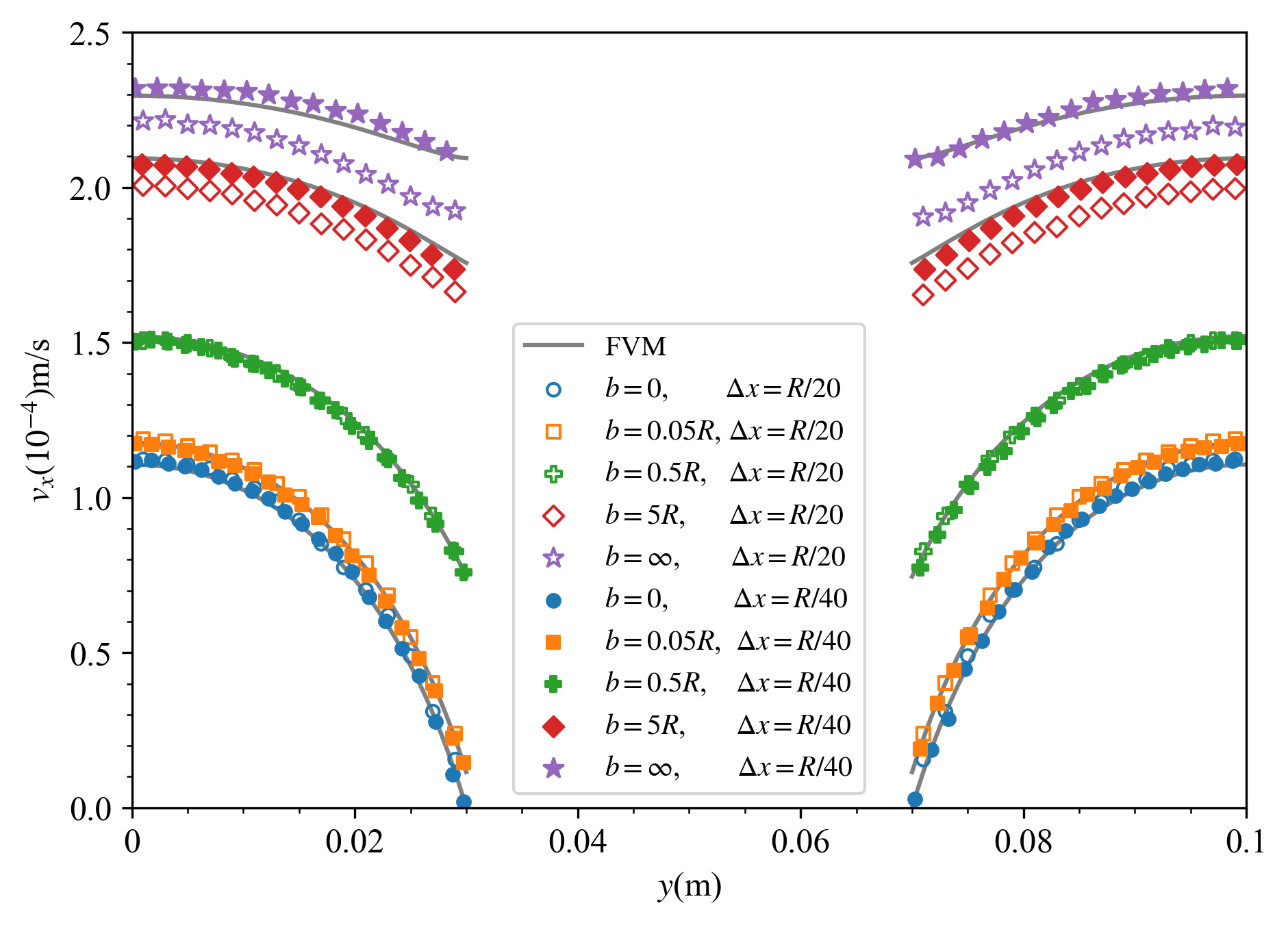}}\quad 
    \subfigure[Path 2]{\includegraphics[scale=0.48]{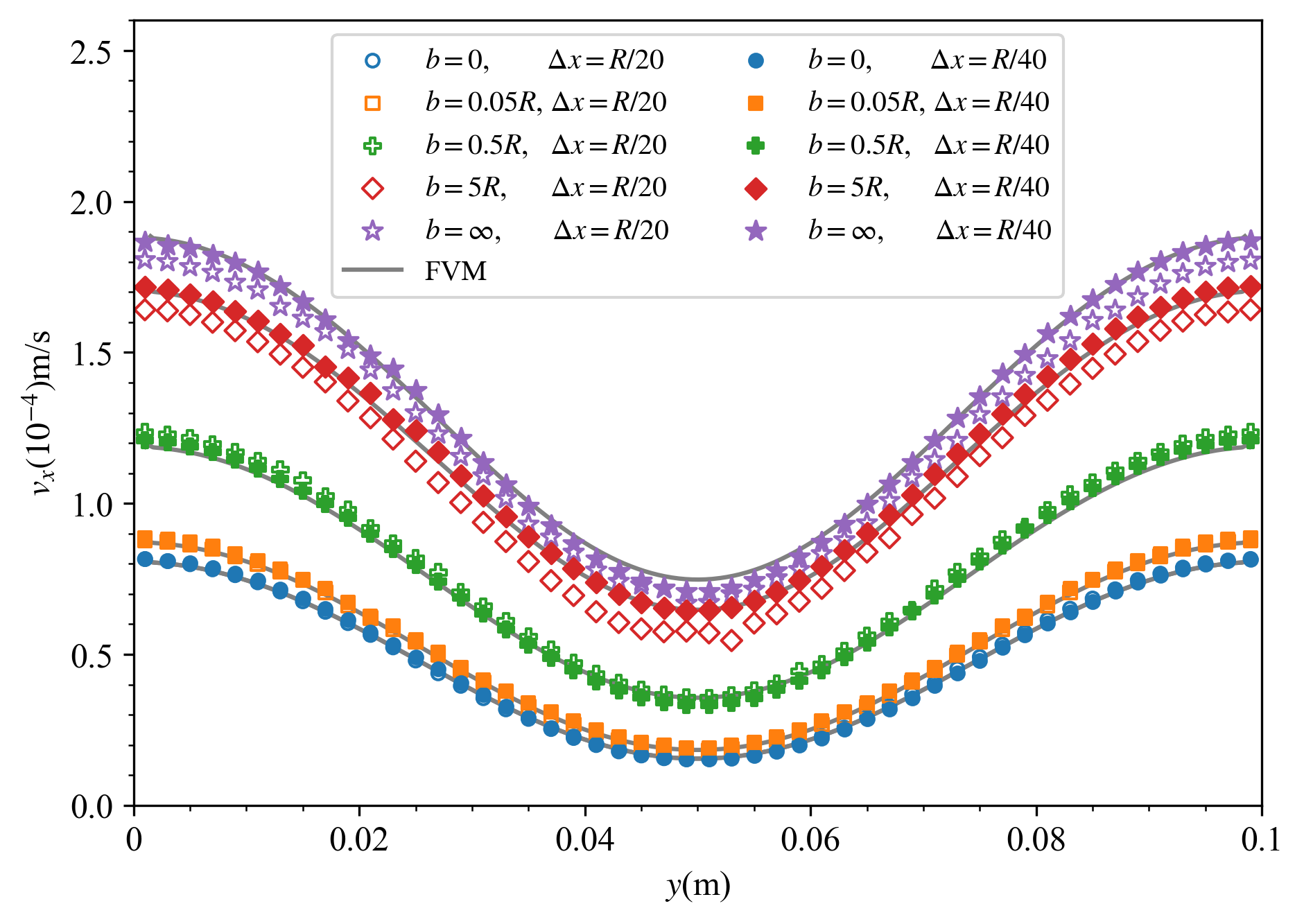}}} 
    \caption{Convergence study on SPH particle resolution with $r_c=3\Delta x$.}\label{cylinder_convergence}
\end{figure}

We also perform simulations with lower SPH resolution at $\Delta x=R/20$, and show the results in Fig.~\ref{cylinder_convergence}. We observe that when the slip length is small ($b \leq 0.5R$), a low resolution is sufficient; when the slip length is substantial ($b \geq 5R$), a high resolution is necessary.

\begin{figure}[h!]
    \centering \mbox{ \subfigure[SPH]{\includegraphics[width=60mm]{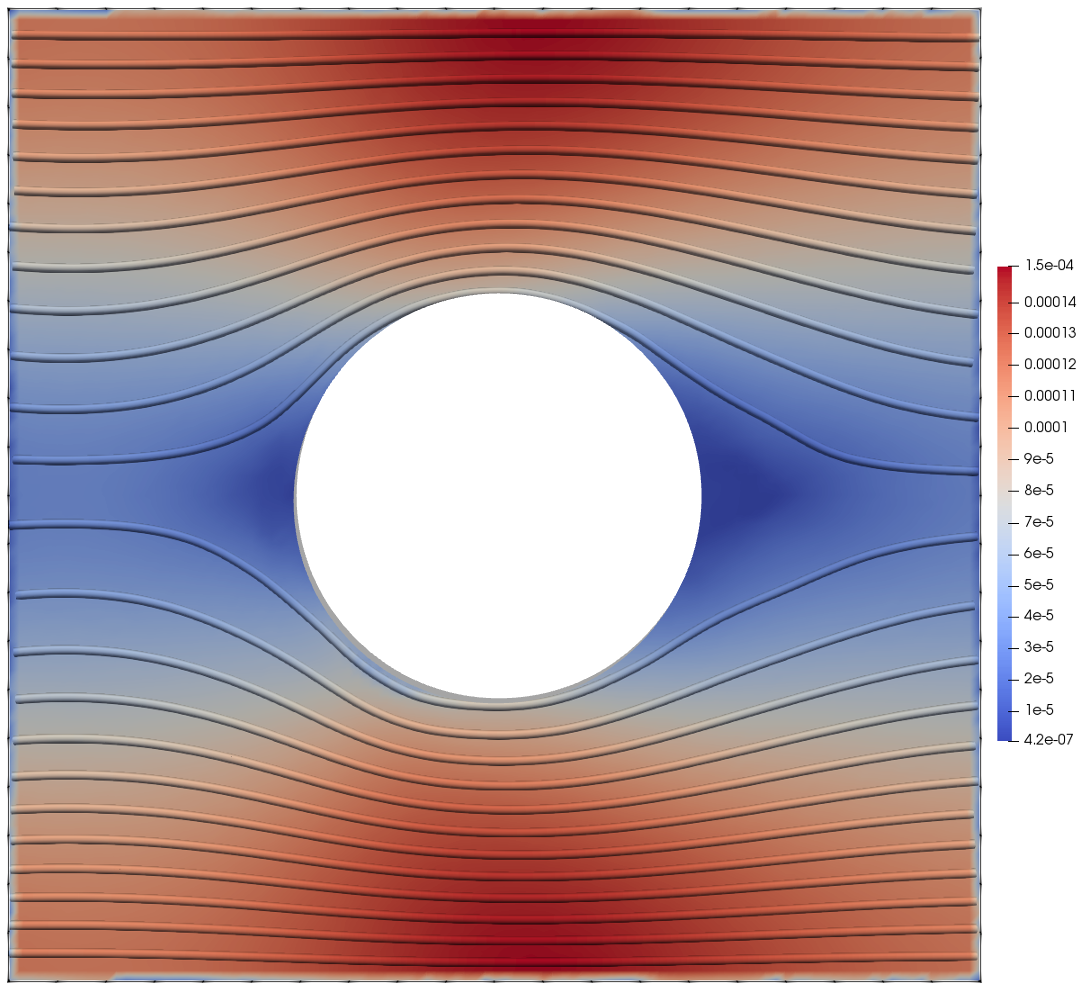}}\quad \subfigure[FVM]{\includegraphics[width=65mm]{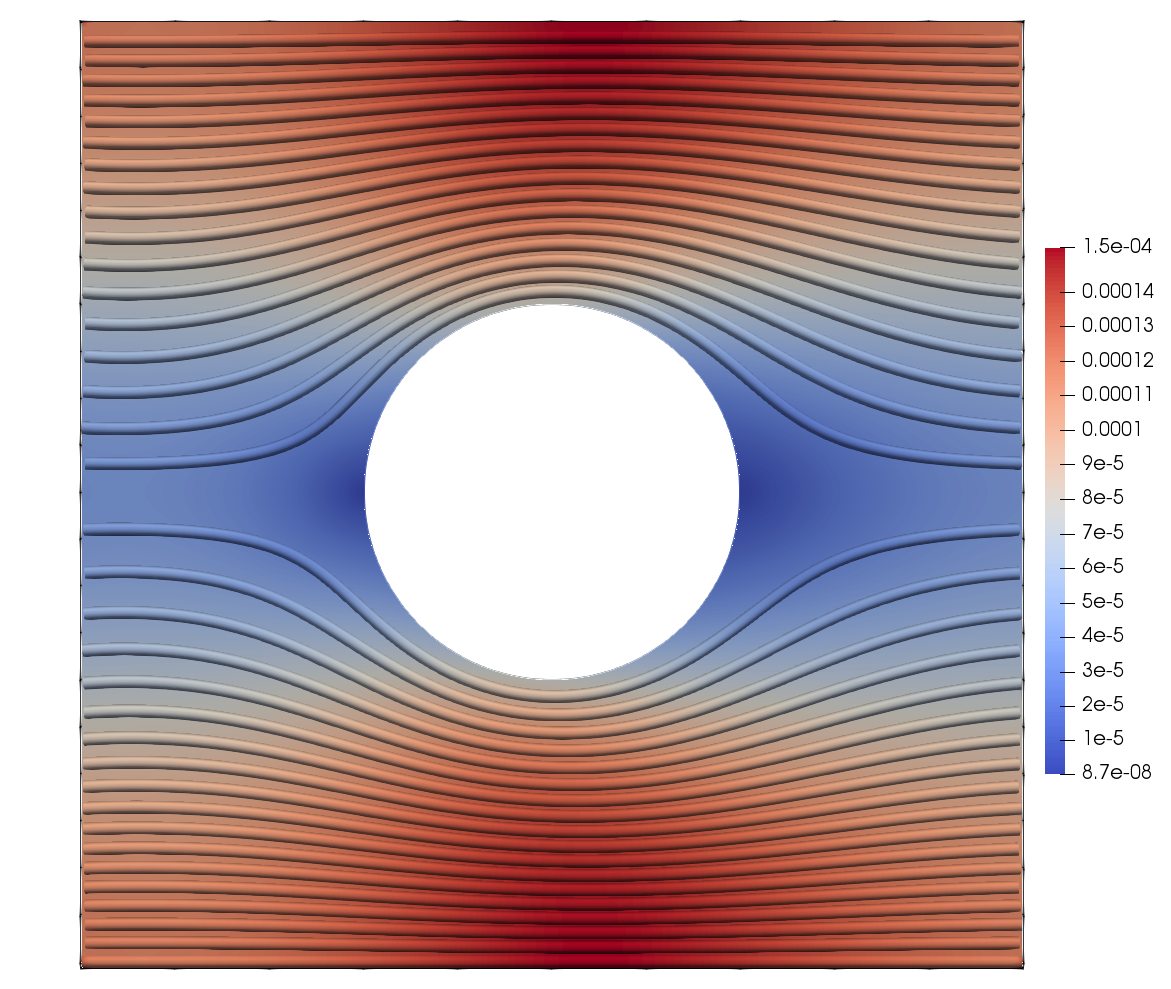}}} 
    \caption{Velocity field and streamlines for flow around cylinder: comparison between SPH with FVM for slip length $b=0.5R$ on the solid surface.}\label{streamline_cylinder}
\end{figure}

Furthermore, in Fig.~\ref{streamline_cylinder} we compare flow field and streamlines between results of SPH and FVM, where slip length $b=0.5R$ on the solid surface.
The SPH results are obtained by interpolation of particles onto a Cartesian grid using kernel functions. 
We observe that results of SPH simulations agree well with those of FVM.

\begin{figure}[h!]
\centering \includegraphics[width=80mm]{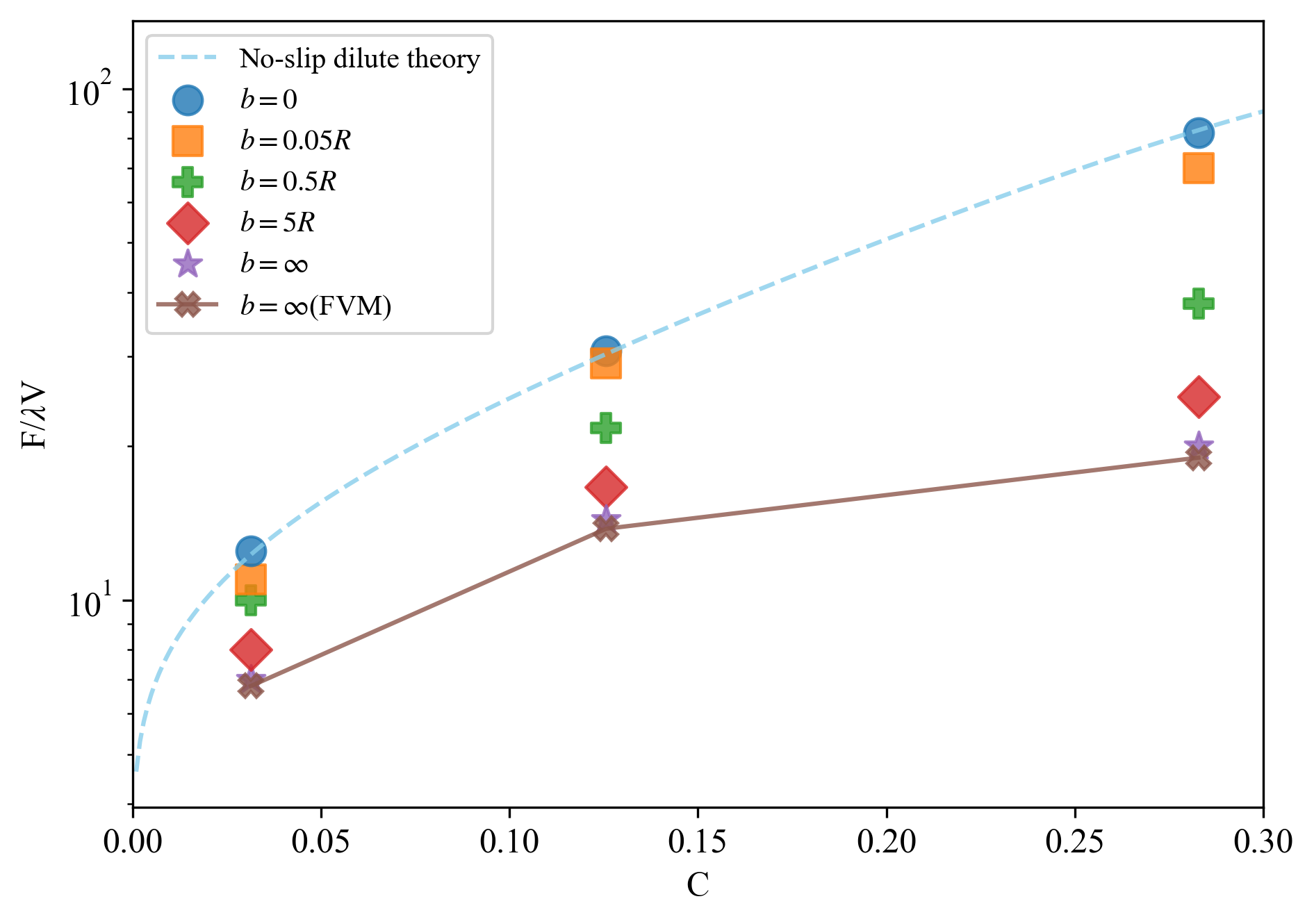} \caption{Drag coefficients of flow around cylinder by SPH simulations. Concentration $C = \pi R^2 / L^2$. For the no-slip boundary condition,
the reference is the dilute theory. As the slip length increases, the drag coefficient decreases. 
For the free-slip boundary condition, the reference is taken from FVM simulations.
}\label{drag_coeff_cylinder}
\end{figure}

Finally, we calculate drag coefficients for the cylinder with different radius and 
with different slip lengths on the surface. 
The drag coefficient is defined as
\begin{equation}
    \lambda = \frac{F}{\rho \nu V},
\end{equation}
where $F$ is the drag on the cylinder, $V$ is a far-field velocity. 
In Fig.~\ref{drag_coeff_cylinder}, we present the drag coefficients at different slip lengths for radii of $R = 1\times 10^{-2}\mathrm{m}$, $2\times 10^{-2}\mathrm{m}$, and $3\times 10^{-2}\mathrm{m}$, respectively, corresponding to different solid concentrations of $C = \pi R^2/L^2$ in the horizontal axis. 
For the results of no-slip boundary condition, we refer to previous work~\citep{sangani1982slow, bian_multiscale_2012}.
We observe that the SPH results coincide with the dilute theory for no slip boundary condition.  As the slip length increases, the drag coefficient decreases and converges to specific values.
For the free-slip boundary conditions,
we take the results of FVM simulations as reference and they coincide with those of SPH simulations.

\subsection{Flow through channels described by semi-circle and trigonometric functions}
\begin{figure}[h!]
    \centering \mbox{ 
    \subfigure[Schematic of a semi-circular channel.]{\includegraphics[width=50mm]{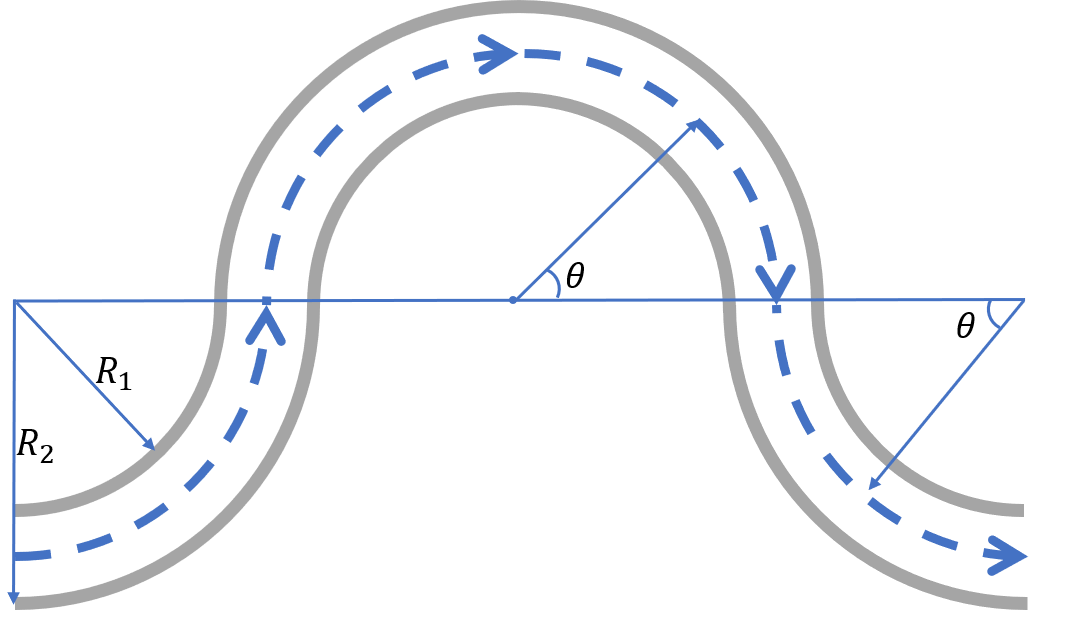}\label{circle_schem1.png}}\quad 
    \subfigure[Velocity magnitude for identical slip length $b=0.1L$ on both walls.]{\includegraphics[width=80mm]{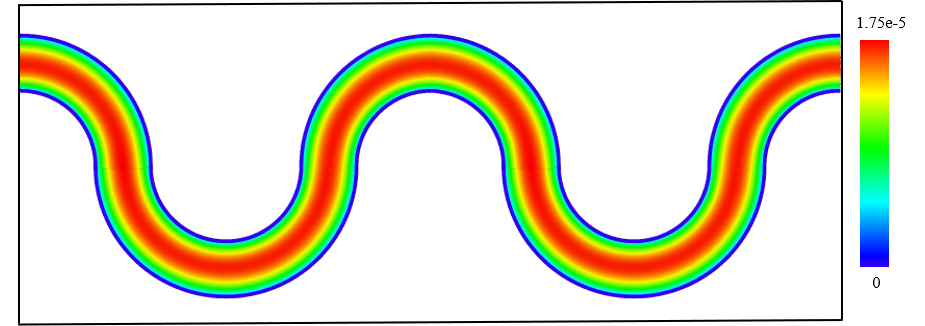}\label{circle_schem2.png}}} 
    \caption{{Periodic flow through a channel of semi-circular shape: $R_1=1.5\times10^{-3}m$, $R_2=2.5\times10^{-3}m$ and the distance between the two walls $L=10^{-3}m.$  The flow is driven by a body force tangential to the direction of the channel. }\label{circle_schem.png}}
\end{figure}

Flow in channel with curvy boundaries is very common. Without loss of generality, we first simulate flow in a channel, axis of which is described by semi-circles connected in reverse directions, as depicted in Fig \ref{circle_schem1.png}. 
Therefore, a periodic boundary condition may be applied in the direction of the flow
where a period of full circle ends.
The inner radius formed by one wall is $R_1=1.5\times 10^{-3}\mathrm{m}$ and outer radius defined by the other wall is  $R_2=2.5\times 10^{-3}\mathrm{m}$.
Therefore, the corresponding width of the channel is always $L=1\times 10^{-3}\mathrm{m}$.
The particle resolution is taken as $\Delta x=L / 50$,
which leads to $50$ particles across the channel. 
The density of the fluid is taken as $\rho=1000\mathrm{kgm^{-3}}$ and kinematic viscosity $\nu=10^{-6}\mathrm{m^2s^{-1}}$. A body force $F=10^{-4}\mathrm{N{kg}^{-1}}$ tangential to the direction of the channel is applied everywhere to drive the flow. 
A particular cross section in the channel may be referred to by the angle $\theta$ in a cylindrical coordinate, center of which coincides with the center of the curvature of the inner wall.

\begin{figure}[h!]
\centering 
\subfigure[no-slip.]{\includegraphics[width=60mm]{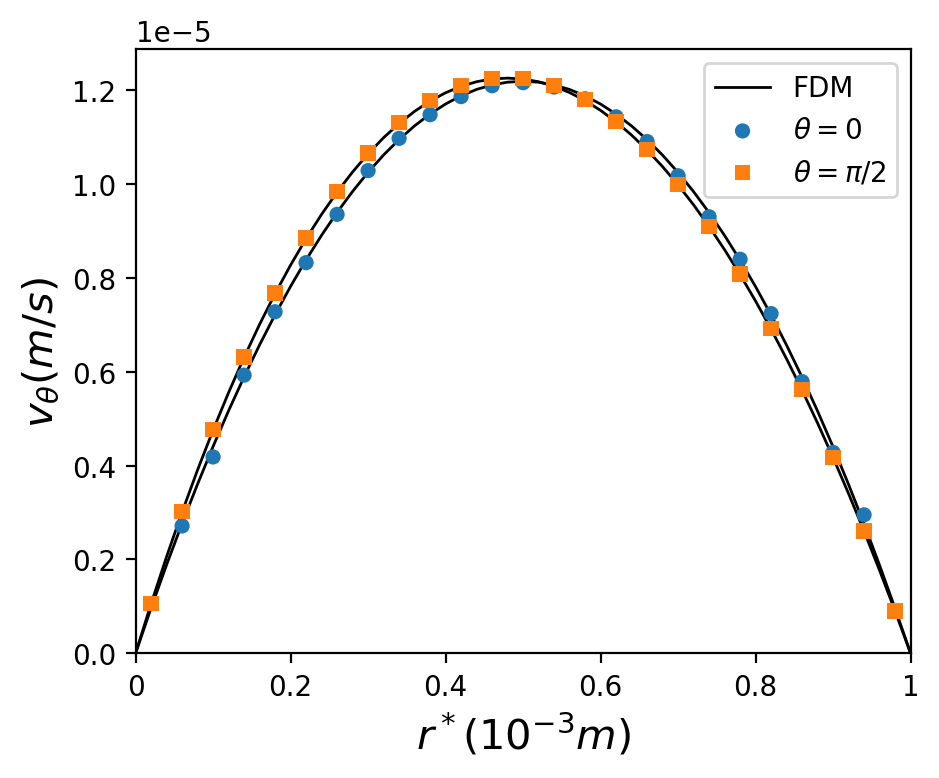}}\quad 
\subfigure[$b = 2\times 10^{-5}m$.]{\includegraphics[width=60mm]{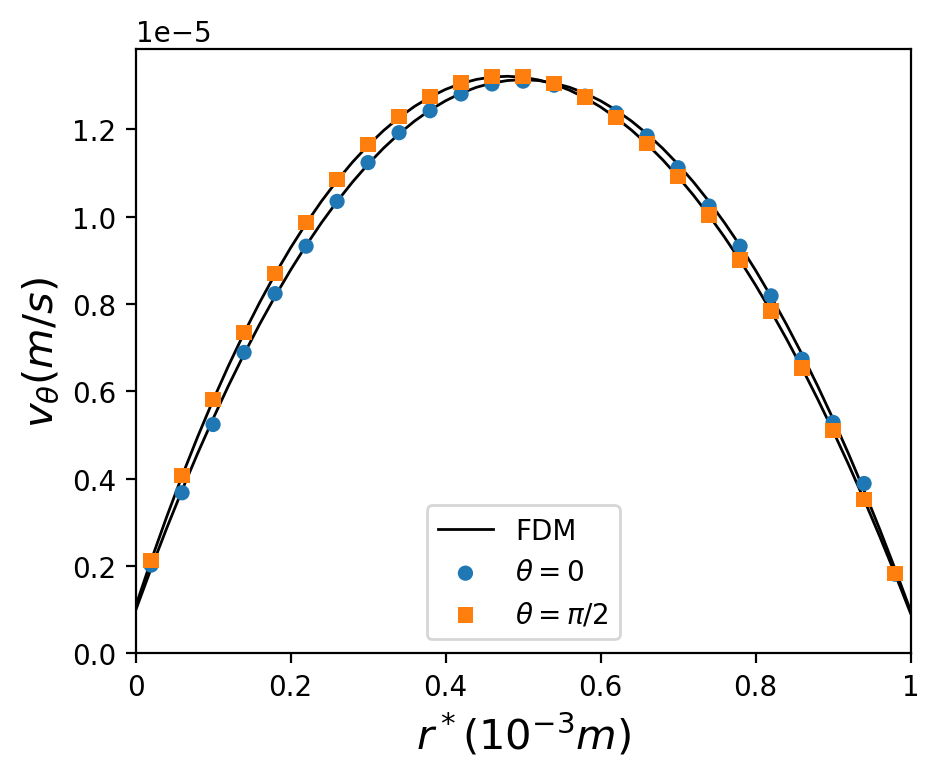}}\quad
\subfigure[$b = 1\times 10^{-4}m$.]{\includegraphics[width=60mm]{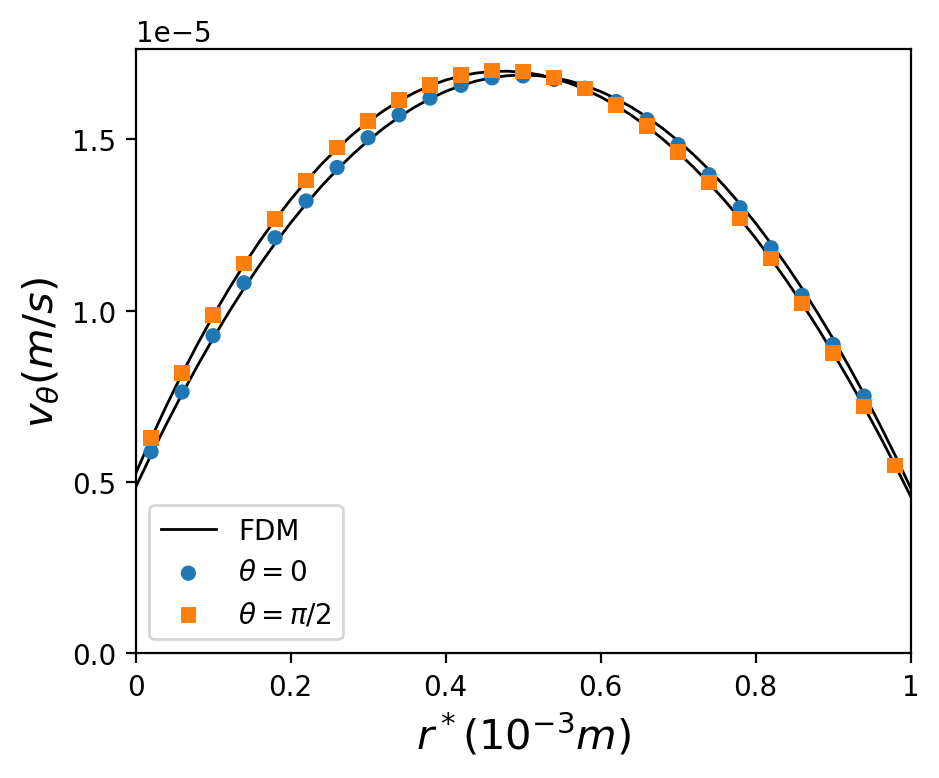}}\quad
\subfigure[$b = 2\times 10^{-4}m$.]{\includegraphics[width=60mm]{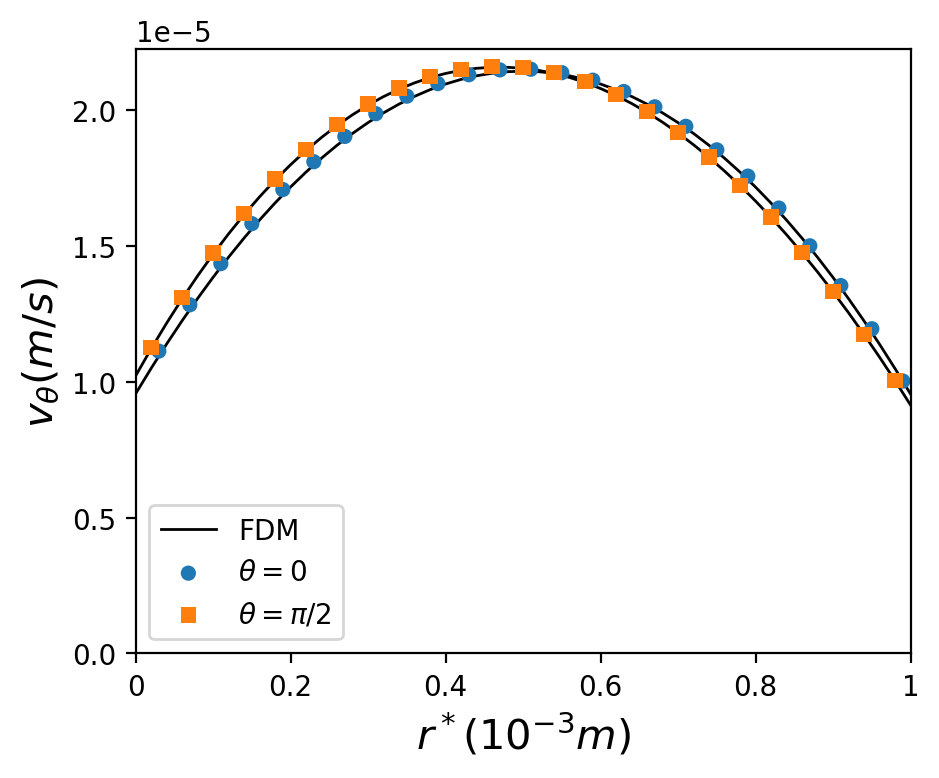}}
\caption{Velocity profiles of SPH simulations for flows in a channel of semi-circular shape: two cross sections at $\theta = 0$ and $\pi/2$ are selected. 
Both walls have identical slip lengths.
The results of FDM are taken as reference.
SPH adopts $50$ particles across the channel
while FDM also employs $50$ grid points across the same distance.
Half of the data are shown for clarity.}\label{circle_slip_vel}
\end{figure}

Fig.~\ref{circle_schem2.png} shows a snapshot of the velocity field
as an example of an identical slip length $b = 0.1L$ on both walls. 
Here blue color represents the static boundary with $v=0$, and red color indicates the maximum velocity of $v=1.75\times 10^{-5}\mathrm{ms^{-1}}$ in the middle of the channel. The velocity distribution is very similar to the Poiseuille flow or Dean flow mentioned earlier.
There are apparent slips for the velocity near the fluid-solid interfaces. 
Furthermore, we consider various slip lengths on both walls
and compare SPH results with those of FDM.
The procedure for the FDM solution is given in \ref{appendixD}.
In particular, we select two representative cross sections 
at $\theta=0$ and $\theta=\pi$ for the comparison.
The results of SPH simulations coincide with those of FDM,
as shown in Fig.~\ref{circle_slip_vel} for four slip lengths.
We emphasize that results of this flow are different from that of Dean flow, as the velocity profiles are asymmetrical due to the reverse connections of semi-circular channels.
We note that the delicate asymmetry of the flow profiles
is well captured by the SPH simulations.


\begin{figure}[h!]
    \centering 
    \subfigure[Schematic of a channel described by a cosine function.]{\includegraphics[width=60mm]{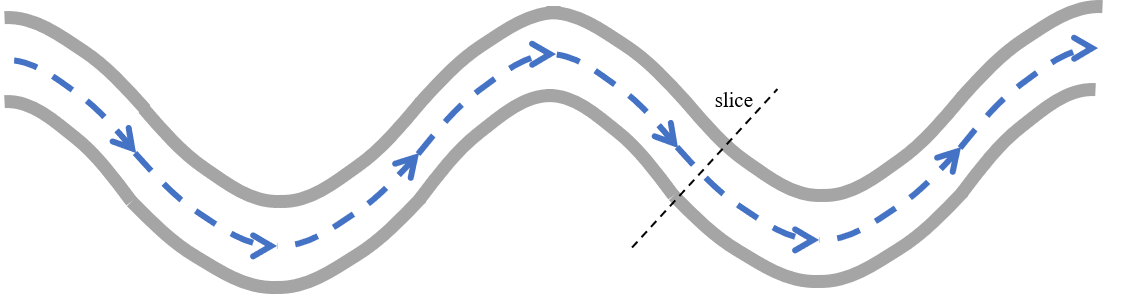}\label{cosfun1}}
    \subfigure[Velocity magnitude for identical slip length $b=0.1L$ on both walls.]{\includegraphics[width=80mm]{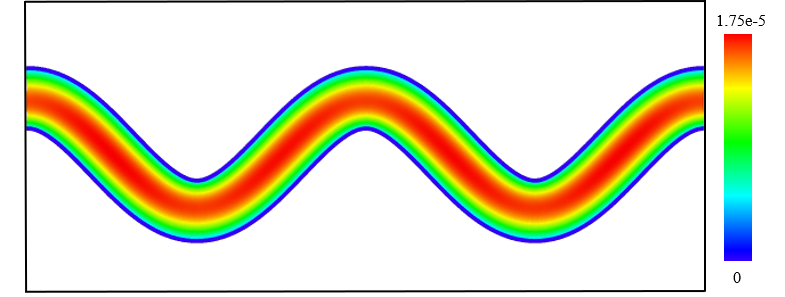}\label{cosfun2}}
    \caption{{Periodic flow through a channel described by a cosine function.
    The flow is driven by a body force tangential to the direction of the channel.}\label{cosfun}}
\end{figure}

\begin{figure}[h!]
\centering \includegraphics[width=80mm]{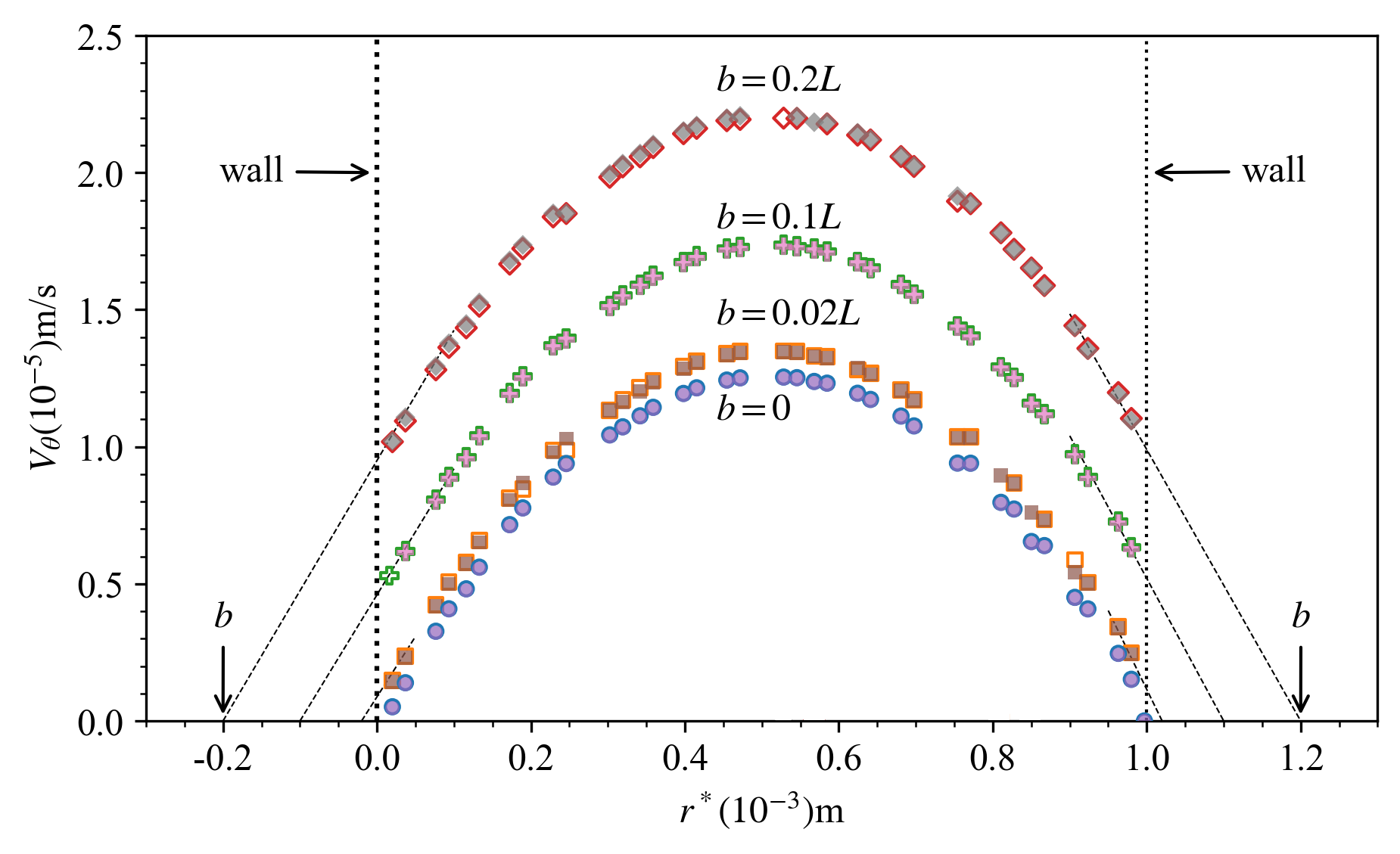} \caption{Velocity profiles of SPH simulations for flows in a channel described by a cosine function. Both walls have identical slip lengths. A cross section as drawn on Fig.~\ref{cosfun1} is considered.
The hollow symbols are from the fluid-particle-centric method while the solid symbols are from the boundary-particle-centric method. }\label{vel_cos}
\end{figure}

Next, we construct a channel described by a trigonometric function, as shown in Fig.~\ref{cosfun1}. The center-line of the channel is described a cosine function with amplitude $A=1\times10^{-3}m$ and period $T=6\times10^{-3}m$. The width of the channel is $L=1\times10^{-3}m$. The fluid's properties, and the body force are the same as the previous case of a semi-circular channel. Fig.~\ref{cosfun2} presents a snapshot of the velocity magnitude as an example of identical slip length $b=0.1L$ on both walls. 
We further consider other slip lengths on the walls
and present results from both the fluid-particle-centric (FPC)
and boundary-particle-centric (BPC) methods in Fig.~\ref{vel_cos}.
In particular, we consider the velocity profiles at the cross section of $\theta=2\pi$ as indicated in Fig.~\ref{cosfun1}.
The linear extrapolations of the velocity profiles into the wall
define the slip lengths of the simulations.
Overall the results from the FPC and BPC methods follow each other very closely,
which give us the confidence of their accuracy.

\begin{figure}[h!]
\centering \mbox{ 
\subfigure[Channels of semi-circular shape.]{\includegraphics[width=60mm]{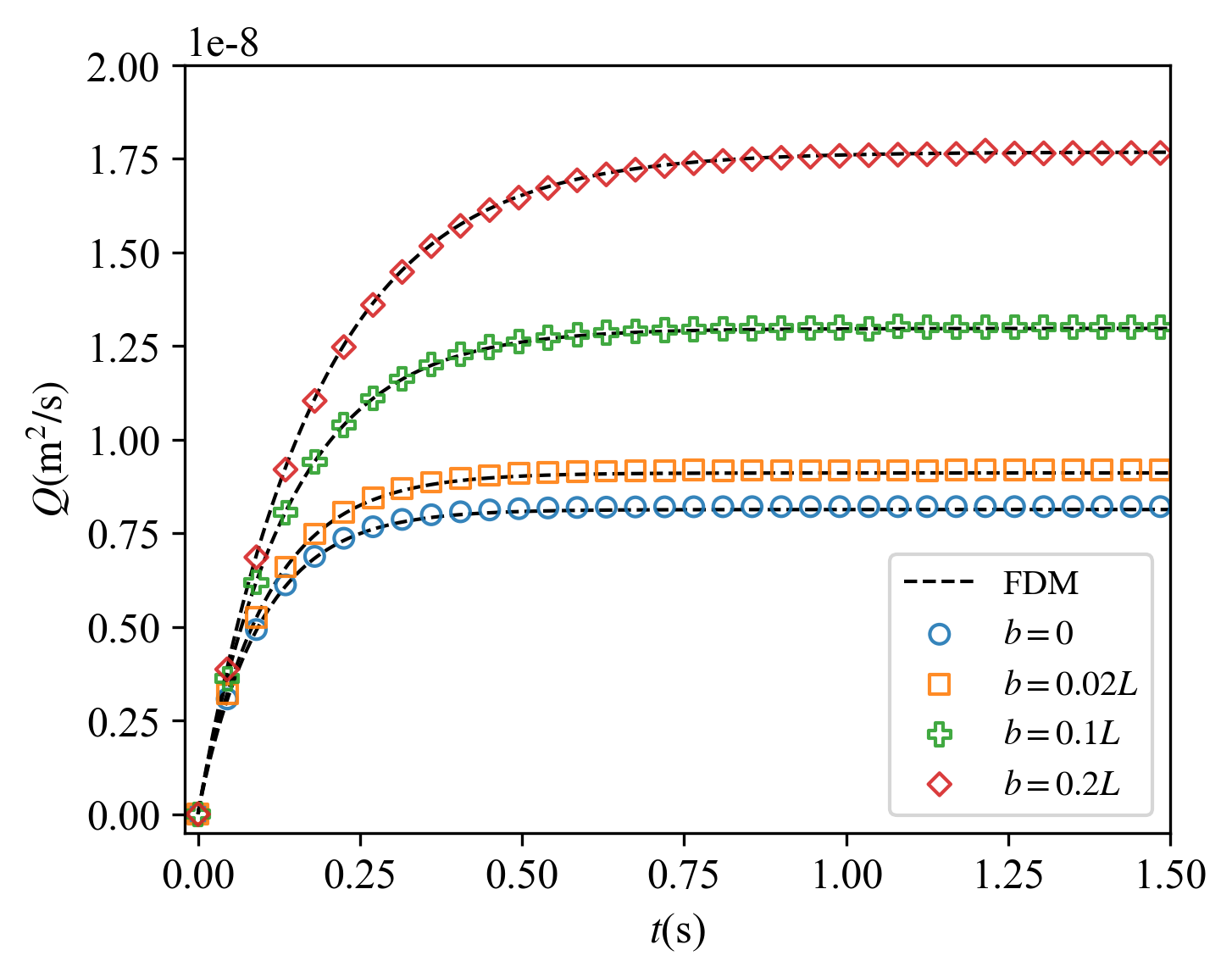}}\quad 
\subfigure[Channels described by a cosine function.]{\includegraphics[width=60mm]{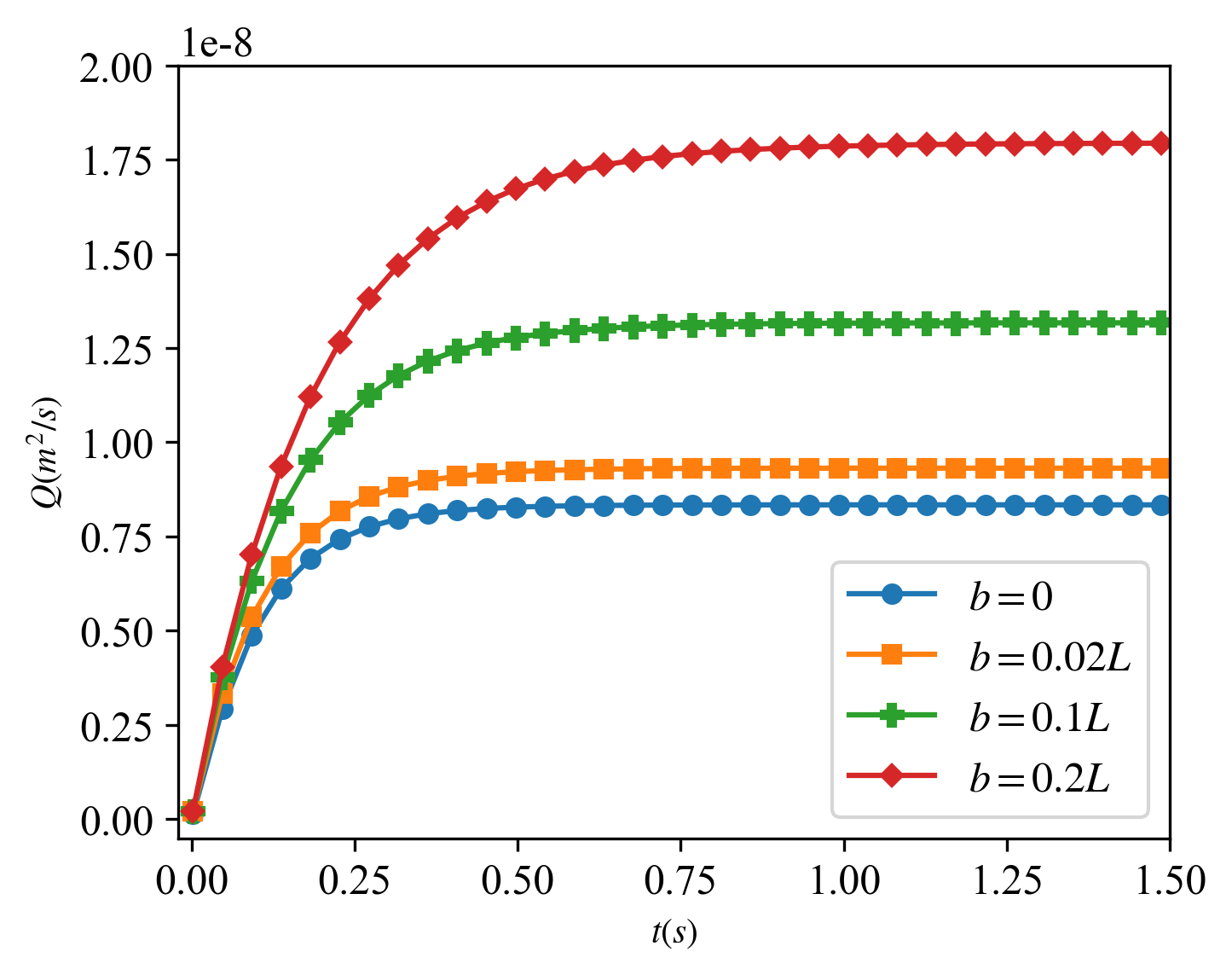}}} 
\caption{Flow volume rate for the flows in a curvy channel.}\label{flow_volume_rate}
\end{figure}

The slip boundaries enhance the flow volume rate significantly,
as shown in Fig.~\ref{flow_volume_rate}, for the time-dependent channel flows. 
For the semi-circular channels, the SPH results are in  agreement with those of FDM. 
For channels described by a cosine function,
results of the FPC and BPC methods overlap each other,
and therefore, we only present the latter.

\subsection{Flow through channels of arbitrary geometry}
\begin{figure}[h!]
\centering \includegraphics[scale=0.3]{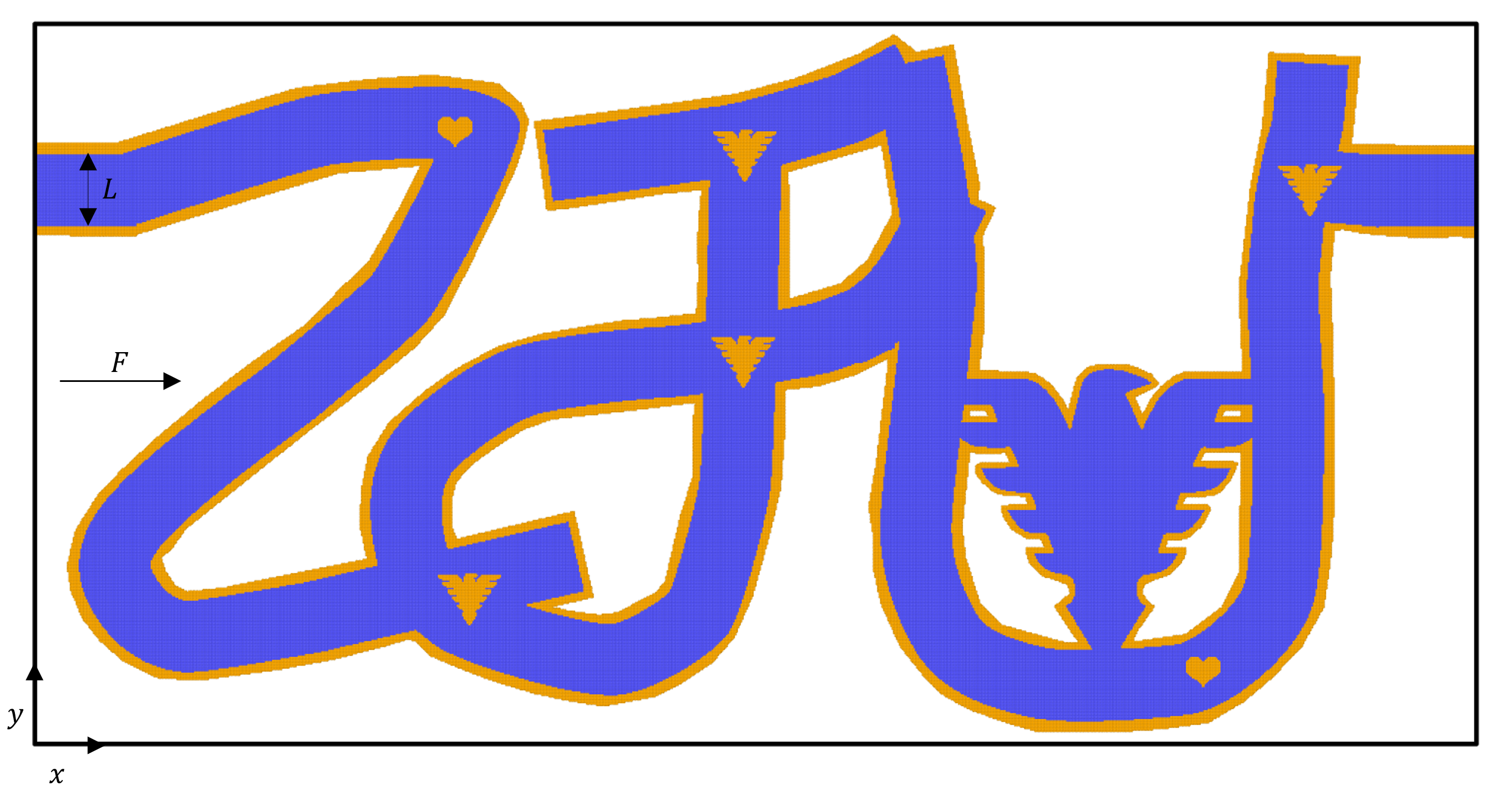} \caption{Schematic of a flow through channels of arbitrary geometry. Periodic boundary conditions are applied in $x$-direction. The yellow particles represent the wall and the blue ones represent the fluid. A constant body force is applied to the fluid in the $x$ direction.}\label{zju_sche}
\end{figure}

We further consider a flow through channels of arbitrary geometry. 
The frame of the geometry is drawn by a brush in an ordinary paint software,
as shown in Fig.~\ref{zju_sche},
where three letters of "ZJU" are written and deliberately connected 
periodically in $x$ direction.
In addition, we inlaid arbitrarily five "Seeking Truth Eagles", which are the emblem of the University, and two hearts in the channel.
The size of the entire rectangular box is $B_x\times B_y=0.02 \mathrm{m}\times 0.01\mathrm{m}$. The width of the channel is set universally as $L = 10^{-3}\mathrm{m}$, and the particle resolution is $\Delta x = L/50 = 2\times10^{-5}\mathrm{m}$, corresponding to a total of $230769$ SPH particles. The density and kinematic viscosity of the fluid are set to $\rho = 10^{3}\mathrm{kgm^{-3}}$ and $\nu = 10^{-6}\mathrm{m^2s^{-1}}$, and a constant body force $F=8\times 10^{-3}\mathrm{N{kg}^{-1}}$ is applied to the entire fluid in $x$ direction.
If there are no slip boundary conditions on the walls,
the characteristic velocity of the flow field is about $v = 10^{-4}ms^{-1}$
and the corresponding Reynolds number is $Re=vL/\nu=0.1$.
\begin{figure}[h!]
\centering 
\subfigure[no-slip.]{\includegraphics[scale=0.15]{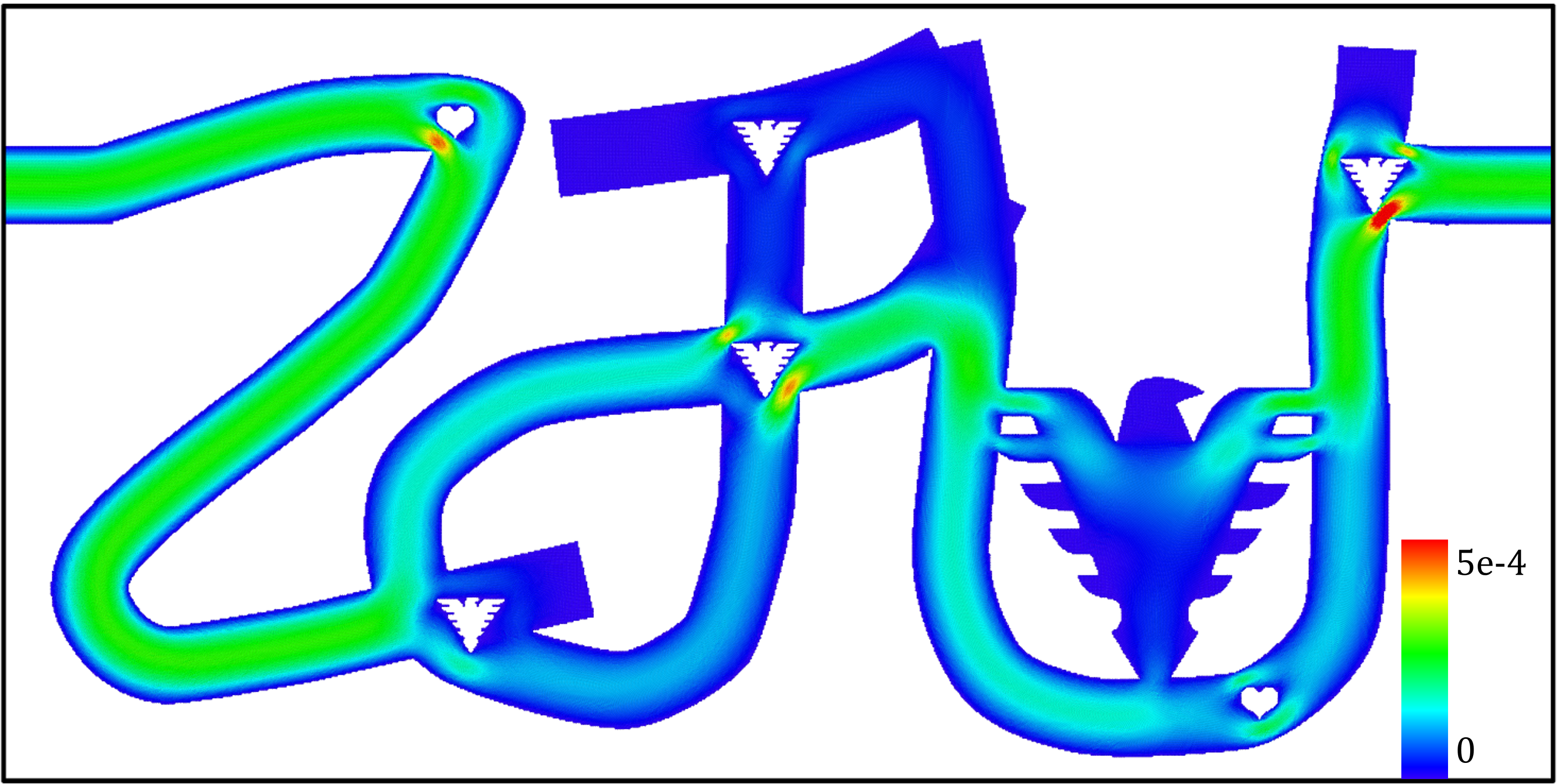}}
\subfigure[$b = 0.2L$.]{\includegraphics[scale=0.15]{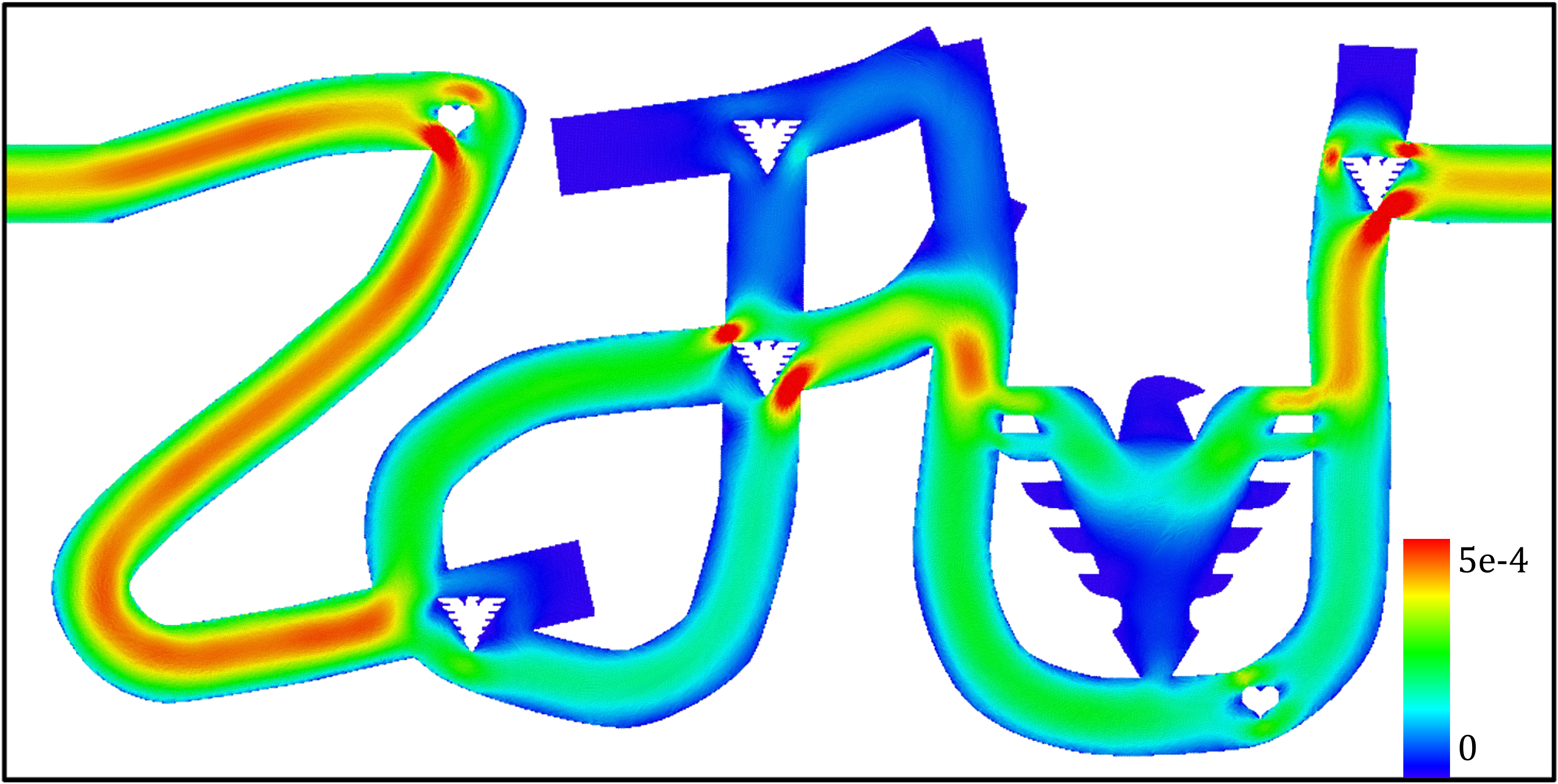}}
\subfigure[$b = 0.5L$.]{\includegraphics[scale=0.15]{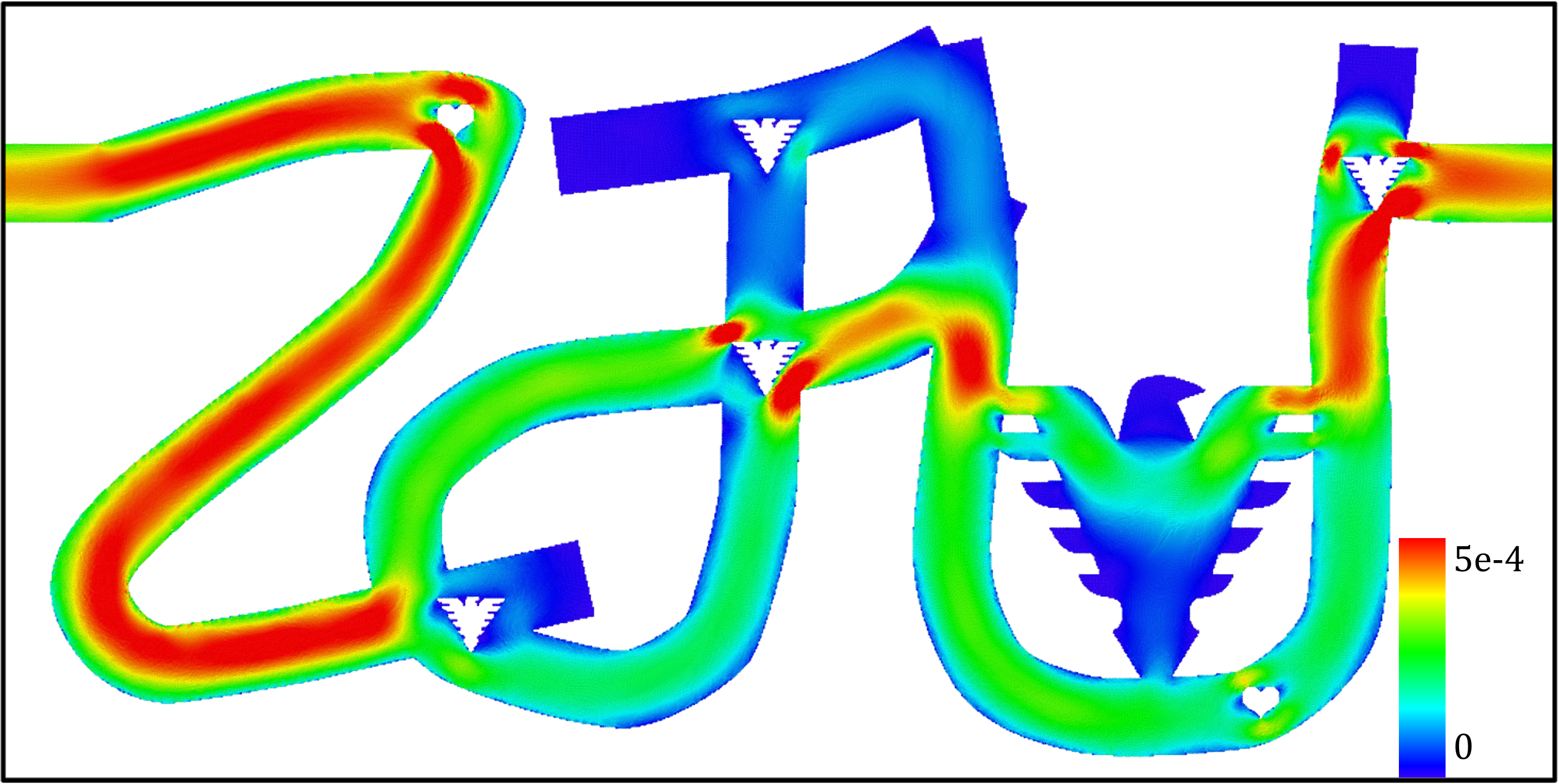}}
\subfigure[$b = L$.]{\includegraphics[scale=0.15]{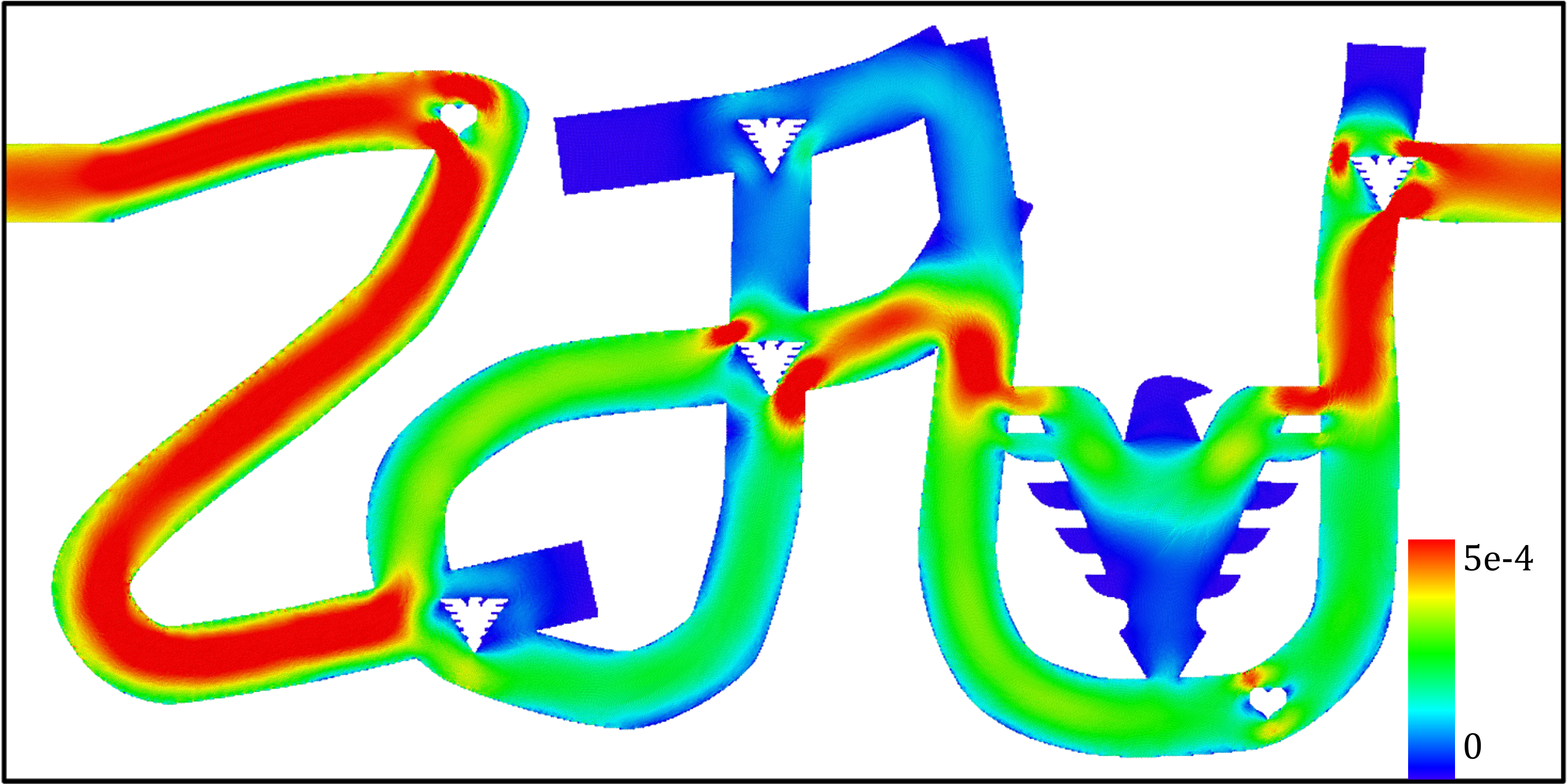}}
\caption{Velocity magnitude for different slip lengths on the wall. Colour bars 
are set to be $[0, 5\times10^{-4}\mathrm{ms^{-1}}]$.}\label{zju}
\end{figure}

The results of SPH simulations for various slip lengths are shown in Fig.~\ref{zju}. If we compare the velocity distribution of the flow field under different slip lengths,
the overall flow field is obviously elevated for a larger slip length on the walls.
The velocity distribution is different from those in the previous examples due to the complexity of the channel and also the fact that the body force is in $x$-direction, not necessarily along the channel.
Nevertheless, as long as some part of the channel is along $x$-direction,
its local velocity distribution is generally consistent with that of the Poiseuille flow.
The results from FPC and BPC methods are almost identical, 
therefore we present only the latter.

\subsection{Dynamics of an ellipsoid in Hagen-Poiseuille flow}
\begin{figure}[h!]
\centering \mbox{ 
\subfigure[An ellipsoid in a tube.]{\includegraphics[width=60mm]{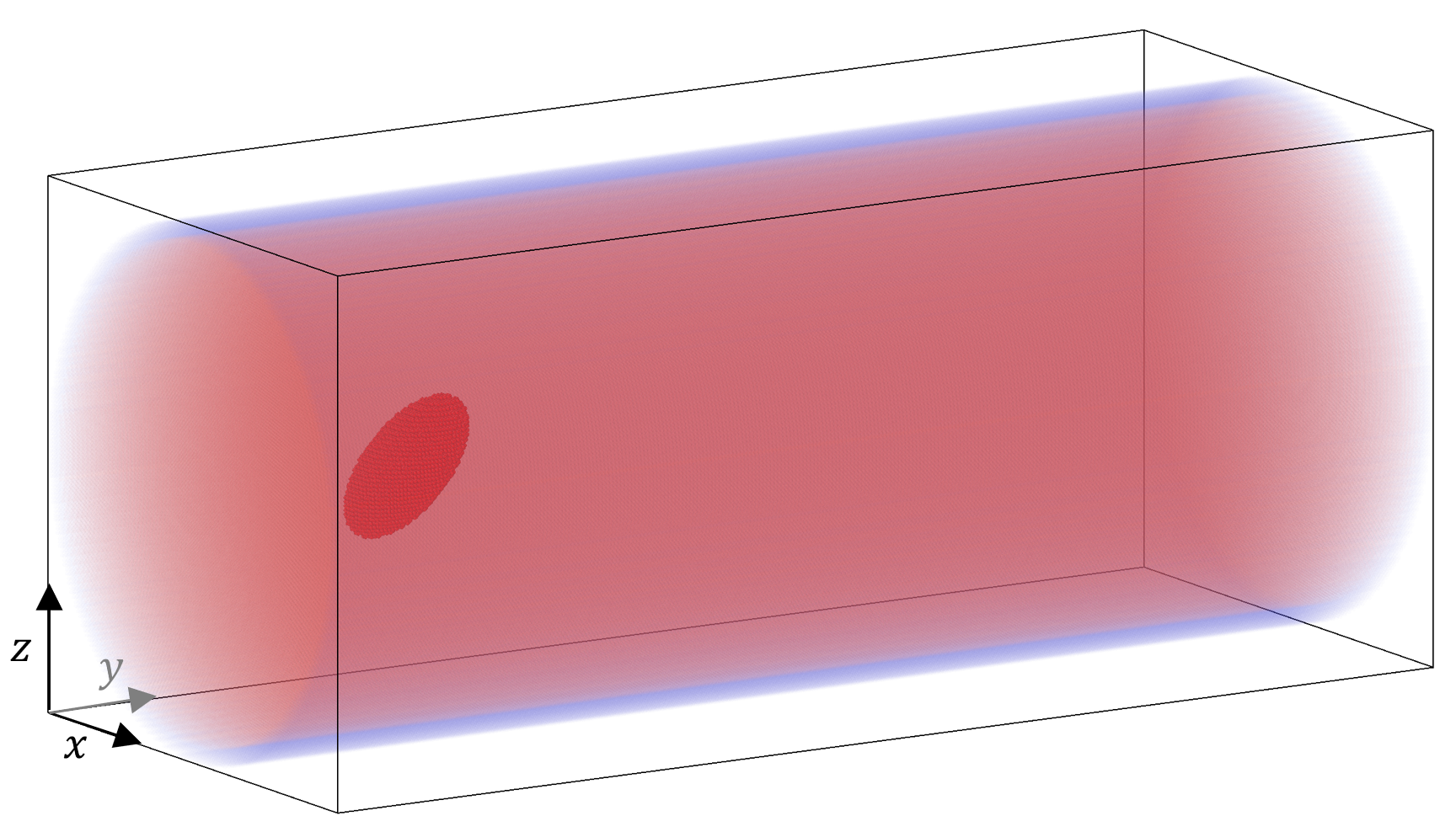}\label{ellipsoid_schematic1}}\quad 
\subfigure[A cross section in $y$-$z$ plane passing the center of the ellipsoid.]{\includegraphics[width=80mm]{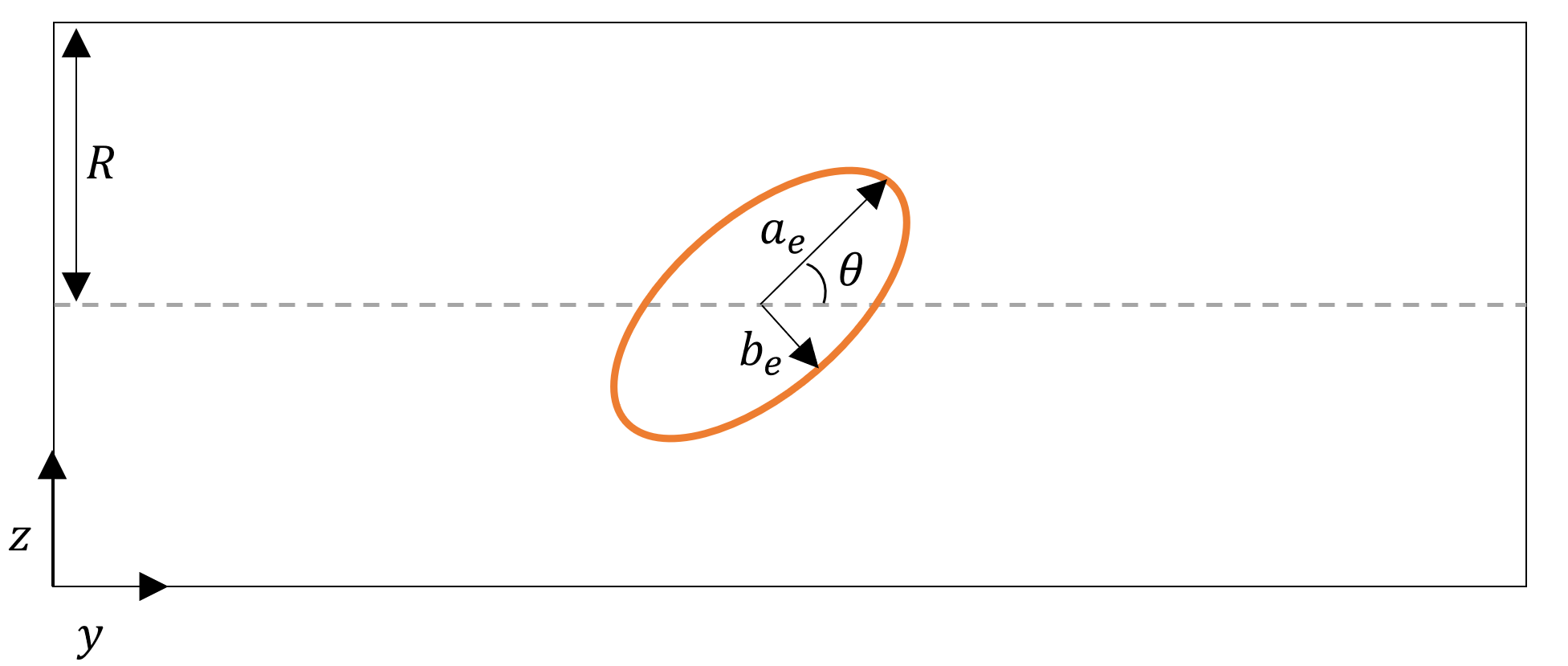}\label{ellipsoid_schematic2}}} 
\caption{Schematic and configuration for an ellipsoid in Hagen-Poiseuille flow}\label{ellipsoid_schematic}
\end{figure}

So far we have considered complex flows inside static solid boundaries.
We further examine the methods for a moving boundary.
This is represented by the dynamics of a three-dimensional ellipsoid in Hagen-Poiseuille flow, which is sketched in Fig.~\ref{ellipsoid_schematic1}. The flow inside the tube is driven by a body force in $y$-direction
and has periodic boundary conditions in the same direction. A neutrally buoyant ellipsoid particle is initially placed inside the tube
with its center at the axis of the tube.
We consider a special case for the three semi-axes as $a_e>b_e=c_e$, 
therefore the ellipsoid is a prolate.
A cross section of the prolate in $y$-$z$ plane presents an ellipse,
as shown in \ref{ellipsoid_schematic2},
where the long axis $a_e$ defines an angel $\theta$ along $y$ direction.

In following simulations, we specify the long axis of the ellipsoid as $2a_e=10^{-4}\mathrm{m}$ and the axis ratio as $a_e:b_e:c_e=2:1:1$. 
We consider two cases for the radius of the tube as $R=1.5a_e$ and $3a_e$,
and tube length as $L=5R$ so that periodic effects are negligible.
The density and kinematic viscosity are $\rho=10^{3} \mathrm{kgm^{-3}}$ and $\nu=10^{-6}\mathrm{m^2s^{-1}}$, respectively.
In a Hagen-Poiseuille flow without the ellipsoid, 
if the body force $F=4\nu U_c/R^2$, 
the maximum velocity of the flow at steady state is $U_c=10^{-3}\mathrm{ms^{-1}}$,
which defines a Reynolds number as $Re = 2a_eU_c/\nu=0.1$. 
We adopt a SPH resolution of $\Delta x = a_e/20$ and $r_c=3\Delta x$. 

\begin{figure}[h!]
\centering \mbox{ 
\subfigure[$R/a_e = 1.5$]{\includegraphics[width=60mm]{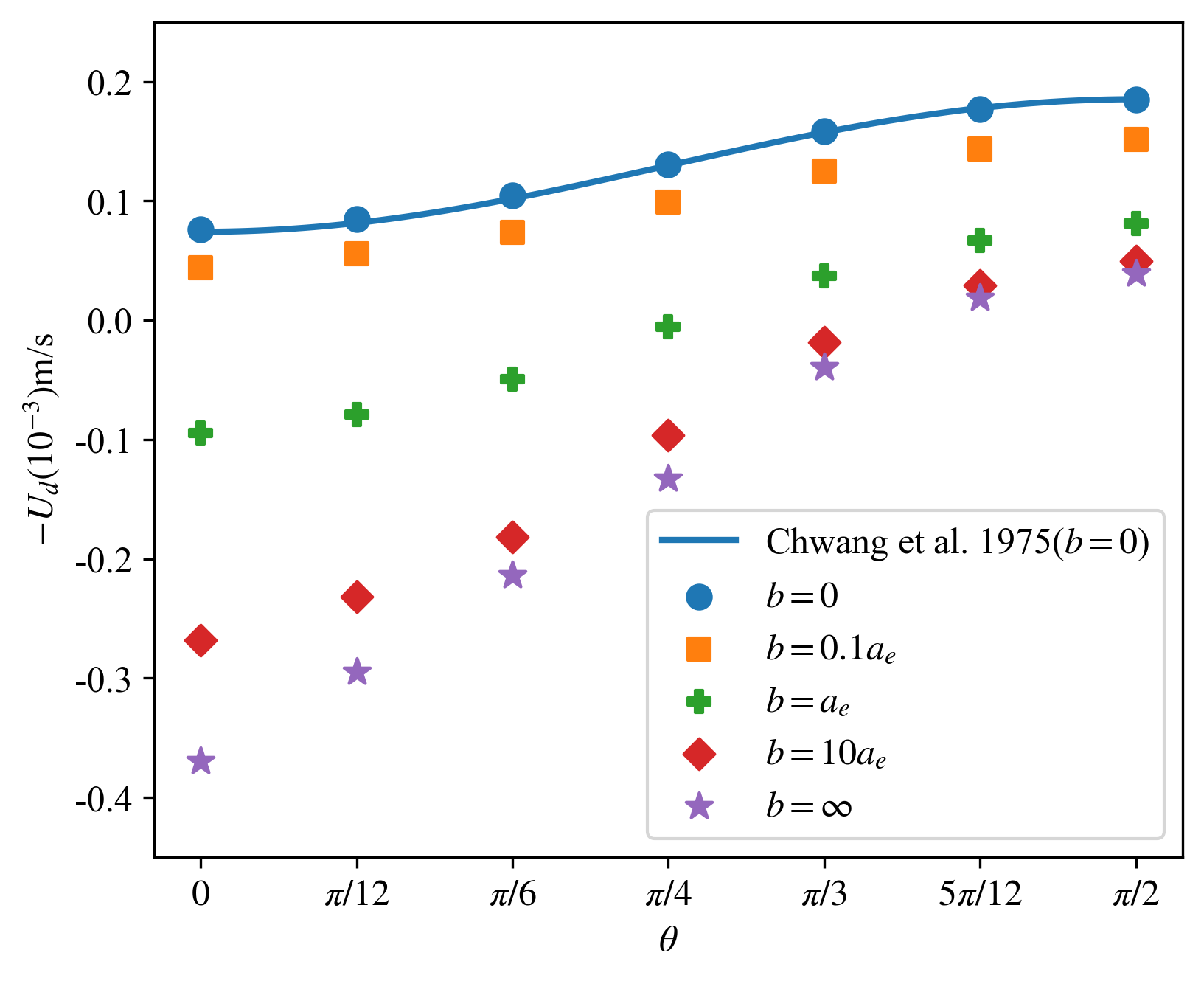}}\quad 
\subfigure[$R/a_e = 3$]{\includegraphics[width=60mm]{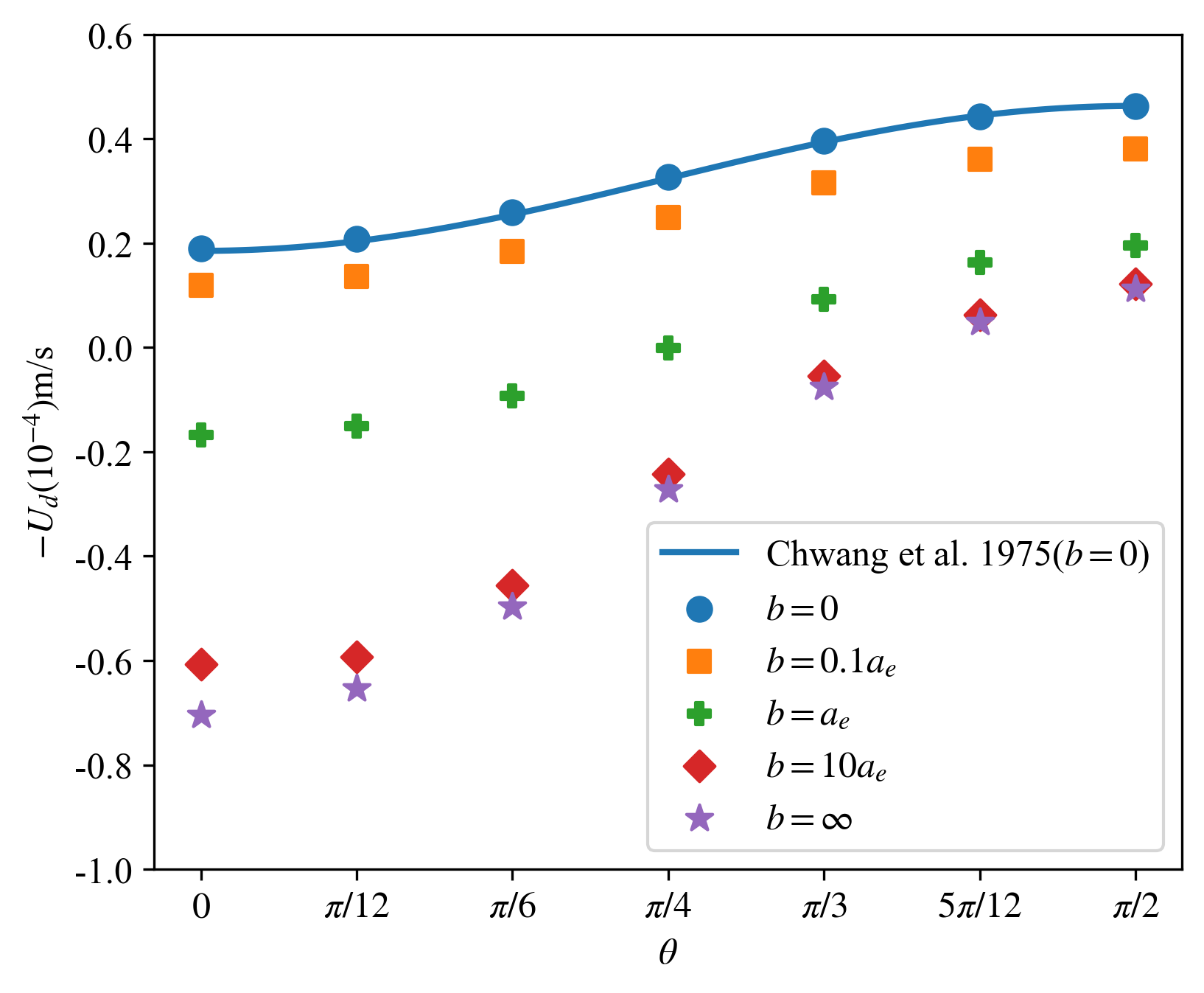}}} 
\caption{Distinct velocity for an ellipsoid in Hagen-Poiseuille flow with different tube radius.}\label{elliptical_vel}
\end{figure}

Due to hydrodynamic interactions between the ellipsoid and the tube wall,
the ellipsoid has different translating velocity from the flow.
In particular, a distinct velocity\footnote{It is called slip velocity in other references.
Since slip velocity relates to boundary condition here, we rename it as distinct velocity.} can be defined as $U_d=U_{e}-U_c$, 
where $U_e$ is the velocity component of the ellipsoid in $y$-direction.
For a typical no-slip boundary on both the ellipsoid and the wall, the distinct velocity is always negative and has a larger magnitude as the tilt angle $\theta$ increases.
We present SPH results for two different tube radii in Fig.~\ref{elliptical_vel},
where they agree well with those of Chwang et al.~\citep{chwang1975} for no-slip boundary conditions.
If a slip boundary takes place on the ellipsoid, its velocity increases significantly.
For slip length $b > a_e$ on the ellipsoid surface, it moves even faster than
the maximum flow velocity when $\theta < \pi /4$,
as shown in Fig.~\ref{elliptical_vel}.

\subsection{Couette flow at mesoscale with thermal fluctuations}
We further consider mesoscopic flows, where thermal fluctuations are present.
These can be simulated by SDPD method.
Given a $3D$ cubic box with length $L=10^{-5}\mathrm{m}$ and the mass density of fluid $\rho = 10^{3}\mathrm{kgm^{-3}}$, the total mass $M=L^3\rho=10^{-12}\mathrm{kg}$;
the kinematic viscosity of the fluid $\nu=10^{-6}\mathrm{m^{2}s^{-1}}$.
There are two walls in $y$ direction: the upper wall with no-slip boundary moves with $v_{up}=0.01\mathrm{ms^{-1}}$, which defines a Reynolds number $Re=v_{up}L/\nu=0.1$;
the lower wall with various slip lengths remains still.
The temperature $T=293.15\mathrm{K}$ and the Boltzmann constant $k_B=1.38\times 10^{-23}\mathrm{m^2kgs^{-2}K^{-1}}$.

To avoid round-off errors due to small physical quantities such as $k_B$,
we adopt non-dimensional numbers for mesoscopic simulations.
In particular, length, energy and mass are taken to be unity, that is, $L^*=1$, $(k_BT)^*=1$ and $M^*=1$. Therefore, density $\rho^*=1$, kinematic viscosity $\nu^*=\nu/(k_BTL^2/M)^{1/2}=1572.2$, and time $t^*$ is in units of $(M(L)^2/(K_BT))^{1/2}$. 
In the following, we omit the $^*$ for simplicity.

\begin{figure}[h!]
\centering 
\subfigure[$T=293.13K$]{\includegraphics[scale=0.48]{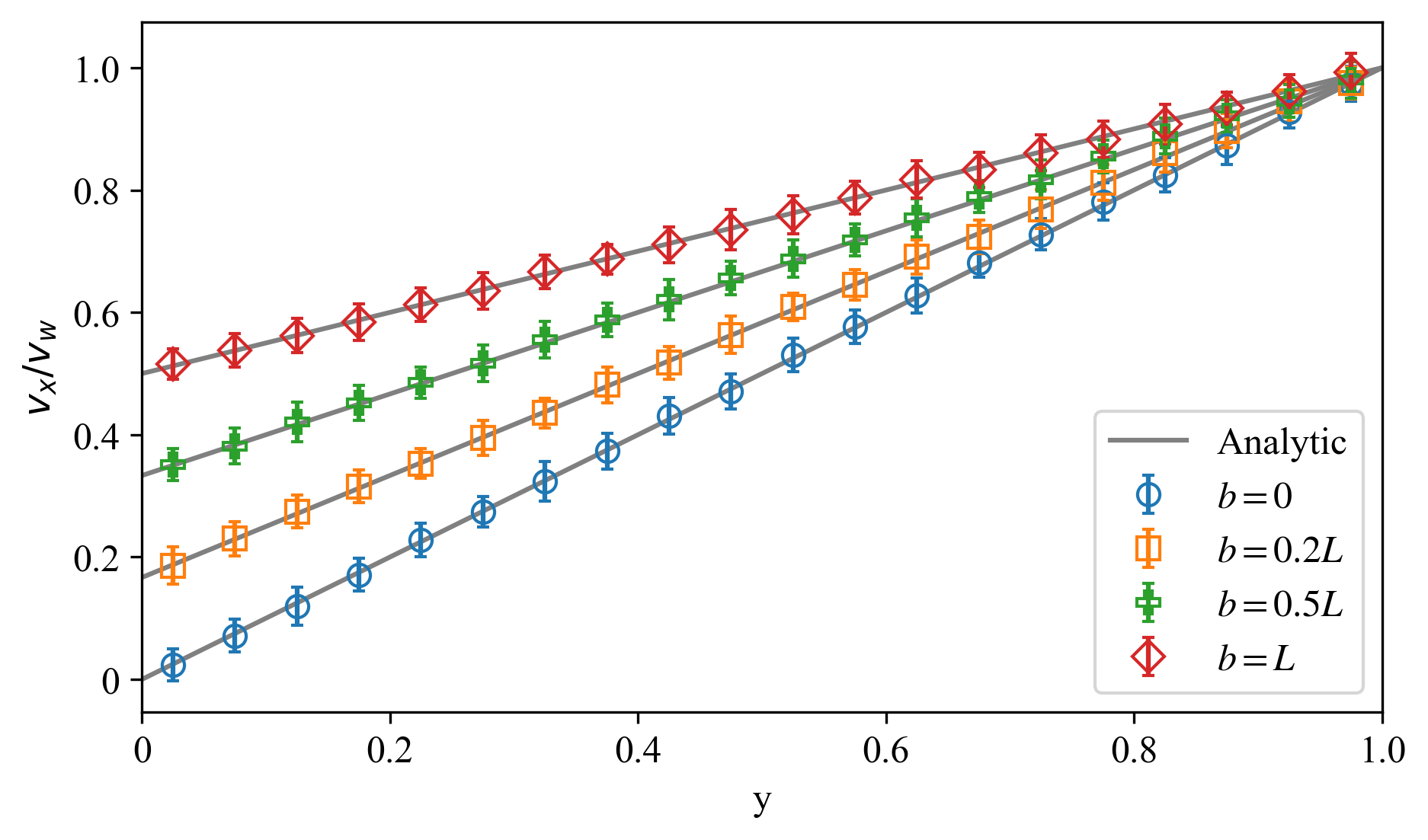}}\quad
\subfigure[$T=313.13K$]{\includegraphics[scale=0.48]{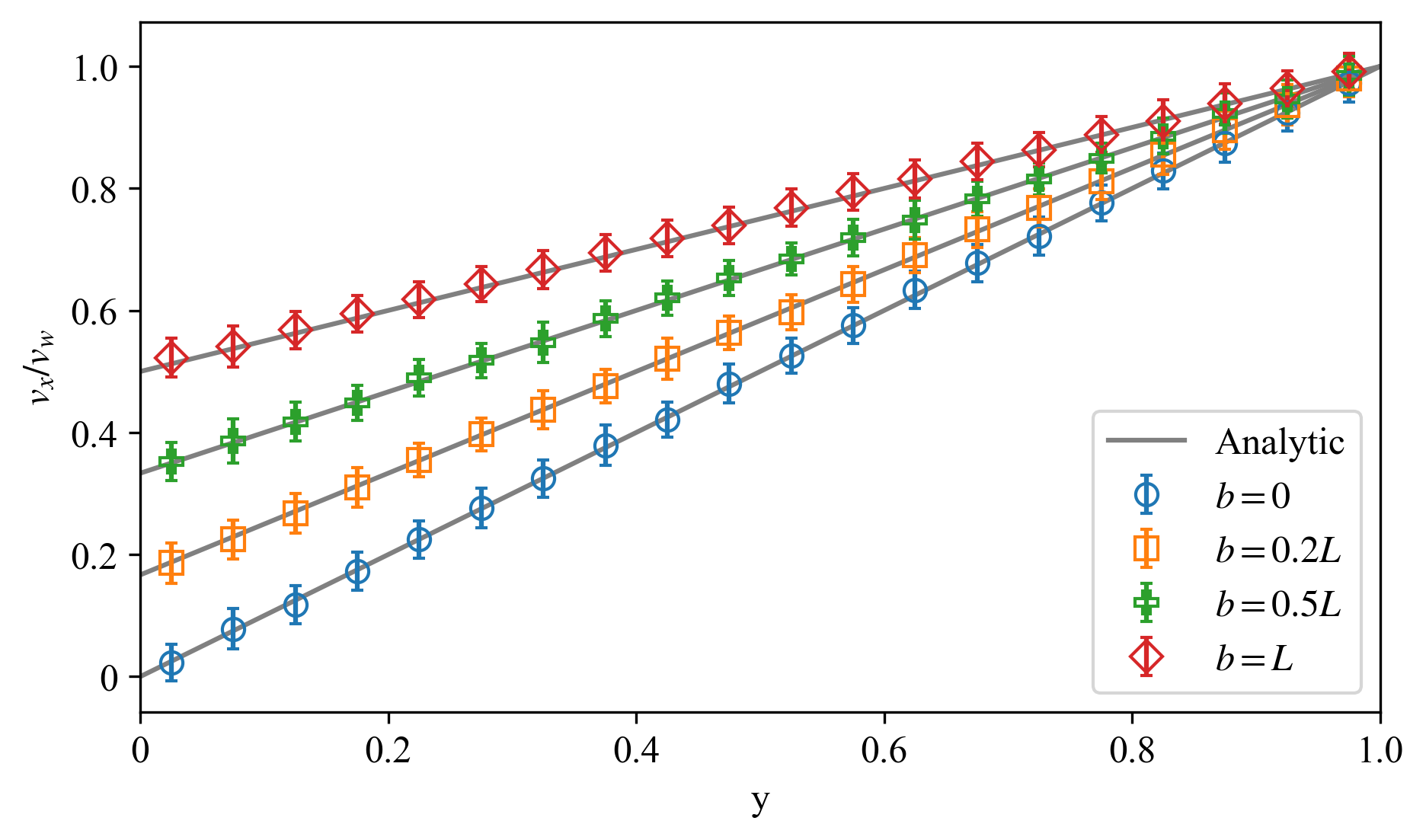}}
\caption{Averaged velocity profiles of SDPD simulations for Couette flow at mesoscale.}\label{sdpd_couette}
\end{figure}

We present averaged velocity profiles of SDPD simulations with different slip lengths on the lower wall for two temperatures in Fig.~\ref{sdpd_couette}. 
The error bars indicate standard deviation of $100$ steady-state moments. 
There are $20$ particles across the channel, which are sufficient to reproduce the analytical solutions. 

\begin{figure}[h!]
\centering 
\subfigure[No-slip on both walls.]{\includegraphics[width=60mm]{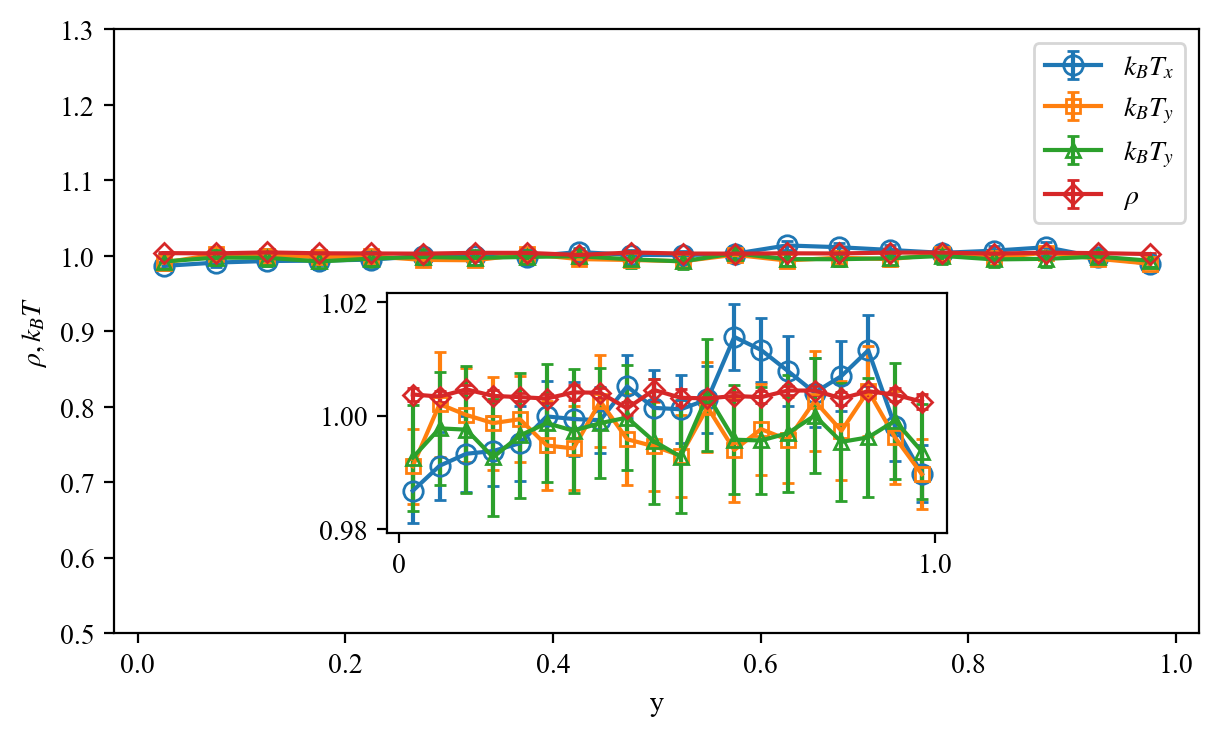}}\quad
\subfigure[Slip length $b = 0.2L$ on lower wall and no-slip on upper wall.]{\includegraphics[width=60mm]{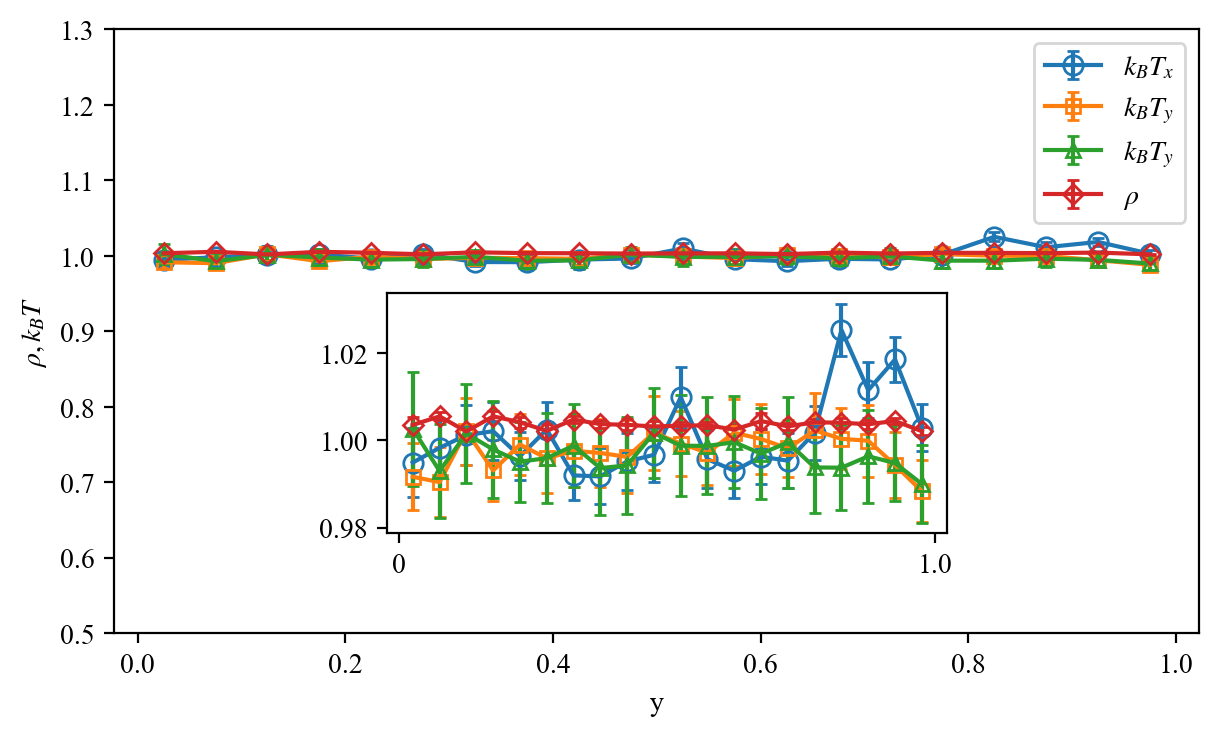}}
\caption{Temperature and average density profiles of SDPD simulations for Couette flow at mesoscale.}\label{density_temperature}
\end{figure}

We may define directional temperature as the averaged kinetic energy due to random motions of particles as follows,
\begin{eqnarray}
    k_BT_x &=& <m(v_x-v_x^{bg})^2>,\\
    k_BT_y &=& <mv_y^2>,\\
    k_BT_z &=& <mv_z^2>,
\end{eqnarray}
where $\mathbf{v}^{bg}=(v_x^{bg},0,0)$ is the linear profile of the background velocity for Couette flow. We present temperature and density of SDPD simulations across the channel with no-slip and slip boundary conditions in Fig.~\ref{density_temperature},
where there is no numerical artifact from the algorithms
for the boundary condition.

\subsection{Brownian sphere}
\begin{figure}[h!]
\centering 
\subfigure[$b=0$]{\includegraphics[scale=0.48]{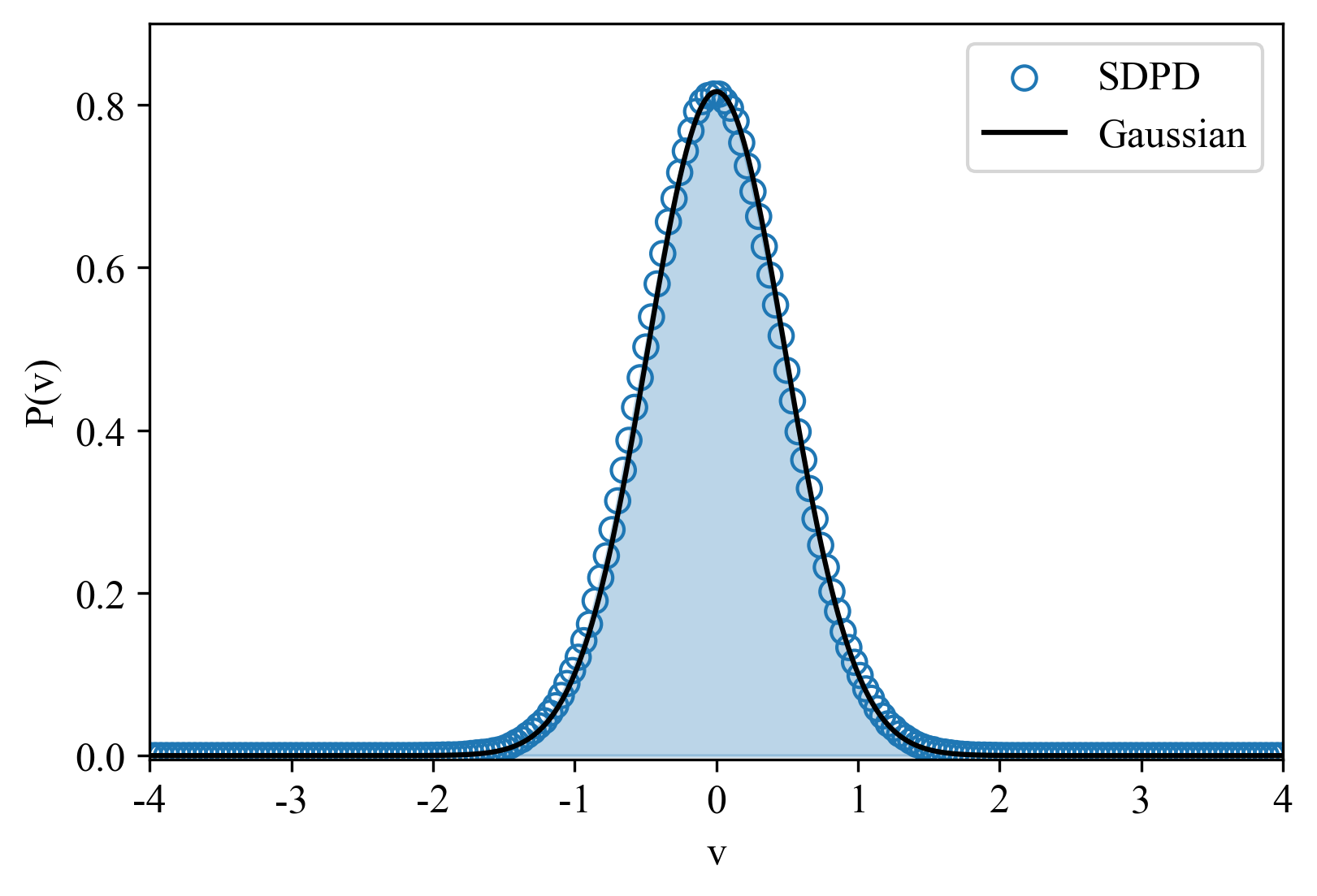}}
\subfigure[$b=0.5R$]{\includegraphics[scale=0.48]{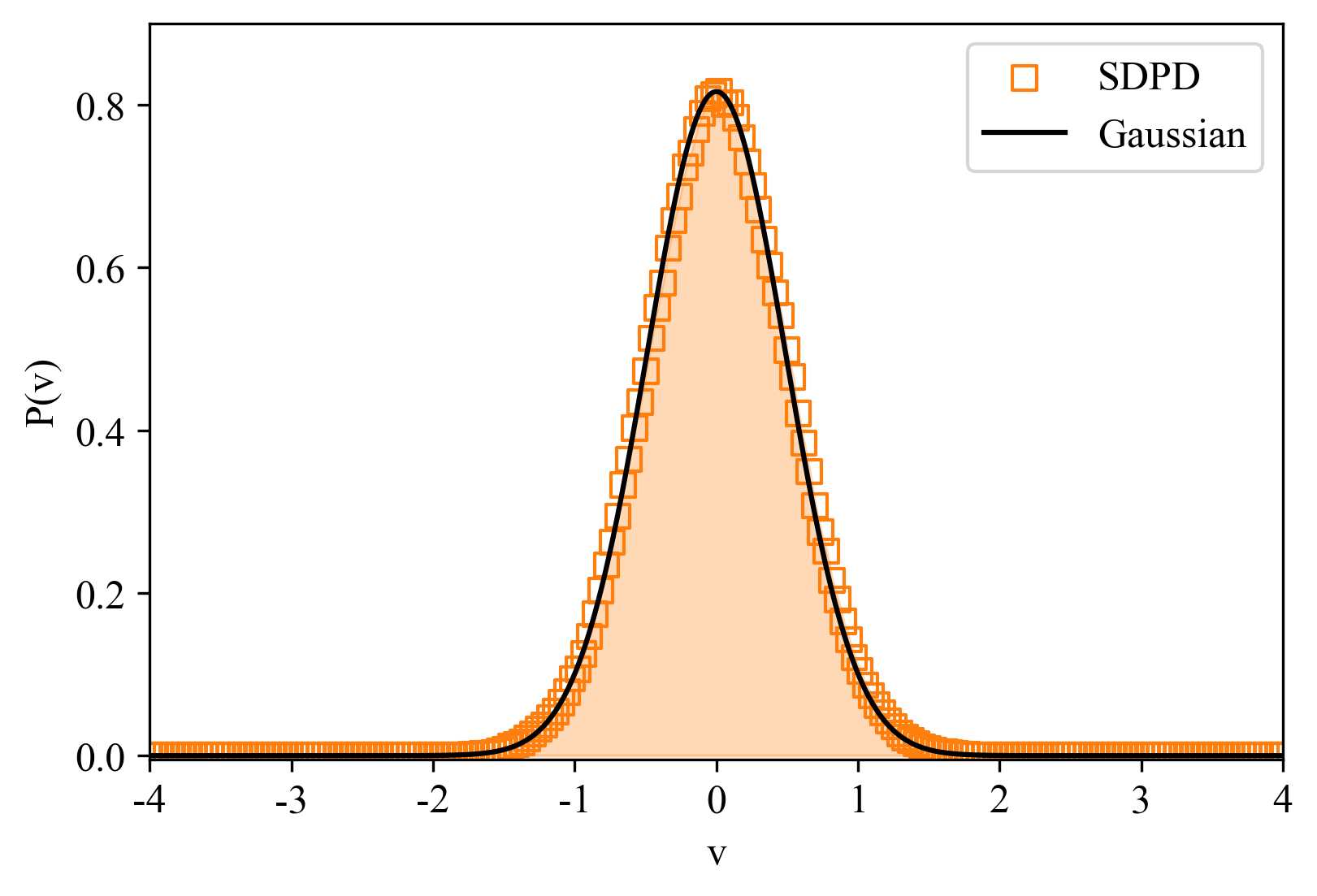}}
\caption{Probability distribution function for the velocity of sphere.}\label{pdf_sdpd}
\end{figure}

\begin{figure}[h!]
\centering 
\subfigure[$b=0$]{\includegraphics[scale=0.48]{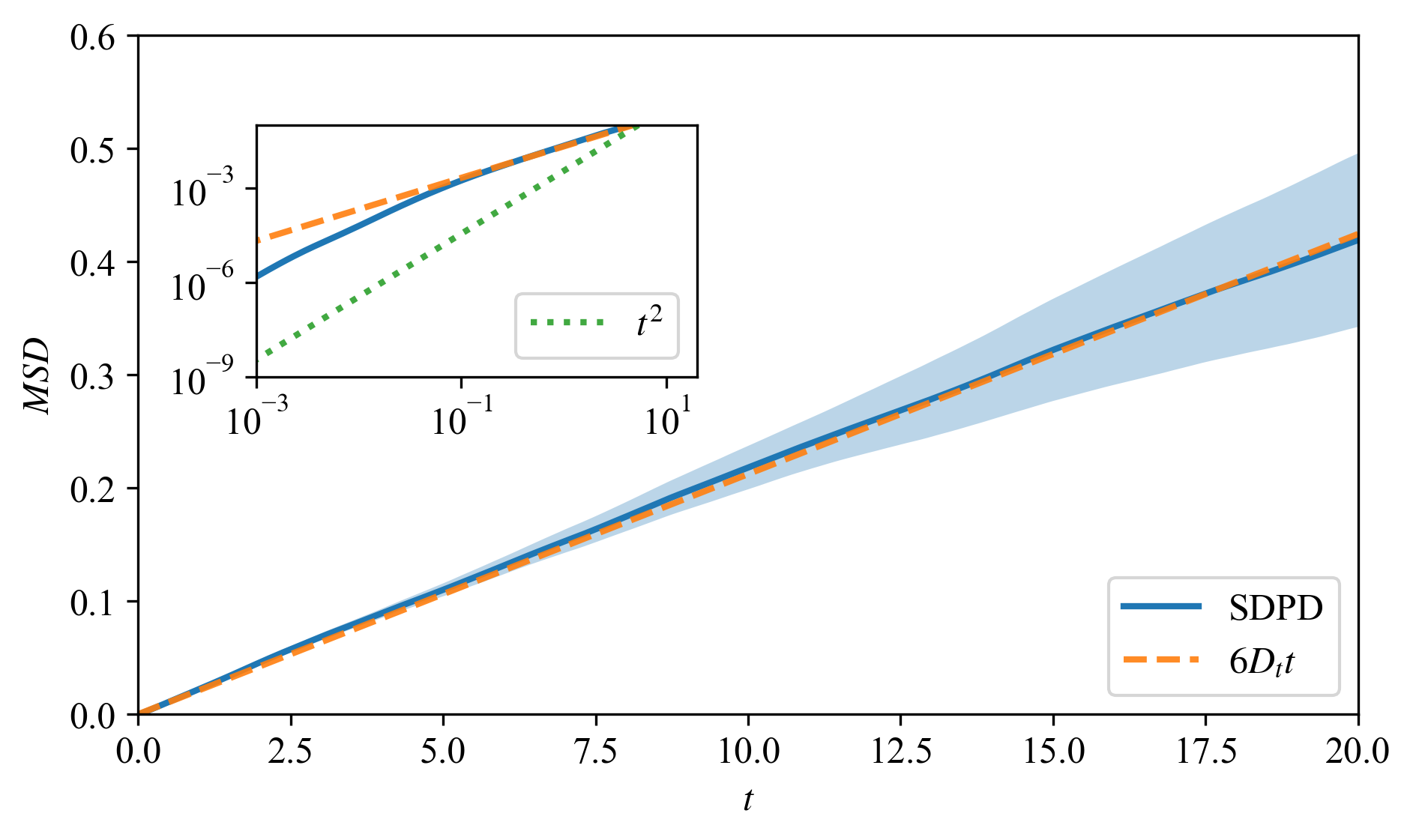}}
\subfigure[$b=0.5R$]{\includegraphics[scale=0.48]{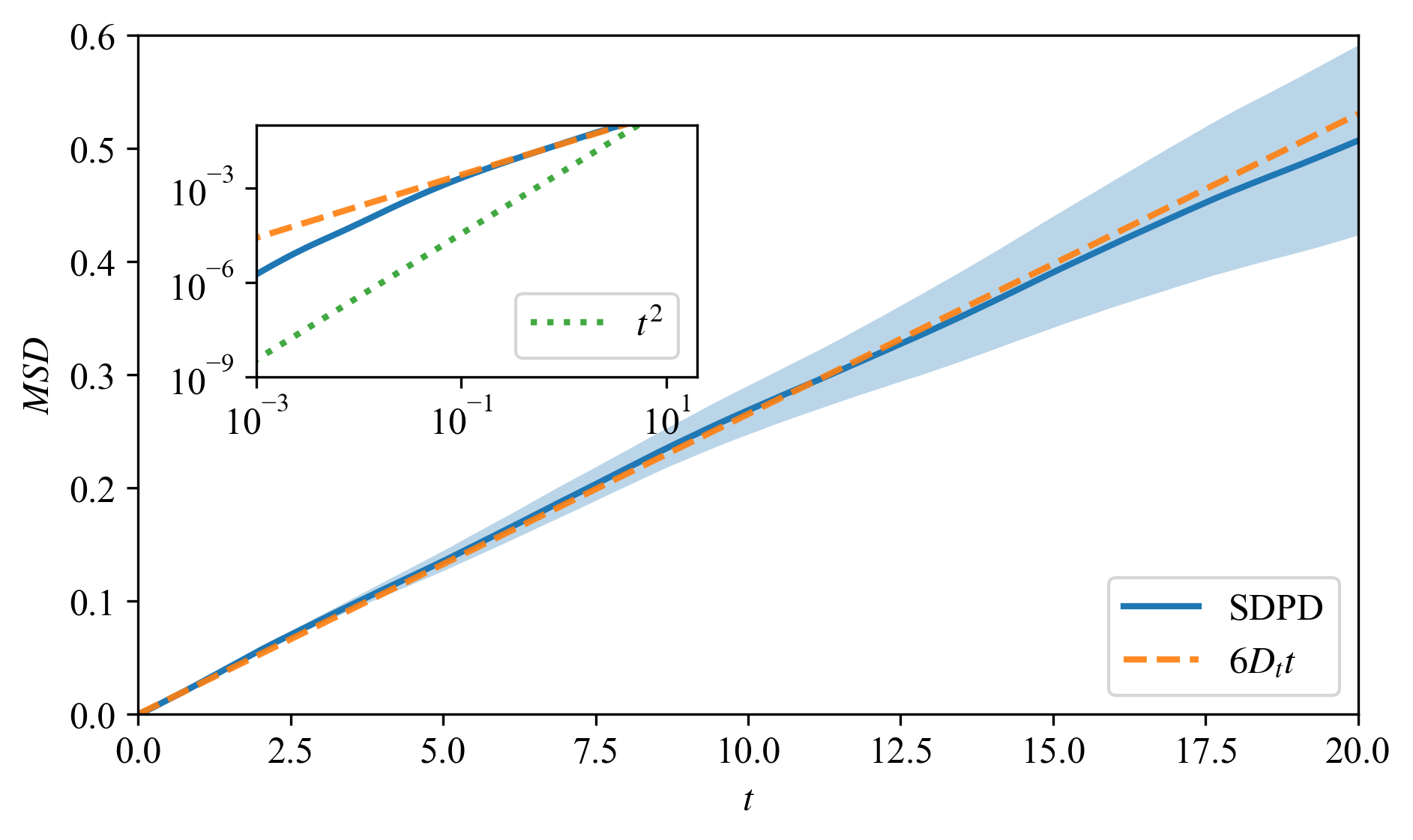}}
\caption{Translational mean square displacement of a sphere: the slope represents the diffusion
coefficient and shadows indicate standard deviations of $20$ simulations. The inset shows the transition from ballistic to diffusive regime in logarithmic-logarithmic scale.}\label{msd_sdpd}
\end{figure}

We simulate the random motion of a neutral buoyant sphere in a solvent. We set the radius of the sphere $R$, fluid density $\rho$, and temperature $k_BT$ to be unity. The kinematic viscosity $\nu=15$. 
We adopt a cubic box with length $L =15 R$ and apply periodic boundary conditions
in all three directions.
We employ three different resolutions $\Delta x = R/5 = 0.2$, $R/8=0.125$ and $R/10=0.1$ to verify convergence of the results. 
The velocity scale is  $v_t = \sqrt{3k_BT/m}$ and the sound speed is taken as $c=20v_t$.
To obtain a smooth statistical average, we perform $20$ simulations using different random seeds.
The probability distribution functions (PDF) of the sphere's velocity with different slip lengths
are shown in Fig.~\ref{pdf_sdpd}. 
The velocity PDF of the sphere is not altered by a slip boundary on its surface
and remains the same as for the case of a no-slip surface.
They both follow the Maxwell-Boltzmann distribution as
\begin{eqnarray}
    P(v) = \sqrt{\frac{m}{2\pi k_BT}}exp(\frac{-mv^2}{2k_BT}).
\end{eqnarray}
However, the mobility of the sphere depends strongly on the boundary condition of its surface.
According to Stokes’ law, the mobility of a sphere in an incompressible fluid at steady state is \cite{bian2016111}
\begin{equation}
    \mu = \frac{1}{6\pi\rho\nu R}\frac{1+3\rho\nu/\kappa R}{1+2\rho\nu/\kappa R} = \frac{1}{6\pi\rho\nu R}\frac{1+3bR}{1+2b R}.
\end{equation}
Further combining with Einstein's relation, we obtain
\begin{equation}
    D_t=\mu k_BT = \frac{k_BT}{6\pi\rho\nu R}\frac{1+3bR}{1+2b R},
\end{equation}
where $D_t$ is the translational diffusion coefficient of the Brownian sphere.
Practically we can verify the diffusion coefficient by measuring the mean square displacement (MSD) of the sphere for two different slip lengths, as shown in Fig.~\ref{msd_sdpd}. 
The SDPD results are in agreement with the theoretical solutions 
within statistical uncertainties,
where the mobility of the Brownian sphere is enhanced by a slip boundary condition.

\begin{figure}[h!]
\centering 
\subfigure[$b=0$]{\includegraphics[scale=0.48]{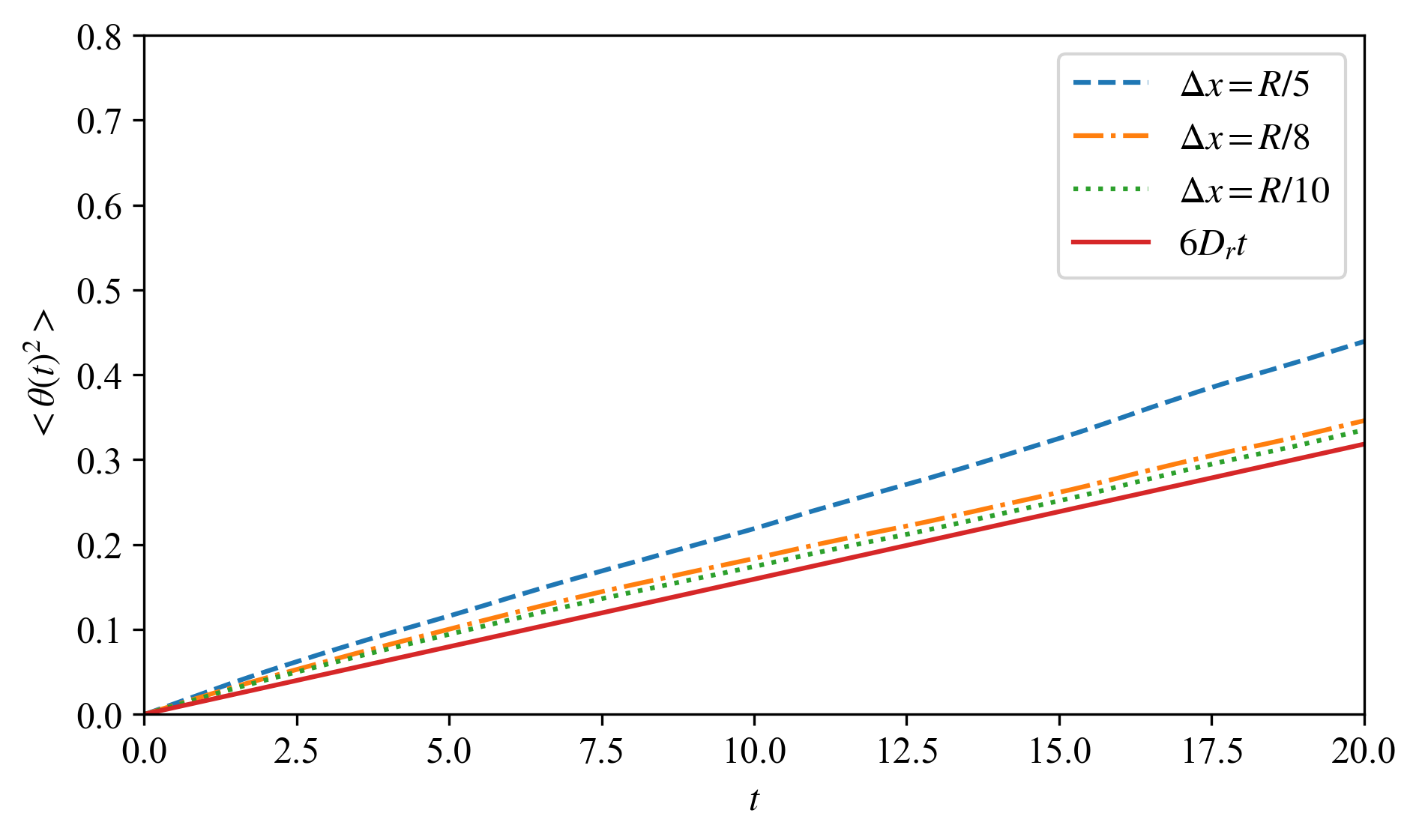}}
\subfigure[$b=0.5R$]{\includegraphics[scale=0.48]{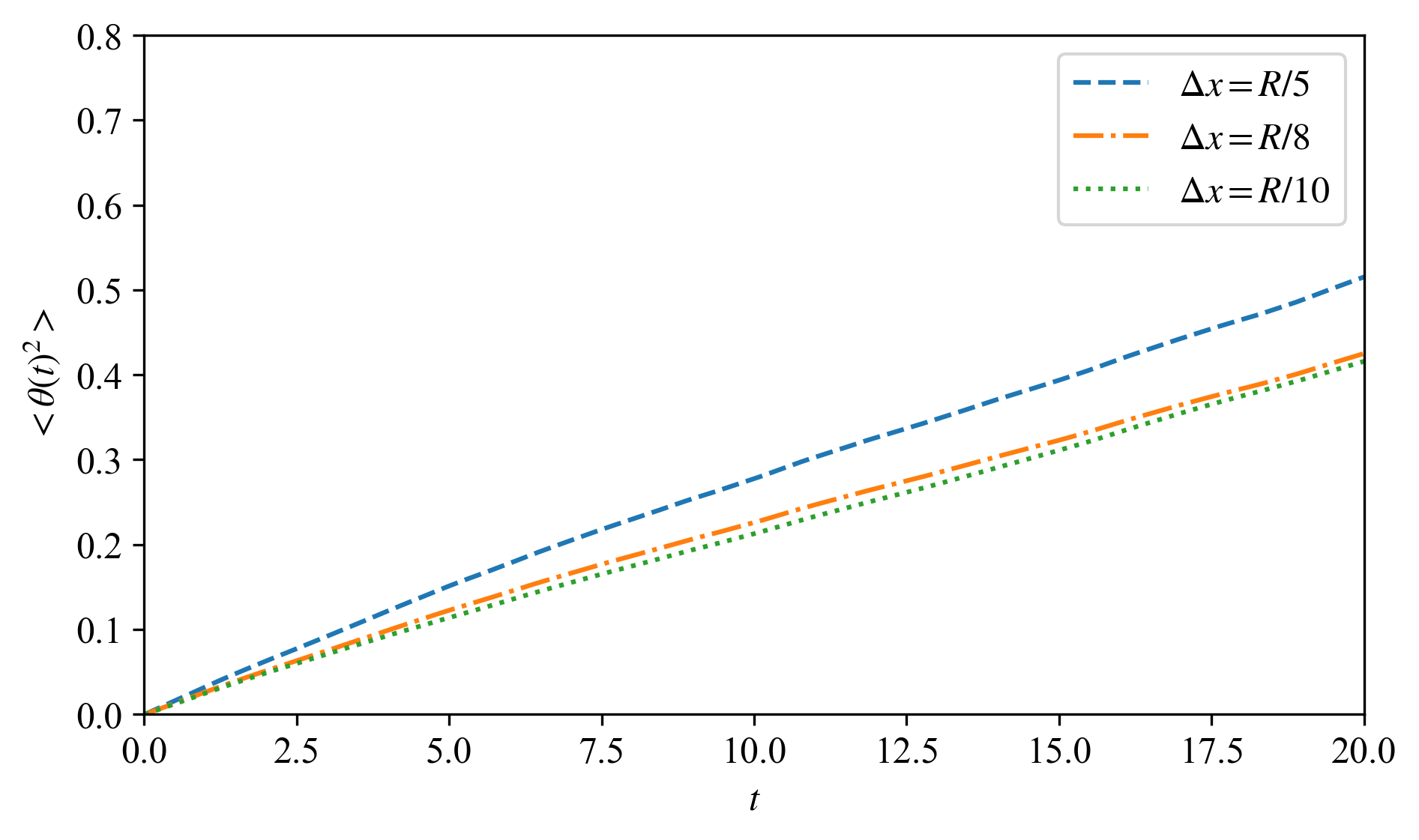}}\\
\caption{Rotational mean square displacement of a sphere with slip length: resolution study.}\label{msa_sdpd}
\end{figure}

Similarly, the rotational diffusion coefficient of the sphere with the no-slip boundary condition is 
\begin{eqnarray}
    D_r = \frac{k_BT}{8\pi \rho \nu R^3}.
\end{eqnarray}
In Fig.~\ref{msa_sdpd}, we can observe that $8$ and $10$ particles across
the sphere radius lead to a negligible difference in the rotational motion of the sphere
and a slip boundary also enhances the rotational mobility of the sphere.

\subsection{Dynamics of an elliptical cylinder in microvessels of arbitrary geometry}
\begin{figure}[h!]
\centering \includegraphics[scale=0.3]{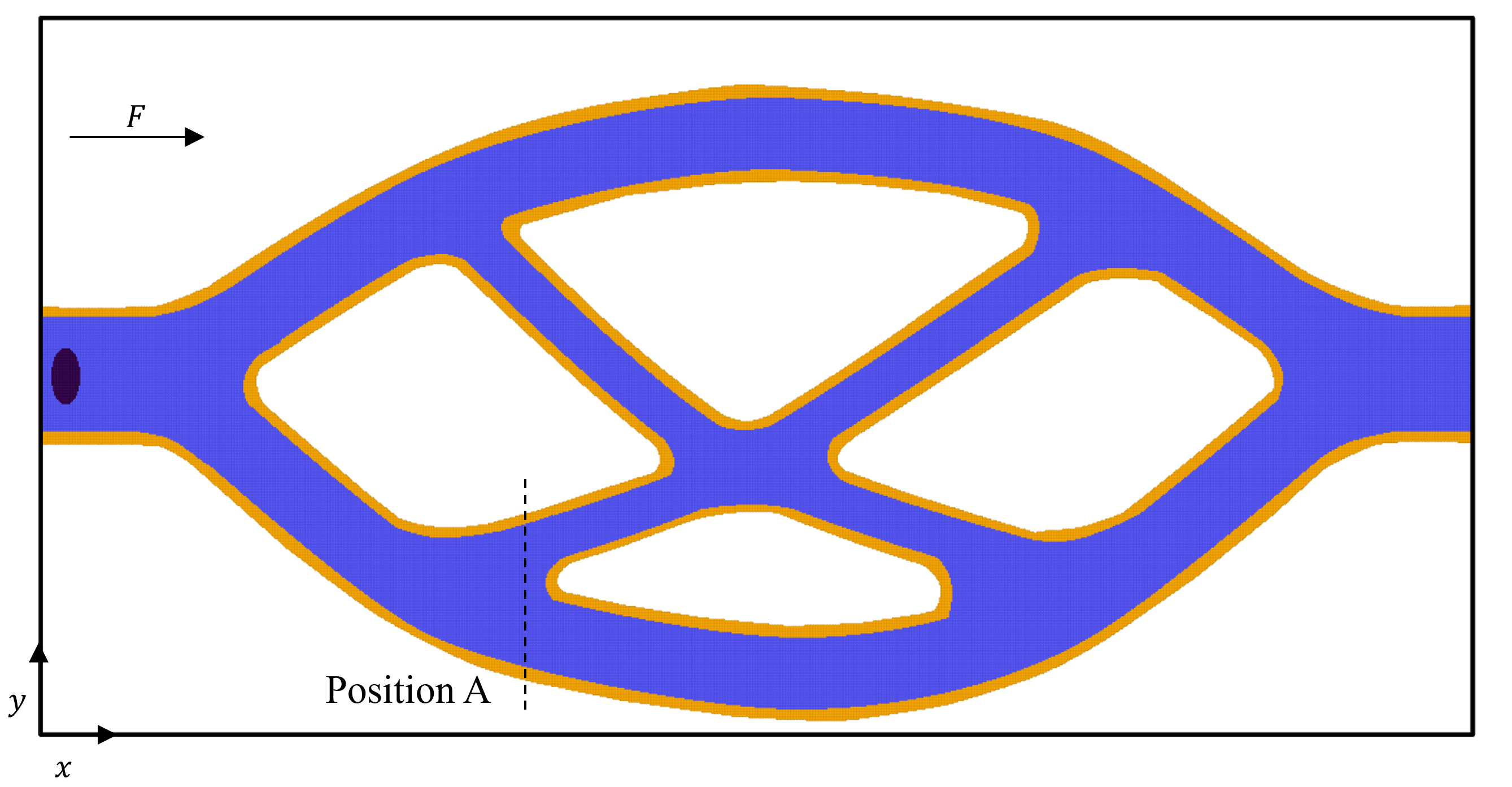} \caption{Schematic of an elliptical cylinder in microvessels of arbitrary geometry at the initial moment. The yellow, blue and purple particles represent the wall, fluid and elliptical cylinder, respectively. A constant body force is applied to the fluid in the $x$ direction. Periodic boundary conditions are applied in $x$-direction. Position A is a branching point for the trajectory of the elliptical cylinder.}\label{ellipse_sche}
\end{figure}

Finally, we simulate dynamics of an elliptical cylinder in an artificial network of microvessels,
as sketched in Fig.~\ref{ellipse_sche}. To facilitate comparison of SPH and SDPD simulations, we employ dimensionless units in both methods.
The size of the rectangular simulation domain is $2\times 1$, 
the width of the channel $L$ varies from $0.08$ to $0.16$, and the particle resolution $\Delta x = 2\times 10^{-3}$. We set density $\rho^*=1$, kinematic viscosity $\nu^*=1572.2$, and temperature $(k_BT)^*=1$. The semi-major and semi-minor axes of the ellipse are $a_e=2b_e=0.02$. 
A body force $F=1.98 \times 10^{10}$ is applied in $x$ direction.
For a no-slip boundary on the wall, 
the maximum velocity of the flow is around $v=15722$, and the corresponding Reynolds number is about $Re = 1$. 

\begin{figure}[h!]
\centering 
\subfigure[$b^e = 0$, $b^w=0$]{\includegraphics[width=60mm]{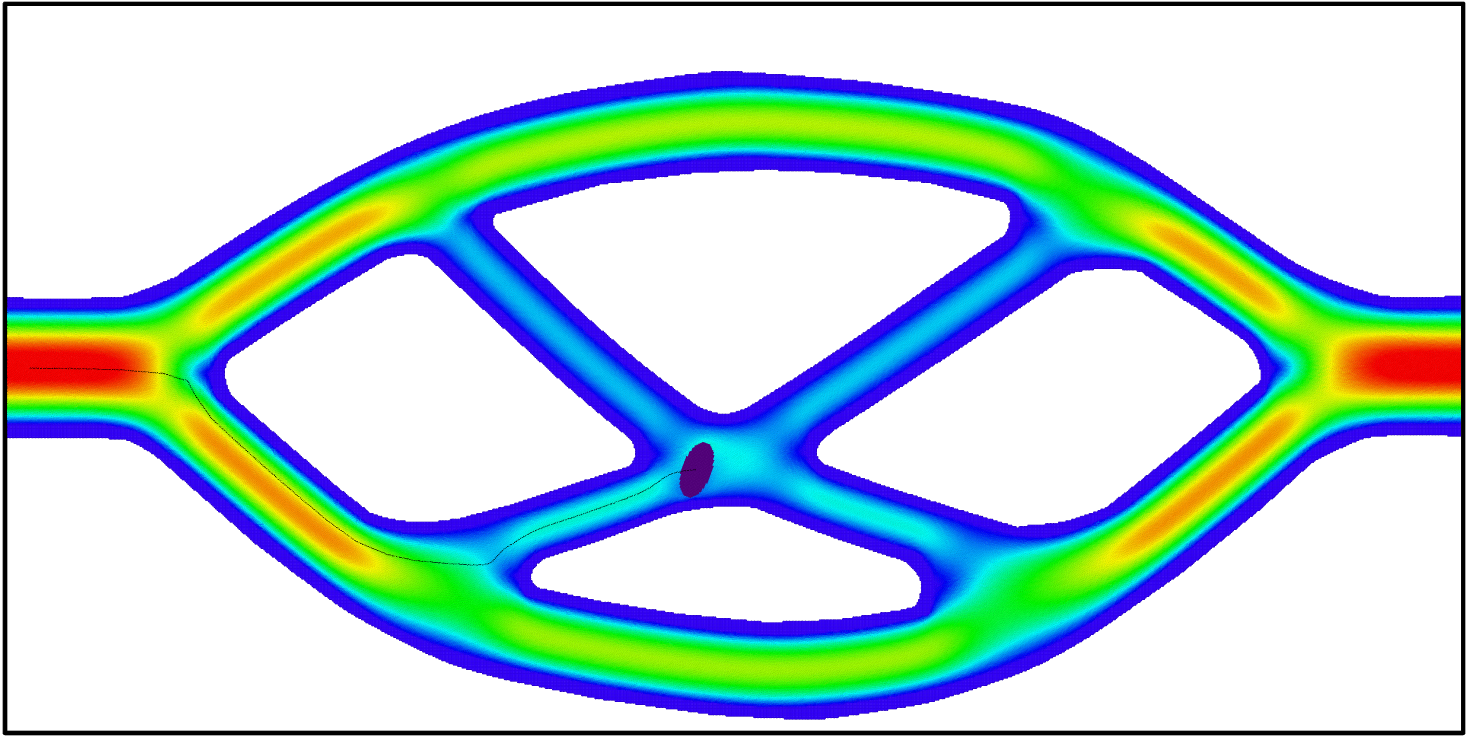}\label{sph_vessel1}}
\subfigure[$b^e = 0.2a_e$, $b^w=0$]{\includegraphics[width=60mm]{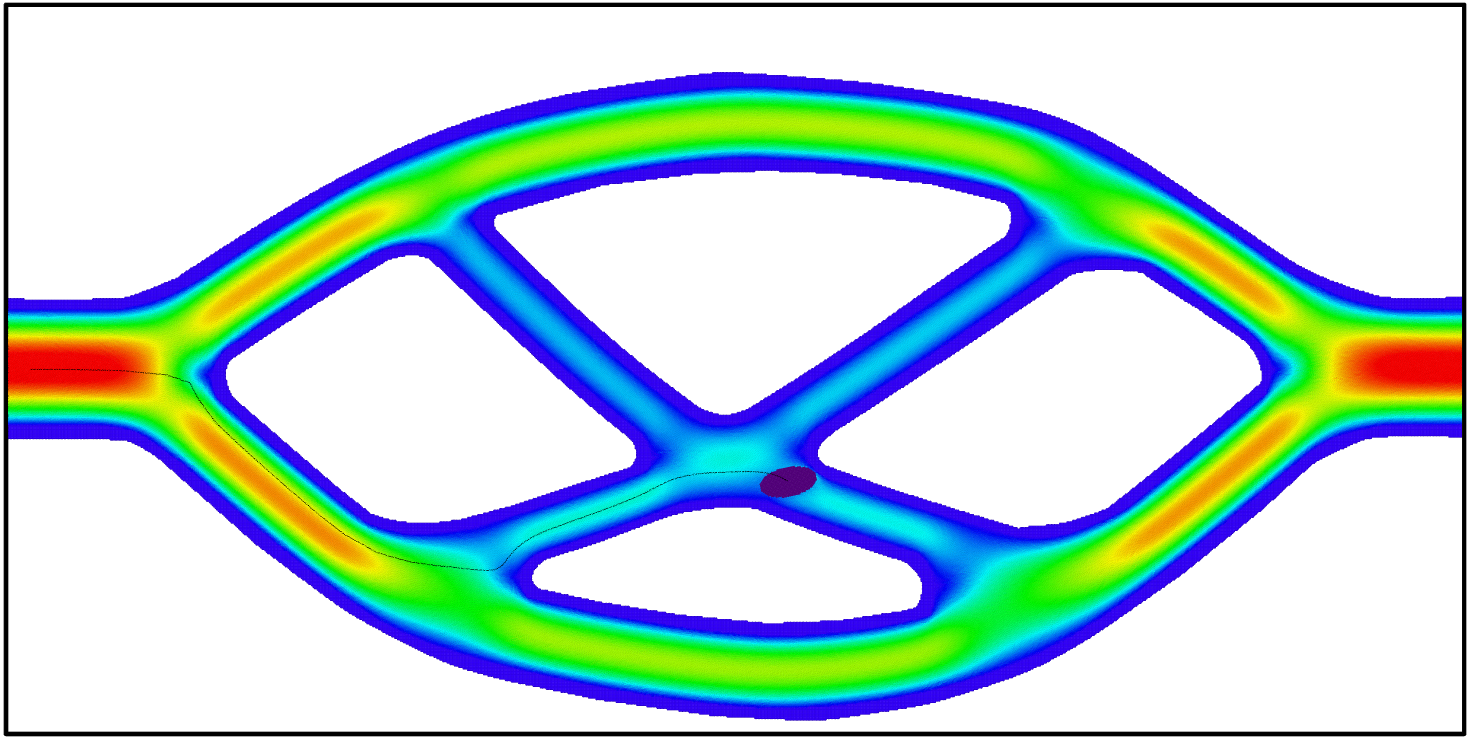}\label{sph_vessel2}}
\subfigure[$b^e = 0.5a_e$, $b^w=0$]{\includegraphics[width=60mm]{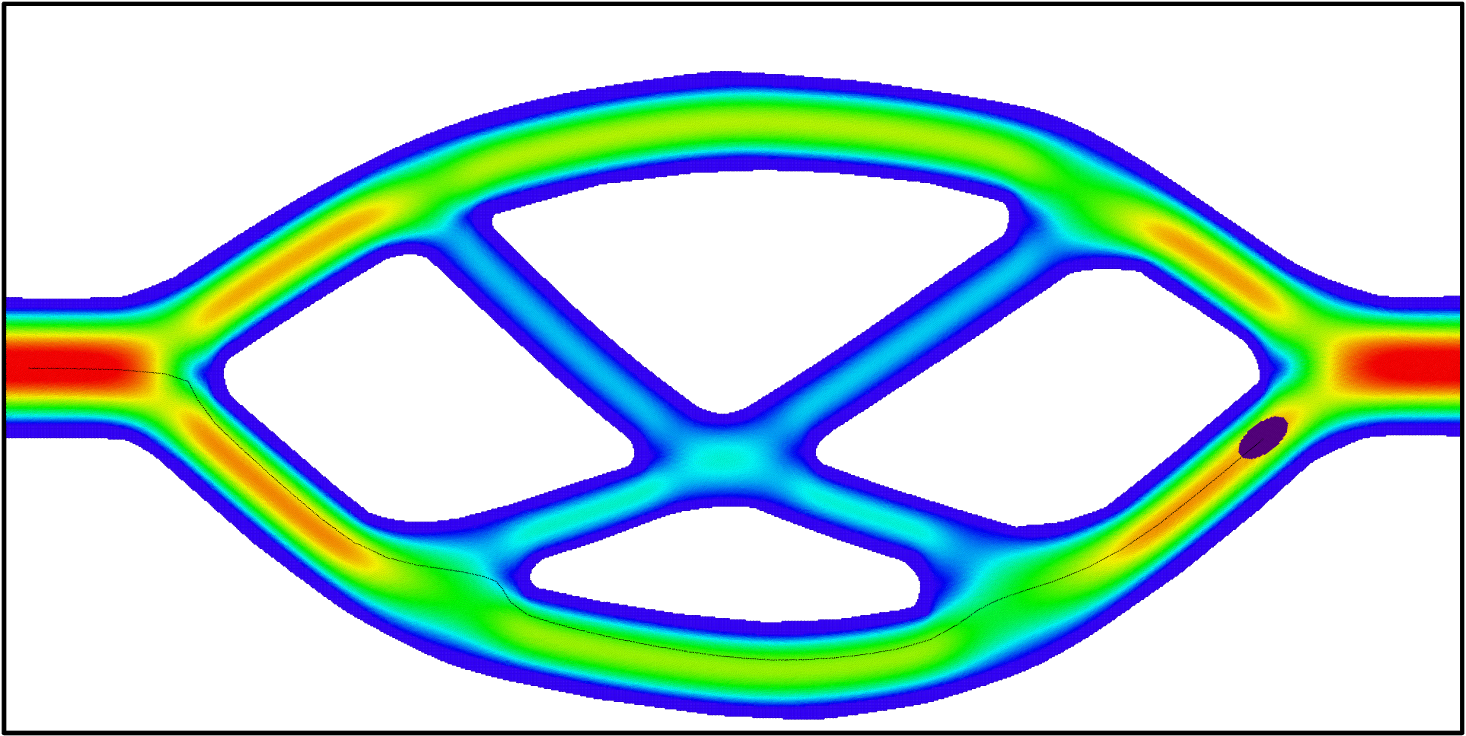}\label{sph_vessel3}}
\subfigure[$b^e = a_e$, $b^w=0$]{\includegraphics[width=60mm]{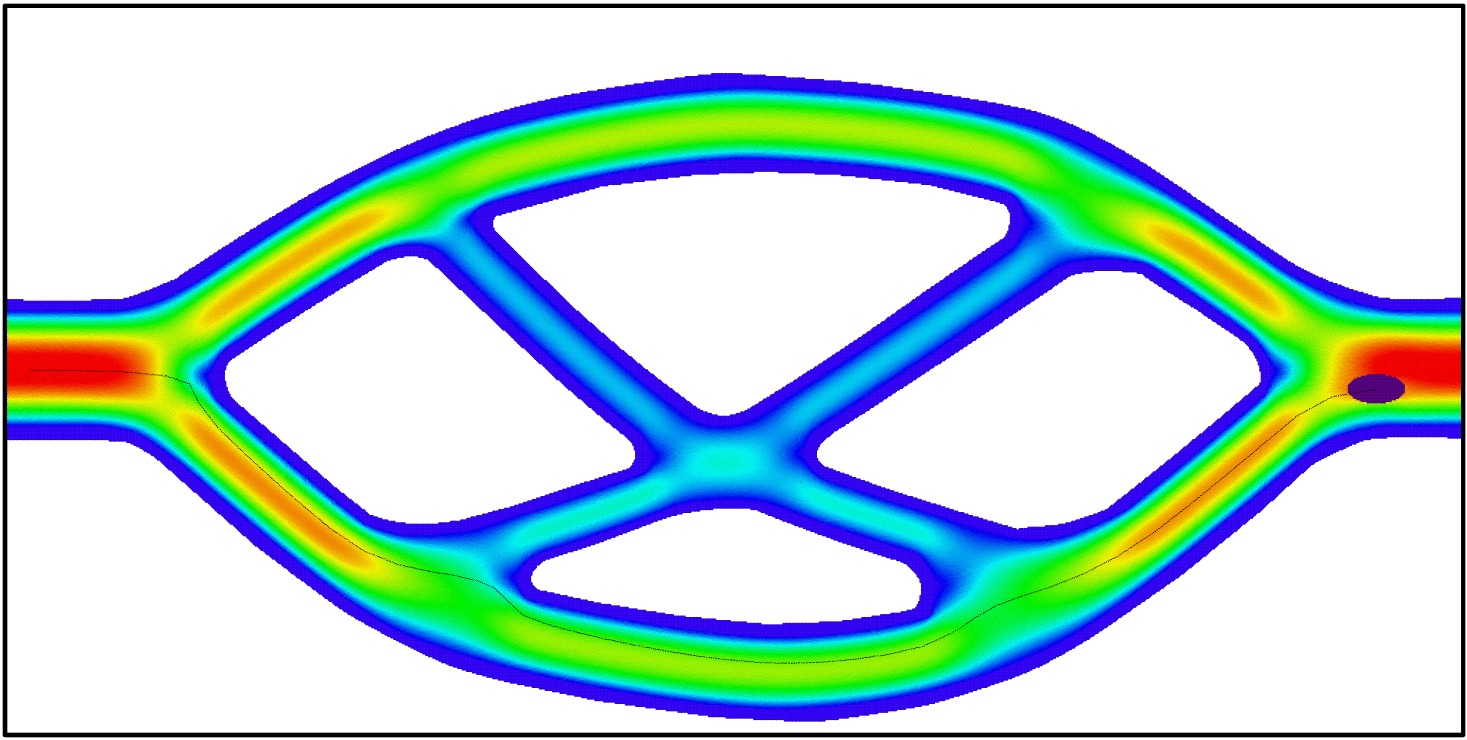}\label{sph_vessel4}}
\subfigure[$b^e = 0$, $b^w=0.2L$]{\includegraphics[width=60mm]{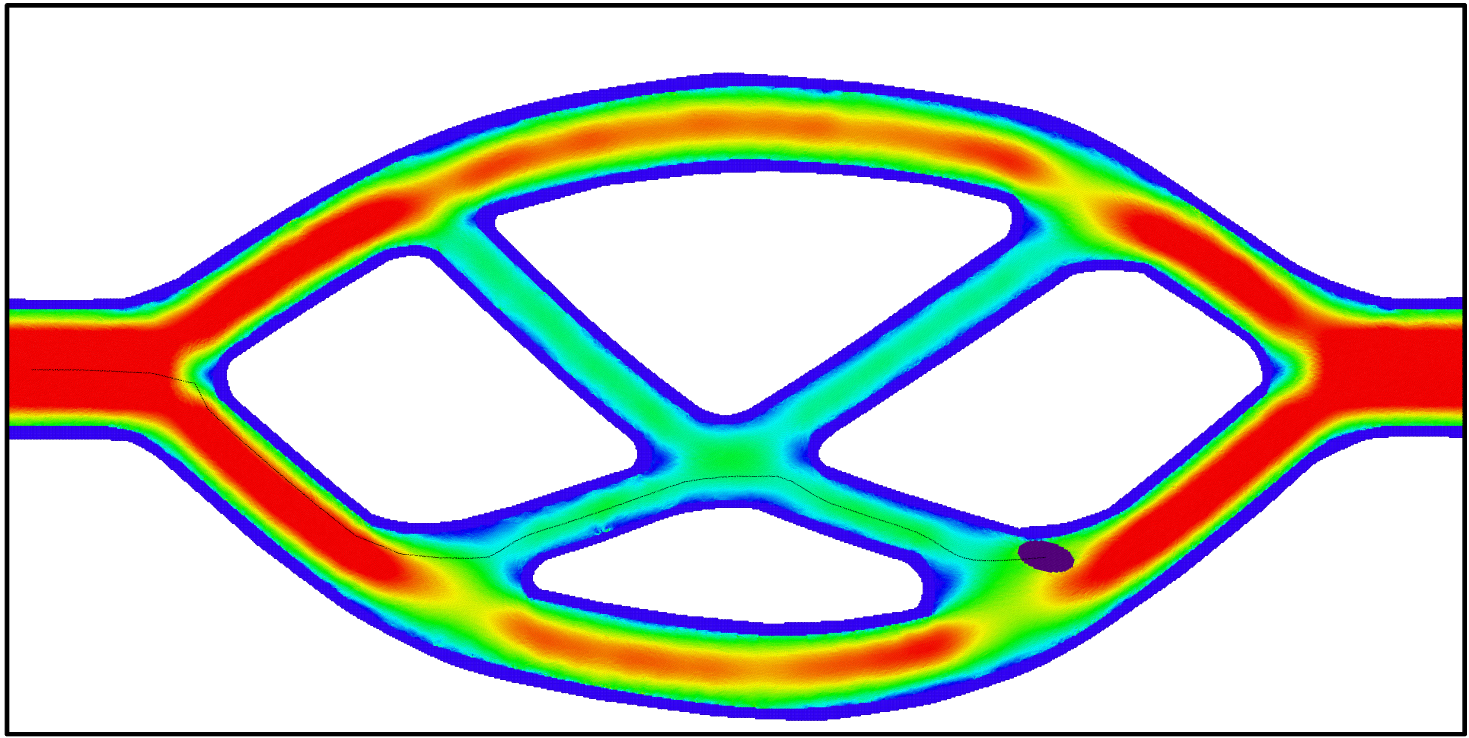}\label{sph_vessel5}}
\subfigure[$b^e = 0.2a_e$, $b^w=0.2L$]{\includegraphics[width=60mm]{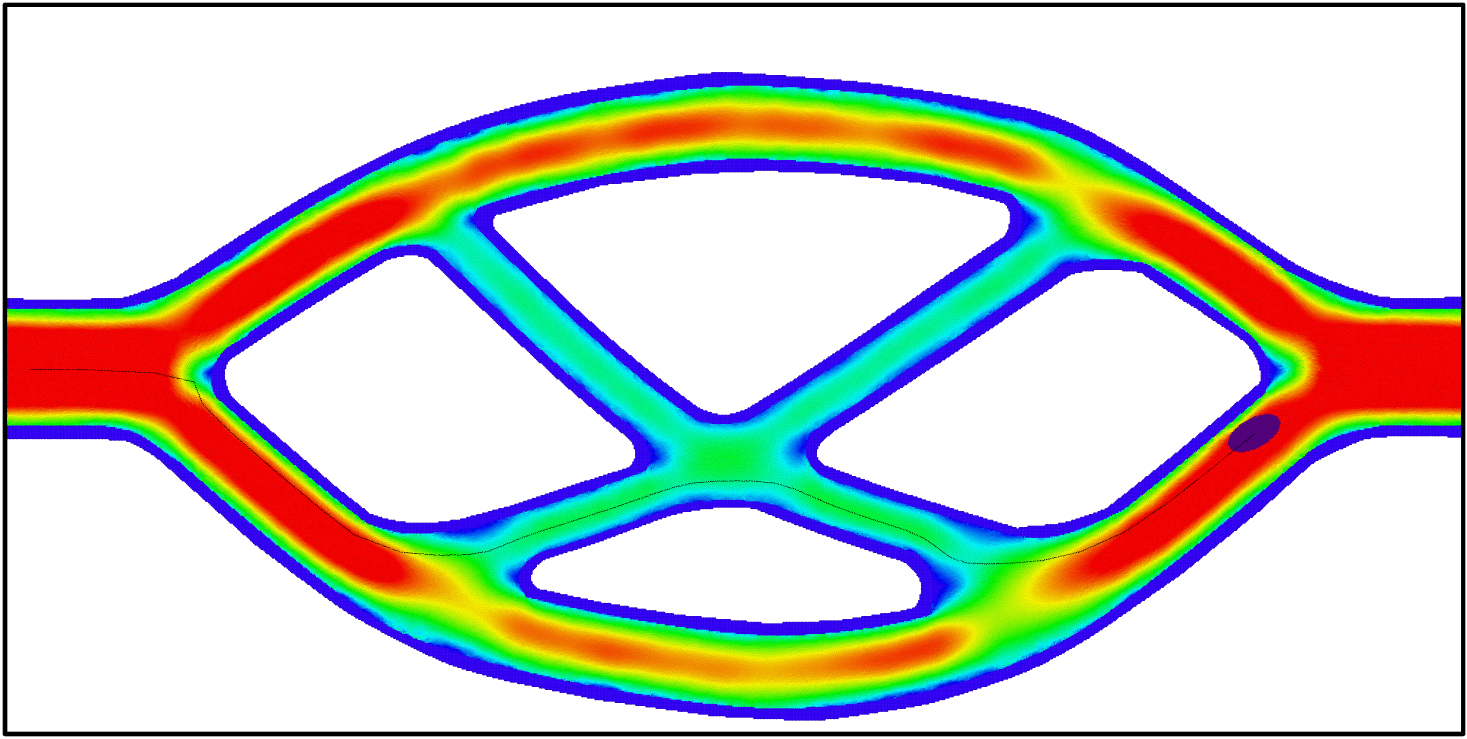}\label{sph_vessel6}}
\caption{Snapshots of an ellipse in microvessels at the same moment $t=0.225$ using SPH method: $b^e$ and $b^w$ represent the slip lengths of ellipse and wall, respectively. The black lines indicate the trajectories of the ellipse. }\label{sph_vessel}
\end{figure}

\begin{figure}[h!]
\centering 
\subfigure[$X$]{\includegraphics[width=70mm]{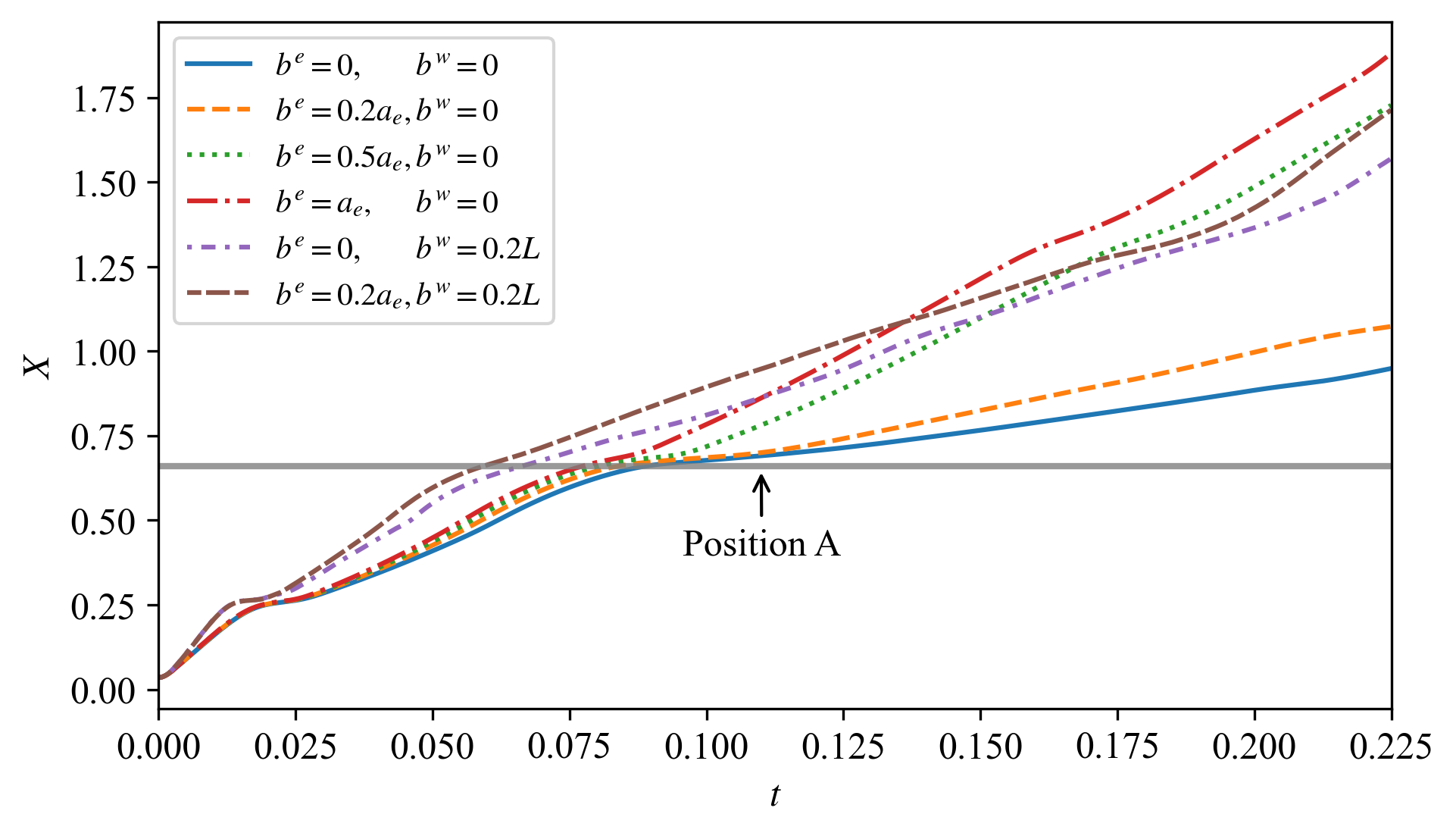}}
\subfigure[$cos\theta$]{\includegraphics[width=70mm]{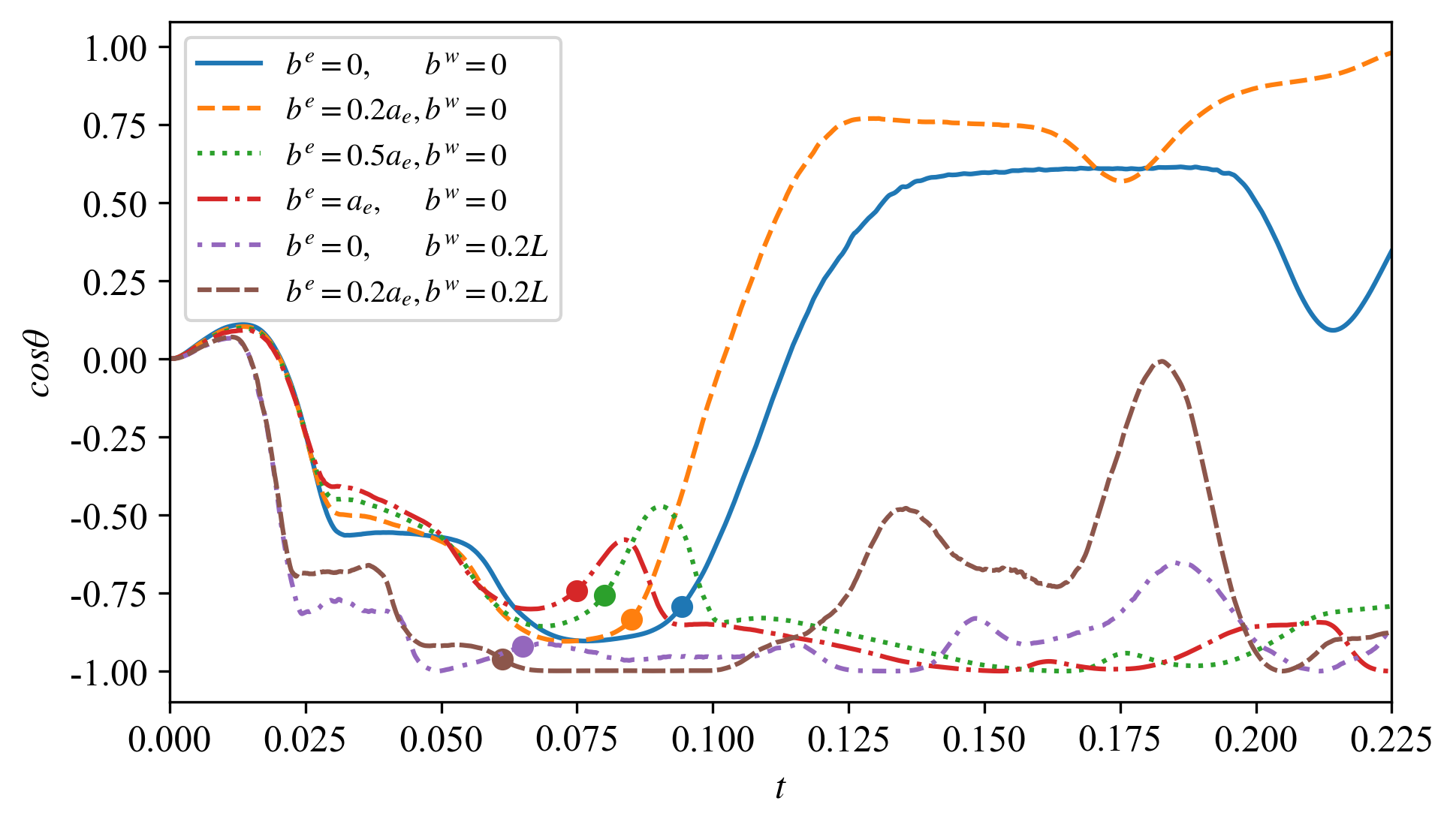}}
\caption{Dynamics of the elliptical cylinder in microvessels with different slip lengths at interfaces. Left: the distance travelled in the $x$-direction by the center of mass of the elliptical cylinder as function of time.
Right: cosine of the angle for the semi-major axis with respect to $x$-direction, as a function of time. At the initial moment, $X=0.035$ and the semi-major axis points to $y$-direction. The branching position A is $X = 0.66$, and the time at which the elliptical cylinder first passes ranges from $t = 0.6$ to $0.09$ approximately.
}\label{ellipse_X_cos_time}
\end{figure}

We present six snapshots at the same selective moment of $t=0.225$ of SPH simulations for various boundary conditions in Fig.~\ref{sph_vessel}. The color map indicates the velocity magnitude of the flow
and its maximum and minimum values are set to be $22800$ and $0$, respectively. 
We apply different slip lengths to the surfaces of the elliptical particle and wall, 
represented by $b^e$ and $b^w$, respectively. 
Among Figs. \ref{sph_vessel1}, \ref{sph_vessel2}, \ref{sph_vessel3}, and \ref{sph_vessel4},
we impose no-slip boundary on the vessel walls, but various slip lengths on the surface of the ellipse.
Although there are similar flow fields among these four cases,
the mobility of the ellipse is significantly different due to its distinct surface properties.
In particular, the ellipse with slip length of $b^e=a_e$ on its surface almost arrives at the outlet
while the one with no-slip surface of $b^e=0$ still tumbles around halfway.
Moreover, due to possibility of branching, the slip length on the surface of the ellipse may also alter its trajectory completely. 
For example, the ellipse has different orientations for
$b^e < 0.5a_e$ and  $b^e \ge 0.5 a_e$ at the branching position A, 
as shown in Fig.~\ref{ellipse_X_cos_time},
therefore, this lead to different trajectories at downstream.
When there is slip at the wall, the flow field is enhanced,
as shown in Figs. \ref{sph_vessel5} and \ref{sph_vessel6}.
Both the velocity and trajectory of the ellipse may be
altered as compared to the case of no-slip at interfaces in Fig.~\ref{sph_vessel1}.

\begin{figure}[h!]
\centering 
\subfigure[$b^e = 0.2a_e$, $b^w=0$]{\includegraphics[width=60mm]{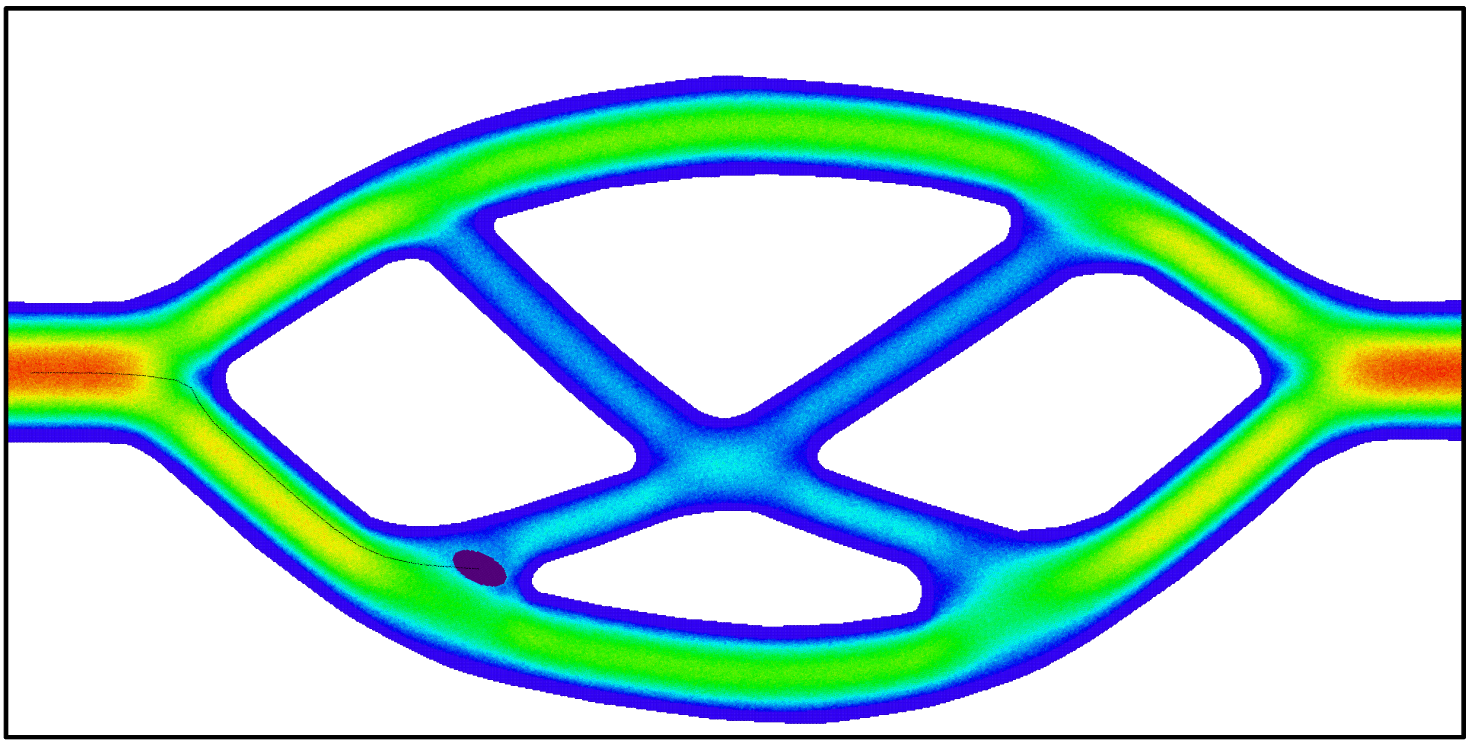}}
\subfigure[$b^e = 0.5a_e$, $b^w=0$]{\includegraphics[width=60mm]{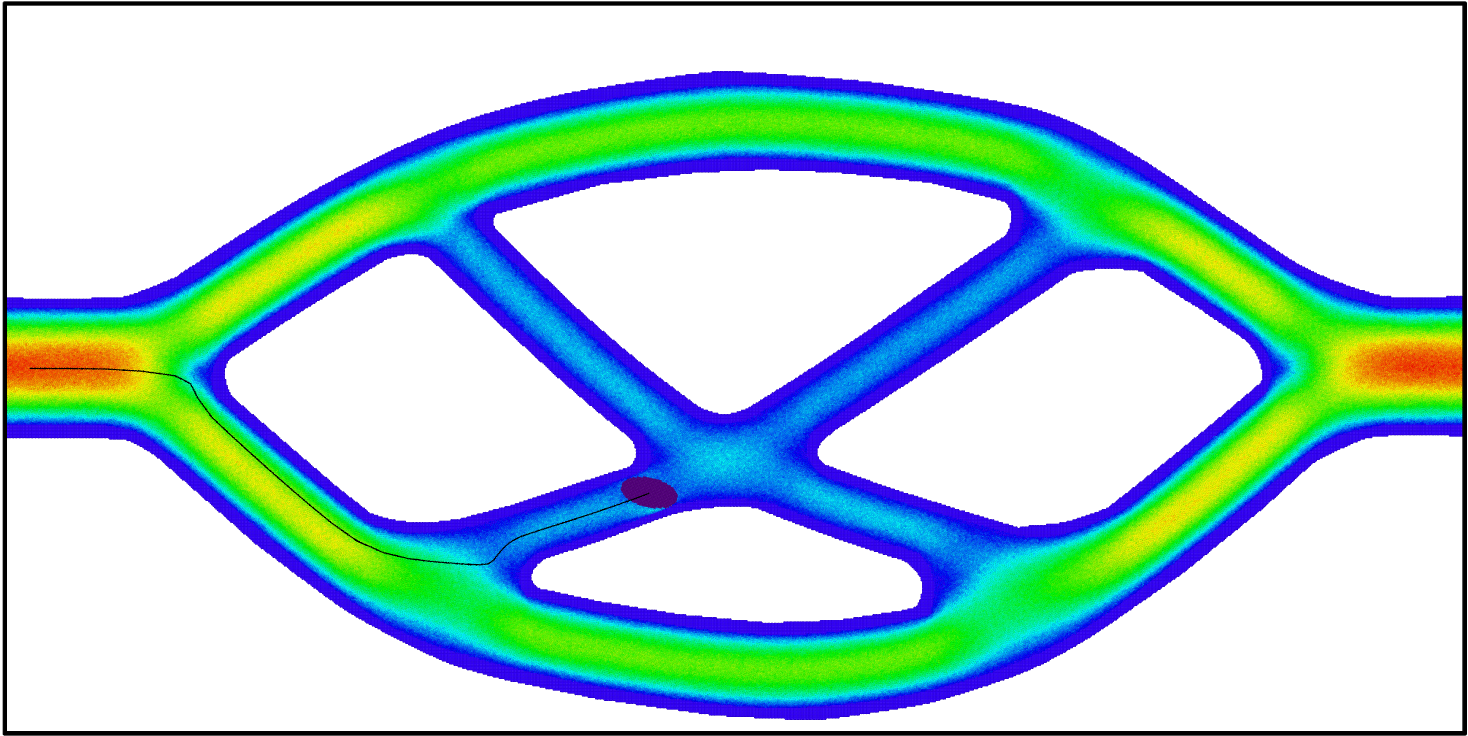}}
\caption{The snapshot of an ellipse in the blood vessel at the same moment $t=0.225$ using SDPD method}\label{sdpd_vessel}
\end{figure}

We further switch on the thermal fluctuations of the fluid in the same microvessels,
and results of SDPD are presented in Fig.~\ref{sdpd_vessel}.
Compared with results of SPH in Figs. \ref{sph_vessel2} and \ref{sph_vessel3}
for the same slip lengths, the Brownian motion of the ellipse
induces more diffusive trajectories and therefore,
leads to shorter distances traveled along the channel.

\section{Conclusions}
\label{section_conclusion}
In this work, we propose two algorithms to achieve an arbitrary slip length at fluid-solid interface of an arbitrary geometry in SPH and SDPD simulations: fluid-particle-centric method (FPC) and boundary-particle-centric (BPC) method.
The two algorithms concern the calculations of pairwise dissipative force and random force between fluid particles and boundary particles near the interface.
The FPC method treats a fluid particle as center and further determines an intersection point on the surface, where the desired slip length/velocity is prescribed.
Thereafter, artificial velocities for the interacting boundary particles are calculated.
The BPC method takes a boundary particle as center and further determines an intersection point on the surface, where the desired slip length/velocity is specified.
Correspondingly, a virtual particle with average position/velocity from the interacting fluid particles
is created and thereafter an artificial velocity for the boundary particle is calculated.
The FPC method recovers the work of Morris et al.~\cite{morris_modeling_1997}
for a no-slip boundary condition at static interfaces of simple geometry.
It is accurate, but requires frequent calculations for distances
of fluid particles from the interface.
It is an effortless task for interface of plane and other simple geometries such as sphere,
but becomes tedious and costly for a complex geometry.
The BPC method resembles the work of Adami et al.~\cite{morris_modeling_1997}
for a no-slip boundary condition.
By taking an average effects of the neighboring fluid particles around the boundary particle, this method appears to be less accurate than the former.
However, the BPC method only involves SPH interpolations and is readily accomplished
even for complex geometry.
After applying both algorithms in a series of flow problems in channels of various geometries, the BPC method has negligible errors in comparison with analytical solutions, references of finite difference/volume methods, and results of FPC method.

\section*{Acknowledgments}
X. Cai and X. Bian acknowledge the national natural science foundation of China under grant number: 12172330.
X. Bian also received the starting grant from 100 talents program of Zhejiang University.

\bibliographystyle{elsarticle-num}
\biboptions{sort&compress}
\bibliography{main}

\appendix

\section{Boundary volume fraction method}
\label{appendixB}
\begin{figure}[h!]
    \centering 
    \includegraphics[scale=0.5]{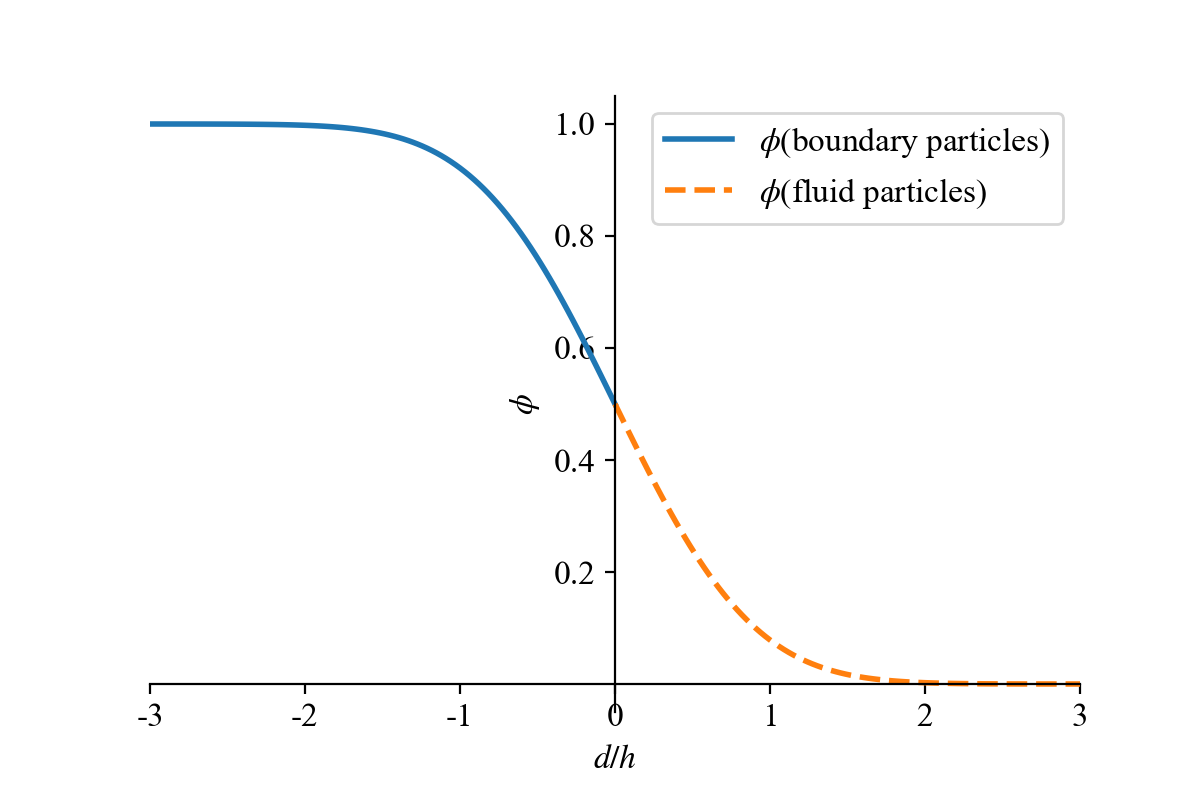} 
    \caption{Relation between the boundary volume fraction $\phi$ and $d/h$. We calculate $\phi$ of a particle by Eq.~(\ref{bvf_phi}) and then the distance from the particle to the interface by Eq.~(\ref{distance_boundary}). For a fluid particle, the value of $\phi$ ranges from $0-0.5$, and $d/h$ calculated by Eq.~(\ref{distance_boundary}) is positive; for a boundary particle, the value of $\phi$ ranges from $0.5-1$, so $d/h$ is negative.}
    \label{bvf}
\end{figure}

For an arbitrarily shaped solid object, the distance $d$ of a fluid/boundary particle to the interface can be calculated by exploiting the normalization properties of the SPH kernel~\citep{holmes2011smooth, li2018dissipative}. In this approach, a boundary volume fraction (BVF)~\citep{li2018dissipative} is defined as
\begin{eqnarray}
\phi_i = \frac{1}{\sigma_i}\sum_{j}W(\mathbf{r}-\mathbf{r}', h) \label{bvf_phi}.
\end{eqnarray}
where $i$, $j$ represent a centric particle and a neighboring boundary particle, and $\sigma_i$ is the number density of the centric particle defined in Eq.~(\ref{sigma_eq}). For both theoretical and computational convenience, we may adopt a Gaussian function as the kernel
\begin{eqnarray}
W(\mathbf{r}-\mathbf{r}', h)=C_D\frac{1}{h^D}e^{-\frac{(\mathbf{r} - \mathbf{r}')^2}{h^2}}
\end{eqnarray}
where $C_D = \frac{1}{\pi^{D/2}}$ is the normalization coefficient.
When the interface is flat or its radius of curvature is much larger than $r_c$, we can calculate $\phi$ by
\begin{eqnarray}
\phi_i = \varphi(d) = \int_{-\infty}^{+\infty}\int_d^{+\infty}W(r_{ij}, h)dxdy = \frac{1}{2}\operatorname{erfc}(\frac{d}{h}) \label{phi}.
\end{eqnarray}
Here, $\operatorname{erfc}(\frac{d}{h})$ is the complementary error function defined as
\begin{eqnarray}
\operatorname{erfc}(z) = \frac{2}{\sqrt{\pi}}\int_z^{+\infty}e^{-t^2}dt.
\end{eqnarray}
Furthermore, we can obtain an approximate expression for $\operatorname{erfc}(z)$~\citep{winitzki2008handy}
\begin{eqnarray}
\operatorname{erfc}(z) \approx 1 - \operatorname{sgn}(z){\sqrt  {1-\exp \left(-z^{2}{\frac  {4/\pi +az^{2}}{1+az^{2}}}\right)}},
\end{eqnarray}
and its inverse function
\begin{eqnarray}
\operatorname{erfc}^{-1}(z) \approx \operatorname{sgn}(1-z)\sqrt{\sqrt{(\frac{2}{a\pi}+\frac{\operatorname{ln}z(2-z)}{2})^2 - \frac{\operatorname{ln}z(2-z)}{a}} - (\frac{2}{a\pi}+\frac{\operatorname{ln}z(2-z)}{2})} \label{erfc_inv}.
\end{eqnarray}
Here, constant $a = 0.147$. 
In Eq.~(\ref{phi}), $\phi_i = 0$ when the distance from a particle $i$ to the interface is larger than $r_c$; $\phi_i = 0.5$ when the particle is on the interface. 
Given a particle configuration, we calculate $\phi_i$ by SPH interpolation in Eq.~(\ref{bvf_phi}), and then employ the inverse function of Eq.~(\ref{phi}) to calculate the distance of particle $i$ to the interface, i.e.
\begin{eqnarray}
d / h = \varphi^{-1}(\phi_i) \approx \operatorname{erfc}^{-1}(2\phi_i). \label{distance_boundary}
\end{eqnarray}
Fig.~(\ref{bvf}) shows the relationship between $d/h$ and $\phi_i$. Since the smooth length $h$ is constant in the whole simulation, $d$ only depends on the value of $\phi_i$.
In three dimensions, we follow the same idea and obtain
\begin{eqnarray}
\phi_i = \varphi(d) = \int_{d}^{+\infty}\int_0^{+\infty}2\pi W(r_{ij}, h)rdrdz = \frac{1}{2}\operatorname{erfc}(\frac{d}{h}), \label{phi_3d}
\end{eqnarray}
which has the same expression as in two dimensions.
Therefore, the distance $d$ is still calculated by Eqs.~(\ref{erfc_inv}) and (\ref{distance_boundary}).

\section{Rigid body dynamics}
\label{appendixA}
Translation and rotation of a rigid body can be described by 
the classical Newton's equation and Euler's equation as
\begin{eqnarray}
M\mathbf{\dot{U}} =\mathbf{F}\notag\\
\mathbf{I}\cdot \mathbf{\dot{\Omega}} + \mathbf{\Omega}\times[\mathbf{I}\cdot \mathbf{\Omega}] =\mathbf{T}
\end{eqnarray}
where $\mathbf{U}$, $\mathbf{\Omega}$ are translational and angular velocities for the center of mass, respectively. $M$ is the total mass and $\mathbf{I}$ is the inertial tensor. $\mathbf{F}$ and $\mathbf{T}$ represent the
total force and torque exerted on the rigid body, respectively.  As a rigid body is composed of particles in SPD simulations, the translations of particles can be calculated directly. 
For the rations,  we adopt a unit quaternion~\citep{miller2002symplectic}
\begin{equation}
\mathbf{q}=\{q_0,q_1,q_2,q_3\}
\end{equation}
with $\sum_iq_i=1$. The quaternion can represent the rotation matrix $\mathbf{A(q)}$ from a space-fixed ($\mathbf{r}$) to a body-fixed ( $\mathbf{r}{'}$) coordinate
\begin{eqnarray}
\mathbf{r}{'} &=& \mathbf{A(q)}\mathbf{r}\notag\\ 
\mathbf{r} &=& \mathbf{A(q)}^T\mathbf{r}{'}
\end{eqnarray}
where
\begin{eqnarray}
    \mathbf{A(q)}=\begin{pmatrix}
        q_0^2+q_1^2-q_2-q_3^2 & 2(q_1q_2+q_0q_3) & 2(q_1q_3-q_0q_2)\\
        2(q_1q_2-q_0q_3) & q_0^2-q_1^2+q_2^2-q_3^2 & 2(q_2q_3+q_0q_1) \\
        2(q_1q_3+q_0q_2) & 2(q_2q_3-q_0q_1) & q_0^2-q_1^2-q_2^2+q_3^2
    \end{pmatrix}
\end{eqnarray}
Here, the four components of the unit quaternion are
\begin{eqnarray}
    q_0&=&cos(\frac{\theta}{2})cos(\frac{\phi+\psi}{2})\notag\\
    q_1&=&sin(\frac{\theta}{2})cos(\frac{\phi-\psi}{2})\notag\\
    q_2&=&sin(\frac{\theta}{2})sin(\frac{\phi-\psi}{2})\notag\\
    q_3&=&cos(\frac{\theta}{2})sin(\frac{\phi+\psi}{2})\label{components_quat}
\end{eqnarray}
where ($\phi,\theta,\psi$) are three Euler angles. Therefore, Eq.~(\ref{components_quat}) represents the relationship between the unit quaternion and Euler angles. In the body-fixed coordinate the rigid body is first rotated $\phi$ about the $z$-axis, then $\theta$ about the $x'$-axis, and finally $\psi$ about the $z'$-axis.

In the body-fixed coordinate, the motion of quaternion and rotation are as follow
\begin{eqnarray}
    \begin{pmatrix}
        \dot{q_0}\\
        \dot{q_1}\\
        \dot{q_2}\\
        \dot{q_3}\\
        \dot{q_4}
    \end{pmatrix}
    &=&\begin{pmatrix}
        q_0 & -q_1 & -q_2 &-q_3\\
        q_1 & q_0 & -q_3 & q_2\\
        q_2 & q_3 & q_0 & -q_1\\
        q_3 & -q_2 & q_1 & q_0
    \end{pmatrix}
    \begin{pmatrix}
        0\\
        \Omega_x\\
        \Omega_y\\
        \Omega_z
    \end{pmatrix}\notag\\
    \dot{\Omega_x} &=& \frac{T_x}{I_{xx}}+\frac{I_{yy}-I_{zz}}{I_{xx}}\Omega_y\Omega_z \notag\\
    \dot{\Omega_y} &=& \frac{T_y}{I_{yy}}+\frac{I_{zz}-I_{xx}}{I_{yy}}\Omega_z\Omega_x \notag\\
    \dot{\Omega_z} &=& \frac{T_z}{I_{zz}}+\frac{I_{xx}-I_{yy}}{I_{zz}}\Omega_x\Omega_y
\end{eqnarray}

Here, $I_{xx}$, $I_{yy}$, $I_{zz}$ are the diagonal elements of the rigid inertia. $\Omega_x$, $\Omega_y$, $\Omega_z$ and $T_x$, $T_y$, $T_z$ are the components of angular velocity and torques in the body-fixed coordinate. 

At each time step, the force and torque on the rigid body are accumulated by the force and torque of each boundary particle constituting the solid. Afterwards, the physical quantities of the center of mass is updated first, and then the boundary particles are updated via the quaternion.

\section{Navier-Stokes equations for the transient Taylor-Couette flow}
\label{appendixC}
Taylor-Couette flow consists of two cylinders of different radii $R_1$ and $R_2$, both of which rotate around the same axis with angular velocities $\Omega_1$ and $\Omega_2$.
The NS equation for transient Taylor-Couette flow in cylindrical coordinates is
\begin{eqnarray}
\frac{\partial v_{\theta}}{\partial t} = \nu(\frac{\partial^2 v_{\theta}}{\partial r^2} + \frac{1}{r}\frac{\partial v_{\theta}}{\partial r}-\frac{v_{\theta}}{r^2}) \label{NSe_TaylorCouette}
\end{eqnarray}
where $v_{\theta}$ is the velocity component in $\theta$ direction. We define slip boundary conditions as follows
\begin{eqnarray}
    v_{\theta}-R_1\Omega_1 &=& b^{in}\frac{\partial v_{\theta}}{\partial \theta},\notag \\
    v_{\theta}-R_2\Omega_2 &=& -b^{out}\frac{\partial v_{\theta}}{\partial \theta}.\label{NSe_TaylorCouette_boundary}
\end{eqnarray}
In FDM, we solve Eq.~(\ref{NSe_TaylorCouette}) with second-order discrete accuracy in space using a distribution of $N=201$ spatial points.  We employ $v_{\theta,i}^t$ to denote the velocity of the discrete points at time $t$, where the index $i$ represents the label of the points in the $r$ direction. Then the discrete form is as follows
\begin{eqnarray}
    v_{\theta,i}^{t+1}=v_{\theta,i}^{t}+\Delta t \nu(\frac{v_{\theta,i+1}^t-2v_{\theta,i}^t+v_{\theta,i-1}^t}{{\Delta x}^2}+\frac{1}{r}\frac{v_{\theta,i+1}^t+v_{\theta,i-1}^t}{2\Delta x}-\frac{v_{\theta,i}^t}{r^2}),
\end{eqnarray}
where $1<i<N$. The discrete boundary conditions are
\begin{eqnarray}
    v_{\theta,i=1} &=& \frac{2r\Omega_1\Delta x+(4v_{\theta,i=2}-v_{\theta,i=3})b^{in}}{2\Delta x+3b^{in}},\notag \\
    v_{\theta,i=N} &=& \frac{2r\Omega_2\Delta x+(4v_{\theta,i=N-1}-v_{\theta,i=N-2})b^{out}}{2\Delta x+3b^{out}}.
\end{eqnarray}

\section{Navier-Stokes equations for the transient semi-circle channel flow}
\label{appendixD}
For the transient flow through channels described by semi-circle functions, we adopt the following NS equation in cylindrical coordinates:
\begin{eqnarray}
    \frac{\partial v_{\theta}}{\partial t} + v_r\frac{\partial v_{\theta}}{\partial r} + \frac{v_{\theta}}{r}(\frac{\partial v_{\theta}}{\partial \theta}+v_r) &=& F_{\theta} + \nu(\frac{\partial^2 v_{\theta}}{\partial r^2} + \frac{1}{r^2}\frac{\partial^2 v_{\theta}}{\partial \theta^2} + \frac{1}{r}\frac{\partial v_{\theta}}{\partial r}+\frac{2}{r^2}\frac{\partial v_r}{\partial \theta}-\frac{v_{\theta}}{r^2}) \notag \\
    \frac{\partial v_r}{\partial t} + v_r\frac{\partial v_r}{\partial r} + \frac{v_\theta}{r}(\frac{\partial v_r}{\partial \theta}-v_{\theta}) &=& \nu(\frac{\partial^2v_r}{\partial r^2}+\frac{1}{r^2}\frac{\partial^2v_r}{\partial \theta^2}+\frac{1}{r}\frac{\partial v_r}{\partial r}-\frac{2}{r^2}\frac{\partial v_{\theta}}{\partial \theta} - \frac{v_r}{r^2}) \label{NSe_semi-circle}
\end{eqnarray}
where $F_{\theta}$ denotes the body force parallel to the $\theta$ direction. Let $R_1$ and $R_2$ denote the inner radius and outer radius of the wall. Then we define slip boundary conditions and impermeable conditions as follows
\begin{eqnarray}
    v_{\theta} &=& b\frac{\partial v_{\theta}}{\partial \theta}, r=R_1 \notag \\
    v_{\theta} &=& -b\frac{\partial v_{\theta}}{\partial \theta}, r = R_2\notag \\
    v_r &=& 0, r = R_1, R_2
\end{eqnarray}
where $b$ represents the slip length on the wall. The inlet and outlet boundary conditions and periodic boundary conditions are as follow
\begin{eqnarray}
    v_{\theta}(r) &=& v_{\theta}(R_1+R_2-r), \theta = 0\notag \\
    v_{r}(r) &=& v_{r}(R_1+R_2-r), \theta = 0\notag \\
    v_{\theta}(\theta=\pi) &=& v_{\theta}({\theta=0})\notag \\
    v_{r}(\theta=\pi) &=& v_r(\theta=0)
\end{eqnarray}

In FDM, we take $N_r = 51$  and $N_{\theta} = 301$ points in the $r$ and $\theta$ directions, respectively. We employ $v_{\theta,i,j}^t$ and $v_{r,i,j}^t$ to denote the velocity of the discrete points at time $t$, where the index $i,j$ represent the labels of the points in the $r$ and $\theta$ directions, respectively. We adopt second-order discrete accuracy in space to solve Eq.~(\ref{NSe_semi-circle}) as follows
\begin{eqnarray}
    v_{\theta,i,j}^{t+1} &=& v_{\theta,i,j}^t + F_{\theta}\Delta t \notag\\
    &+& \nu(\frac{v_{\theta,i+1,j}^t-2v_{\theta,i,j}^t+v_{\theta,i-1,j}^t}{(\Delta r)^2} + \frac{1}{r^2}\frac{v_{\theta,i,j+1}^t-2v_{\theta,i,j}^t+v_{\theta,i,j-1}^t}{(\Delta \theta)^2}
    +\frac{1}{r}\frac{v_{\theta,i+1,j}^t-v_{\theta,i-1,j}^t}{2\Delta r} \notag\\
    &+&\frac{2}{r^2}\frac{v_{r,i,j+1}^t-v_{r,i,j-1}^t}{2\Delta \theta}
    -\frac{v_{\theta, i,j}^t}{r^2})\Delta t - v_{r,i,j}^t\frac{v_{\theta,i+1,j}^t-v_{\theta,i-1,j}^t}{2\Delta r}\Delta t-\frac{v_{\theta,i,j}^t}{r}(\frac{v_{r,i,j+1}^t-v_{r,i,j-1}^t}{2\Delta \theta}+v_{r,i,j}^t)\Delta t
\end{eqnarray}

\begin{eqnarray}
        v_{r,i,j}^{t+1} &=& v_{r,i,j}^t+\nu(\frac{v_{r,i+1,j}^t-2v_{r,i,j}^t+v_{r,i-1,j}^t}{(\Delta r)^2} + \frac{1}{r^2}\frac{v_{r,i,j+1}^t-2v_{r,i,j}^t+v_{r,i,j-1}^t}{(\Delta \theta)^2}+
    \frac{1}{r}\frac{v_{r,i+1,j}^t-v_{r,i-1,j}^t}{2\Delta r} \notag\\
    &-& \frac{2}{r^2}\frac{v_{\theta,i,j+1}^t-v_{\theta,i,j-1}^t}{2\Delta \theta}-
    \frac{v_r^t}{r^2})\Delta t - v_{r,i,j}^t\frac{v_{r,i+1,j}^t-v_{r,i-1,j}^t}{2\Delta r} \Delta t - \frac{v_{\theta,i,j}^t}{r}(\frac{v_{r,i,j+1}^t-v_{r,j-1}^t}{2\Delta \theta}-v_{\theta,i,j}^t)\Delta t
\end{eqnarray}
Here, the index $1<i<N_r$, $1<j<N_{\theta}$. For slip boundary and impermeable conditions, the discrete form are as follow
\begin{eqnarray}
    v_{\theta,i=1,j}^t &=& \frac{(4v_{\theta,i=2,j}^t-v_{\theta,i=3,j}^t)b}{2\Delta \theta+3b}\notag \\
        v_{\theta,i=N_r,j}^t &=& \frac{(4v_{\theta,i=N_r-1,j}^t-v_{\theta,i=N_r-2,j}^t)b}{2\Delta \theta+3b}\notag \\
    v_{r,i=0,j}^t&=&0\notag \\
    v_{r,i=N,j}^t &=&0 
\end{eqnarray}
For inlet and outlet boundary conditions and periodic conditions, we have
\begin{eqnarray}
    v_{\theta,i,j=1}^t&=&v_{\theta,i,j=1}^t(R_1+R_2-r)\notag\\
    v_{r,i,j=1}^t &=& v_{r,i,j=1}^t(R_1+R_2-r)\notag\\
    v_{\theta,i,j=N}^t&=&v_{\theta,i,j=1}^t\notag\\
    v_{r,i,j=N}^t &=& v_{\theta,i,j=N}^t
\end{eqnarray}

\end{document}